\documentclass[longauth]{aa} 

\usepackage{graphicx}
\usepackage{txfonts}
\usepackage{hyperref}
\usepackage{booktabs}
\usepackage{academicons}
\usepackage{xcolor}
\usepackage{fontawesome}
\usepackage{xcolor}
\usepackage[normalem]{ulem}
\usepackage{comment}
\usepackage{placeins}

\usepackage{chemformula}

\newcommand{\pyratbay}{\textsc{Pyrat Bay}}
\newcommand{\platon}{\textsc{platon}}
\DeclareSymbolFont{UPM}{U}{eur}{m}{n}
\DeclareMathSymbol{\umu}{0}{UPM}{"16}
\let\oldumu=\umu
\renewcommand\umu{\ifmmode\oldumu\else$\oldumu$\fi}

\newcommand{\micron}{{\umu}m}
\newcommand{\microns}{\micron}

\begin{document}

\title{The KELT-7b atmospheric thermal-inversion conundrum revisited with CHEOPS, TESS, and additional data\thanks{This article uses data from CHEOPS programmes CH\_PR110016 and CH\_PR100047.}}
\subtitle{}

\author{
Z.~Garai\inst{\ref{inst:1},\ref{inst:2},\ref{inst:3}}\,$^{\href{https://orcid.org/0000-0001-9483-2016}{\protect\includegraphics[height=0.19cm]{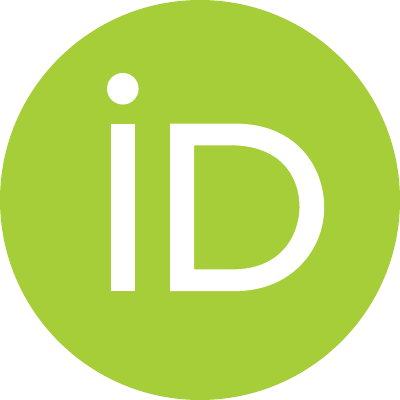}}}$, 
A.~Krenn\inst{\ref{inst:4}}\,$^{\href{https://orcid.org/0000-0003-3615-4725}{\protect\includegraphics[height=0.19cm]{KELT-7/orcid.pdf}}}$, 
P.~E.~Cubillos\inst{\ref{inst:4},\ref{inst:5}}, 
G.~Bruno\inst{\ref{inst:6}}\,$^{\href{https://orcid.org/0000-0002-3288-0802}{\protect\includegraphics[height=0.19cm]{KELT-7/orcid.pdf}}}$, 
A.~M.~S.~Smith\inst{\ref{inst:7}}\,$^{\href{https://orcid.org/0000-0002-2386-4341}{\protect\includegraphics[height=0.19cm]{KELT-7/orcid.pdf}}}$, 
T.~G.~Wilson\inst{\ref{inst:8}}\,$^{\href{https://orcid.org/0000-0001-8749-1962}{\protect\includegraphics[height=0.19cm]{KELT-7/orcid.pdf}}}$, 
A.~Brandeker\inst{\ref{inst:9}}\,$^{\href{https://orcid.org/0000-0002-7201-7536}{\protect\includegraphics[height=0.19cm]{KELT-7/orcid.pdf}}}$, 
M.~N.~Günther\inst{\ref{inst:10}}\,$^{\href{https://orcid.org/0000-0002-3164-9086}{\protect\includegraphics[height=0.19cm]{KELT-7/orcid.pdf}}}$, 
A.~Heitzmann\inst{\ref{inst:11}}\,$^{\href{https://orcid.org/0000-0002-8091-7526}{\protect\includegraphics[height=0.19cm]{KELT-7/orcid.pdf}}}$, 
L.~Carone\inst{\ref{inst:4}}, 
V.~Singh\inst{\ref{inst:6}}\,$^{\href{https://orcid.org/0000-0002-7485-6309}{\protect\includegraphics[height=0.19cm]{KELT-7/orcid.pdf}}}$, 
M.~Lendl\inst{\ref{inst:11}}\,$^{\href{https://orcid.org/0000-0001-9699-1459}{\protect\includegraphics[height=0.19cm]{KELT-7/orcid.pdf}}}$, 
O.~D.~S.~Demangeon\inst{\ref{inst:12},\ref{inst:13}}\,$^{\href{https://orcid.org/0000-0001-7918-0355}{\protect\includegraphics[height=0.19cm]{KELT-7/orcid.pdf}}}$, 
Y.~Alibert\inst{\ref{inst:14},\ref{inst:15}}\,$^{\href{https://orcid.org/0000-0002-4644-8818}{\protect\includegraphics[height=0.19cm]{KELT-7/orcid.pdf}}}$, 
R.~Alonso\inst{\ref{inst:16},\ref{inst:17}}\,$^{\href{https://orcid.org/0000-0001-8462-8126}{\protect\includegraphics[height=0.19cm]{KELT-7/orcid.pdf}}}$, 
J.~Asquier\inst{\ref{inst:10}}, 
T.~Bárczy\inst{\ref{inst:18}}\,$^{\href{https://orcid.org/0000-0002-7822-4413}{\protect\includegraphics[height=0.19cm]{KELT-7/orcid.pdf}}}$, 
D.~Barrado\inst{\ref{inst:19}}\,$^{\href{https://orcid.org/0000-0002-5971-9242}{\protect\includegraphics[height=0.19cm]{KELT-7/orcid.pdf}}}$, 
S.~C.~C.~Barros\inst{\ref{inst:12},\ref{inst:13}}\,$^{\href{https://orcid.org/0000-0003-2434-3625}{\protect\includegraphics[height=0.19cm]{KELT-7/orcid.pdf}}}$, 
W.~Baumjohann\inst{\ref{inst:4}}\,$^{\href{https://orcid.org/0000-0001-6271-0110}{\protect\includegraphics[height=0.19cm]{KELT-7/orcid.pdf}}}$, 
W.~Benz\inst{\ref{inst:15},\ref{inst:14}}\,$^{\href{https://orcid.org/0000-0001-7896-6479}{\protect\includegraphics[height=0.19cm]{KELT-7/orcid.pdf}}}$, 
N.~Billot\inst{\ref{inst:11}}\,$^{\href{https://orcid.org/0000-0003-3429-3836}{\protect\includegraphics[height=0.19cm]{KELT-7/orcid.pdf}}}$, 
L.~Borsato\inst{\ref{inst:20}}\,$^{\href{https://orcid.org/0000-0003-0066-9268}{\protect\includegraphics[height=0.19cm]{KELT-7/orcid.pdf}}}$, 
C.~Broeg\inst{\ref{inst:15},\ref{inst:14}}\,$^{\href{https://orcid.org/0000-0001-5132-2614}{\protect\includegraphics[height=0.19cm]{KELT-7/orcid.pdf}}}$, 
A.~Collier~Cameron\inst{\ref{inst:21}}\,$^{\href{https://orcid.org/0000-0002-8863-7828}{\protect\includegraphics[height=0.19cm]{KELT-7/orcid.pdf}}}$, 
A.~C.~M.~Correia\inst{\ref{inst:22}}\,$^{\href{https://orcid.org/0000-0002-8946-8579}{\protect\includegraphics[height=0.19cm]{KELT-7/orcid.pdf}}}$, 
Sz.~Csizmadia\inst{\ref{inst:7}}\,$^{\href{https://orcid.org/0000-0001-6803-9698}{\protect\includegraphics[height=0.19cm]{KELT-7/orcid.pdf}}}$, 
M.~B.~Davies\inst{\ref{inst:23}}\,$^{\href{https://orcid.org/0000-0001-6080-1190}{\protect\includegraphics[height=0.19cm]{KELT-7/orcid.pdf}}}$, 
M.~Deleuil\inst{\ref{inst:24}}\,$^{\href{https://orcid.org/0000-0001-6036-0225}{\protect\includegraphics[height=0.19cm]{KELT-7/orcid.pdf}}}$, 
A.~Deline\inst{\ref{inst:11}}, 
B.-O.~Demory\inst{\ref{inst:14},\ref{inst:15}}\,$^{\href{https://orcid.org/0000-0002-9355-5165}{\protect\includegraphics[height=0.19cm]{KELT-7/orcid.pdf}}}$, 
A.~Derekas\inst{\ref{inst:2},\ref{inst:51}}, 
B.~Edwards\inst{\ref{inst:25}}, 
J.~A.~Egger\inst{\ref{inst:15}}\,$^{\href{https://orcid.org/0000-0003-1628-4231}{\protect\includegraphics[height=0.19cm]{KELT-7/orcid.pdf}}}$, 
D.~Ehrenreich\inst{\ref{inst:11},\ref{inst:26}}\,$^{\href{https://orcid.org/0000-0001-9704-5405}{\protect\includegraphics[height=0.19cm]{KELT-7/orcid.pdf}}}$, 
A.~Erikson\inst{\ref{inst:7}}, 
J.~Farinato\inst{\ref{inst:20}}\,$^{\href{https://orcid.org/0000-0002-5840-8362}{\protect\includegraphics[height=0.19cm]{KELT-7/orcid.pdf}}}$, 
A.~Fortier\inst{\ref{inst:15},\ref{inst:14}}\,$^{\href{https://orcid.org/0000-0001-8450-3374}{\protect\includegraphics[height=0.19cm]{KELT-7/orcid.pdf}}}$, 
L.~Fossati\inst{\ref{inst:4}}\,$^{\href{https://orcid.org/0000-0003-4426-9530}{\protect\includegraphics[height=0.19cm]{KELT-7/orcid.pdf}}}$, 
M.~Fridlund\inst{\ref{inst:27},\ref{inst:28}}\,$^{\href{https://orcid.org/0000-0002-0855-8426}{\protect\includegraphics[height=0.19cm]{KELT-7/orcid.pdf}}}$, 
D.~Gandolfi\inst{\ref{inst:29}}\,$^{\href{https://orcid.org/0000-0001-8627-9628}{\protect\includegraphics[height=0.19cm]{KELT-7/orcid.pdf}}}$, 
K.~Gazeas\inst{\ref{inst:30}}, 
M.~Gillon\inst{\ref{inst:31}}\,$^{\href{https://orcid.org/0000-0003-1462-7739}{\protect\includegraphics[height=0.19cm]{KELT-7/orcid.pdf}}}$, 
M.~Güdel\inst{\ref{inst:32}}, 
Ch.~Helling\inst{\ref{inst:4},\ref{inst:33}}, 
K.~G.~Isaak\inst{\ref{inst:10}}\,$^{\href{https://orcid.org/0000-0001-8585-1717}{\protect\includegraphics[height=0.19cm]{KELT-7/orcid.pdf}}}$, 
F.~Kerschbaum\inst{\ref{inst:32}}\,$^{\href{https://orcid.org/0000-0001-6320-0980}{\protect\includegraphics[height=0.19cm]{KELT-7/orcid.pdf}}}$, 
L.~L.~Kiss\inst{\ref{inst:34},\ref{inst:35}}, 
J.~Korth\inst{\ref{inst:36}}\,$^{\href{https://orcid.org/0000-0002-0076-6239}{\protect\includegraphics[height=0.19cm]{KELT-7/orcid.pdf}}}$, 
K.~W.~F.~Lam\inst{\ref{inst:7}}\,$^{\href{https://orcid.org/0000-0002-9910-6088}{\protect\includegraphics[height=0.19cm]{KELT-7/orcid.pdf}}}$, 
J.~Laskar\inst{\ref{inst:37}}\,$^{\href{https://orcid.org/0000-0003-2634-789X}{\protect\includegraphics[height=0.19cm]{KELT-7/orcid.pdf}}}$, 
A.~Lecavelier~des~Etangs\inst{\ref{inst:38}}\,$^{\href{https://orcid.org/0000-0002-5637-5253}{\protect\includegraphics[height=0.19cm]{KELT-7/orcid.pdf}}}$, 
D.~Magrin\inst{\ref{inst:20}}\,$^{\href{https://orcid.org/0000-0003-0312-313X}{\protect\includegraphics[height=0.19cm]{KELT-7/orcid.pdf}}}$, 
P.~F.~L.~Maxted\inst{\ref{inst:39}}\,$^{\href{https://orcid.org/0000-0003-3794-1317}{\protect\includegraphics[height=0.19cm]{KELT-7/orcid.pdf}}}$, 
B.~Merín\inst{\ref{inst:40}}\,$^{\href{https://orcid.org/0000-0002-8555-3012}{\protect\includegraphics[height=0.19cm]{KELT-7/orcid.pdf}}}$, 
C.~Mordasini\inst{\ref{inst:15},\ref{inst:14}}, 
V.~Nascimbeni\inst{\ref{inst:20}}\,$^{\href{https://orcid.org/0000-0001-9770-1214}{\protect\includegraphics[height=0.19cm]{KELT-7/orcid.pdf}}}$, 
G.~Olofsson\inst{\ref{inst:9}}\,$^{\href{https://orcid.org/0000-0003-3747-7120}{\protect\includegraphics[height=0.19cm]{KELT-7/orcid.pdf}}}$,
R.~Ottensamer\inst{\ref{inst:32}}, 
I.~Pagano\inst{\ref{inst:6}}\,$^{\href{https://orcid.org/0000-0001-9573-4928}{\protect\includegraphics[height=0.19cm]{KELT-7/orcid.pdf}}}$, 
E.~Pallé\inst{\ref{inst:16},\ref{inst:17}}\,$^{\href{https://orcid.org/0000-0003-0987-1593}{\protect\includegraphics[height=0.19cm]{KELT-7/orcid.pdf}}}$, 
G.~Peter\inst{\ref{inst:41}}\,$^{\href{https://orcid.org/0000-0001-6101-2513}{\protect\includegraphics[height=0.19cm]{KELT-7/orcid.pdf}}}$, 
D.~Piazza\inst{\ref{inst:42}}, 
G.~Piotto\inst{\ref{inst:20},\ref{inst:43}}\,$^{\href{https://orcid.org/0000-0002-9937-6387}{\protect\includegraphics[height=0.19cm]{KELT-7/orcid.pdf}}}$, 
D.~Pollacco\inst{\ref{inst:8}}, 
D.~Queloz\inst{\ref{inst:44},\ref{inst:45}}\,$^{\href{https://orcid.org/0000-0002-3012-0316}{\protect\includegraphics[height=0.19cm]{KELT-7/orcid.pdf}}}$, 
R.~Ragazzoni\inst{\ref{inst:20},\ref{inst:43}}\,$^{\href{https://orcid.org/0000-0002-7697-5555}{\protect\includegraphics[height=0.19cm]{KELT-7/orcid.pdf}}}$, 
N.~Rando\inst{\ref{inst:10}}, 
H.~Rauer\inst{\ref{inst:7},\ref{inst:46}}\,$^{\href{https://orcid.org/0000-0002-6510-1828}{\protect\includegraphics[height=0.19cm]{KELT-7/orcid.pdf}}}$, 
I.~Ribas\inst{\ref{inst:47},\ref{inst:48}}\,$^{\href{https://orcid.org/0000-0002-6689-0312}{\protect\includegraphics[height=0.19cm]{KELT-7/orcid.pdf}}}$, 
N.~C.~Santos\inst{\ref{inst:12},\ref{inst:13}}\,$^{\href{https://orcid.org/0000-0003-4422-2919}{\protect\includegraphics[height=0.19cm]{KELT-7/orcid.pdf}}}$, 
G.~Scandariato\inst{\ref{inst:6}}\,$^{\href{https://orcid.org/0000-0003-2029-0626}{\protect\includegraphics[height=0.19cm]{KELT-7/orcid.pdf}}}$, 
D.~Ségransan\inst{\ref{inst:11}}\,$^{\href{https://orcid.org/0000-0003-2355-8034}{\protect\includegraphics[height=0.19cm]{KELT-7/orcid.pdf}}}$, 
A.~E.~Simon\inst{\ref{inst:15},\ref{inst:14}}\,$^{\href{https://orcid.org/0000-0001-9773-2600}{\protect\includegraphics[height=0.19cm]{KELT-7/orcid.pdf}}}$, 
S.~G.~Sousa\inst{\ref{inst:12}}\,$^{\href{https://orcid.org/0000-0001-9047-2965}{\protect\includegraphics[height=0.19cm]{KELT-7/orcid.pdf}}}$, 
M.~Stalport\inst{\ref{inst:49},\ref{inst:31}}, 
S.~Sulis\inst{\ref{inst:24}}\,$^{\href{https://orcid.org/0000-0001-8783-526X}{\protect\includegraphics[height=0.19cm]{KELT-7/orcid.pdf}}}$, 
Gy.~M.~Szabó\inst{\ref{inst:2},\ref{inst:1}}\,$^{\href{https://orcid.org/0000-0002-0606-7930}{\protect\includegraphics[height=0.19cm]{KELT-7/orcid.pdf}}}$, 
S.~Udry\inst{\ref{inst:11}}\,$^{\href{https://orcid.org/0000-0001-7576-6236}{\protect\includegraphics[height=0.19cm]{KELT-7/orcid.pdf}}}$, 
S.~Ulmer-Moll\inst{\ref{inst:11},\ref{inst:15},\ref{inst:49}}\,$^{\href{https://orcid.org/0000-0003-2417-7006}{\protect\includegraphics[height=0.19cm]{KELT-7/orcid.pdf}}}$, 
V.~Van~Grootel\inst{\ref{inst:49}}\,$^{\href{https://orcid.org/0000-0003-2144-4316}{\protect\includegraphics[height=0.19cm]{KELT-7/orcid.pdf}}}$, 
J.~Venturini\inst{\ref{inst:11}}\,$^{\href{https://orcid.org/0000-0001-9527-2903}{\protect\includegraphics[height=0.19cm]{KELT-7/orcid.pdf}}}$, 
E.~Villaver\inst{\ref{inst:16},\ref{inst:17}}, 
N.~A.~Walton\inst{\ref{inst:50}}\,$^{\href{https://orcid.org/0000-0003-3983-8778}{\protect\includegraphics[height=0.19cm]{KELT-7/orcid.pdf}}}$, 
S.~Wolf\inst{\ref{inst:42}}, 
D.~Wolter\inst{\ref{inst:7}}, and
T.~Zingales\inst{\ref{inst:43},\ref{inst:20}}\,$^{\href{https://orcid.org/0000-0001-6880-5356}{\protect\includegraphics[height=0.19cm]{KELT-7/orcid.pdf}}}$
}

\institute{
\label{inst:1} HUN-REN-ELTE Exoplanet Research Group, Szent Imre h. u. 112, H-9700 Szombathely, Hungary, \email{zgarai@gothard.hu} \and
\label{inst:2} ELTE Gothard Astrophysical Observatory, Szent Imre h. u. 112, H-9700 Szombathely, Hungary, \email{zgarai@gothard.hu} \and
\label{inst:3} Astronomical Institute, Slovak Academy of Sciences, 05960 Tatranská Lomnica, Slovakia, \email{zgarai@ta3.sk} \and
\label{inst:4} Space Research Institute, Austrian Academy of Sciences, Schmiedlstrasse 6, A-8042 Graz, Austria \and
\label{inst:5} INAF, Osservatorio Astrofisico di Torino, Via Osservatorio, 20, I-10025 Pino Torinese To, Italy \and
\label{inst:6} INAF, Osservatorio Astrofisico di Catania, Via S. Sofia 78, 95123 Catania, Italy \and
\label{inst:7} Institute of Planetary Research, German Aerospace Center (DLR), Rutherfordstrasse 2, 12489 Berlin, Germany \and
\label{inst:8} Department of Physics, University of Warwick, Gibbet Hill Road, Coventry CV4 7AL, United Kingdom \and
\label{inst:9} Department of Astronomy, Stockholm University, AlbaNova University Center, 10691 Stockholm, Sweden \and
\label{inst:10} European Space Agency (ESA), European Space Research and Technology Centre (ESTEC), Keplerlaan 1, 2201 AZ Noordwijk, The Netherlands \and
\label{inst:11} Observatoire astronomique de l'Université de Genève, Chemin Pegasi 51, 1290 Versoix, Switzerland \and
\label{inst:12} Instituto de Astrofisica e Ciencias do Espaco, Universidade do Porto, CAUP, Rua das Estrelas, 4150-762 Porto, Portugal \and
\label{inst:13} Departamento de Fisica e Astronomia, Faculdade de Ciencias, Universidade do Porto, Rua do Campo Alegre, 4169-007 Porto, Portugal \and
\label{inst:14} Center for Space and Habitability, University of Bern, Gesellschaftsstrasse 6, 3012 Bern, Switzerland \and
\label{inst:15} Space Research and Planetary Sciences, Physics Institute, University of Bern, Gesellschaftsstrasse 6, 3012 Bern, Switzerland \and
\label{inst:16} Instituto de Astrofísica de Canarias, Vía Láctea s/n, 38200 La Laguna, Tenerife, Spain \and
\label{inst:17} Departamento de Astrofísica, Universidad de La Laguna, Astrofísico Francisco Sanchez s/n, 38206 La Laguna, Tenerife, Spain \and
\label{inst:18} Admatis, 5. Kandó Kálmán Street, 3534 Miskolc, Hungary \and
\label{inst:19} Depto. de Astrofísica, Centro de Astrobiología (CSIC-INTA), ESAC campus, 28692 Villanueva de la Cañada (Madrid), Spain \and
\label{inst:20} INAF, Osservatorio Astronomico di Padova, Vicolo dell'Osservatorio 5, 35122 Padova, Italy \and
\label{inst:21} Centre for Exoplanet Science, SUPA School of Physics and Astronomy, University of St Andrews, North Haugh, St Andrews KY16 9SS, UK \and
\label{inst:22} CFisUC, Departamento de Física, Universidade de Coimbra, 3004-516 Coimbra, Portugal \and
\label{inst:23} Centre for Mathematical Sciences, Lund University, Box 118, 221 00 Lund, Sweden \and
\label{inst:24} Aix Marseille Univ, CNRS, CNES, LAM, 38 rue Frédéric Joliot-Curie, 13388 Marseille, France \and
\label{inst:25} SRON Netherlands Institute for Space Research, Niels Bohrweg 4, 2333 CA Leiden, Netherlands \and
\label{inst:26} Centre Vie dans l’Univers, Faculté des sciences, Université de Genève, Quai Ernest-Ansermet 30, 1211 Genève 4, Switzerland \and
\label{inst:27} Leiden Observatory, University of Leiden, PO Box 9513, 2300 RA Leiden, The Netherlands \and
\label{inst:28} Department of Space, Earth and Environment, Chalmers University of Technology, Onsala Space Observatory, 439 92 Onsala, Sweden \and
\label{inst:29} Dipartimento di Fisica, Università degli Studi di Torino, via Pietro Giuria 1, I-10125, Torino, Italy \and
\label{inst:30} National and Kapodistrian University of Athens, Department of Physics, University Campus, Zografos GR-157 84, Athens, Greece \and
\label{inst:31} Astrobiology Research Unit, Université de Liège, Allée du 6 Août 19C, B-4000 Liège, Belgium \and
\label{inst:32} Department of Astrophysics, University of Vienna, Türkenschanzstrasse 17, 1180 Vienna, Austria \and
\label{inst:33} Institute for Theoretical Physics and Computational Physics, Graz University of Technology, Petersgasse 16, 8010 Graz, Austria \and
\label{inst:34} Konkoly Observatory, Research Centre for Astronomy and Earth Sciences, 1121 Budapest, Konkoly Thege Miklós út 15-17, Hungary \and
\label{inst:35} ELTE E\"otv\"os Lor\'and University, Institute of Physics, P\'azm\'any P\'eter s\'et\'any 1/A, 1117 Budapest, Hungary \and
\label{inst:36} Lund Observatory, Division of Astrophysics, Department of Physics, Lund University, Box 118, 22100 Lund, Sweden \and
\label{inst:37} IMCCE, UMR8028 CNRS, Observatoire de Paris, PSL Univ., Sorbonne Univ., 77 av. Denfert-Rochereau, 75014 Paris, France \and
\label{inst:38} Institut d'astrophysique de Paris, UMR7095 CNRS, Université Pierre \& Marie Curie, 98bis blvd. Arago, 75014 Paris, France \and
\label{inst:39} Astrophysics Group, Lennard Jones Building, Keele University, Staffordshire, ST5 5BG, United Kingdom \and
\label{inst:40} European Space Agency, ESA - European Space Astronomy Centre, Camino Bajo del Castillo s/n, 28692 Villanueva de la Cañada, Madrid, Spain \and
\label{inst:41} Institute of Optical Sensor Systems, German Aerospace Center (DLR), Rutherfordstrasse 2, 12489 Berlin, Germany \and
\label{inst:42} Weltraumforschung und Planetologie, Physikalisches Institut, University of Bern, Gesellschaftsstrasse 6, 3012 Bern, Switzerland \and
\label{inst:43} Dipartimento di Fisica e Astronomia "Galileo Galilei", Università degli Studi di Padova, Vicolo dell'Osservatorio 3, 35122 Padova, Italy \and
\label{inst:44} ETH Zurich, Department of Physics, Wolfgang-Pauli-Strasse 2, CH-8093 Zurich, Switzerland \and
\label{inst:45} Cavendish Laboratory, JJ Thomson Avenue, Cambridge CB3 0HE, UK \and
\label{inst:46} Institut fuer Geologische Wissenschaften, Freie Universitaet Berlin, Maltheserstrasse 74-100,12249 Berlin, Germany \and
\label{inst:47} Institut de Ciencies de l'Espai (ICE, CSIC), Campus UAB, Can Magrans s/n, 08193 Bellaterra, Spain \and
\label{inst:48} Institut d'Estudis Espacials de Catalunya (IEEC), 08860 Castelldefels (Barcelona), Spain \and
\label{inst:49} Space sciences, Technologies and Astrophysics Research (STAR) Institute, Université de Liège, Allée du 6 Août 19C, 4000 Liège, Belgium \and
\label{inst:50} Institute of Astronomy, University of Cambridge, Madingley Road, Cambridge, CB3 0HA, United Kingdom \and
\label{inst:51} HUN-REN-SZTE Stellar Astrophysics Research Group, 6500, Baja, Szegedi \'ut, Kt. 766, Hungary
}

\date{Received September 15, 1996; accepted March 16, 1997}

\abstract
{Early theoretical works suggested that ultrahot Jupiters have inverted temperature-pressure (T-P) profiles in the presence of optical absorbers, such as TiO and VO. Recently, an inverted T-P profile of KELT-7b was detected, in agreement with the predictions. However, the diagnosis of T-P inversions has always been recognized to be a model-dependent process.}
{We used the Characterising Exoplanet Satellite (CHEOPS), the Transiting Exoplanet Survey Satellite (TESS), and additional literature data to characterize the atmosphere of KELT-7b, rederive the T-P profile, provide a precise measurement of the albedo of KELT-7b, and search for a possible distortion in the precise CHEOPS transit light curve of the planet.}
{We first jointly fitted the CHEOPS and TESS data and measured the occultation depths in these passbands. The CHEOPS transits were also fitted with a model including the gravity-darkening effect. Emission and absorption retrievals were performed to characterize the atmosphere of KELT-7b. The albedo of the planet was calculated in the CHEOPS and TESS passbands.}
{When adopting a thermochemical-equilibrium atmospheric composition, the emission retrievals return a non-inverted T-P profile, in contrast with previous results. When adopting a free-chemistry atmospheric parameterization, the emission retrievals return an inverted T-P profile with -- likely unphysically -- high concentrations of TiO and VO. The 3D general circulation model (GCM) supports a TiO-induced temperature inversion. We report for KELT-7b a very low geometric albedo of $A_\mathrm{g} = 0.05 \pm 0.06$, which is consistent with the heat distribution $\epsilon$ being close to zero and also consistent with a 3D GCM simulation, using magnetic drag ($\tau_\mathrm{drag}=10^4\,\mathrm{s}$). Based on the CHEOPS photometry, we are unable to place any meaningful constraint on the sky-projected orbital obliquity.} 
{The choice of a free-chemistry approach or a thermochemical-equilibrium chemistry is the main factor determining the retrieval results. Free-chemistry retrievals generally yield better fits; however, assuming free chemistry risks adopting unphysical scenarios for ultrahot Jupiters, such as KELT-7b. We applied a coherent stellar variability treatment on TESS and CHEOPS observations, commensurate with the known stellar activity of the host star. Other observations of KELT-7b would also benefit from a coherent stellar variability treatment.}


\titlerunning{The KELT-7b atmospheric conundrum revisited}
\authorrunning{Z. Garai et al.}
\maketitle

\section{Introduction}
\label{intro}

Hot Jupiters are giant gaseous planets with short orbital periods and high equilibrium temperatures ($T_\mathrm{eq}$). In particular, ultrahot Jupiters, which have hot ($T_\mathrm{eq} > 2000\,\mathrm{K}$) and extended atmospheres \citep{Bell1}, are ideal targets for both transmission and emission spectroscopy, owing to their atmospheric scale heights and brightness. Transmission spectroscopy, which measures the wavelength-dependent depth of the transit as the planet passes in front of its host star, is primarily sensitive to the atmospheric composition of the planet at the pressures of $\sim 0.1 - 1000\,\mathrm{mbar}$ along the day-night terminator \citep{Seager2, Brown1}. Transmission spectroscopy is also sensitive to clouds and high-altitude hazes, which can mask atmospheric absorption features by acting as a gray opacity source \citep{Fortney1, Helling2019}. This technique and its variants have been used to detect several atomic and molecular species in exoplanetary atmospheres, such as Na \citep{Charbonneau3, Snellen2, Redfield1, Nikolov1}, K \citep{Sing2, Wilson1}, TiO \citep{Sedaghati1}, H$_2$O \citep{Deming2, Kreidberg1}, CO \citep{Snellen3}, CO$_2$ \citep{Swain2, Madhusudhan3, JWST1}, CH$_4$ \citep{Swain1, Swain2, Bell2, Madhusudhan3}, NH$_3$ \citep{MacDonald1}, SO$_2$ \citep{Tsai1, Powell1, Dyrek1}, and H$_2$S \citep{Fu1}.   

We can also characterize the thermal emission spectra of transiting planets \citep{Barman1, Burrows1, Seager3} by measuring the wavelength-dependent depth of the secondary eclipse (occultation) as the planet passes behind its host star (see, e.g., \citealt{Kreidberg1, Stevenson1, Haynes1, Line1, Beatty2017, Evans1, Mikal-Evans1, Mansfield1, Edwards1}, or \citealt{Pluriel1}). Unlike transmission spectroscopy, which probes the atmosphere near the day-night terminator, these emission spectra tell us about the global properties of the planet's dayside atmosphere. They are sensitive to both the dayside composition and the vertical temperature-pressure (T-P) profile, which determines if the molecular absorption features are seen in absorption or emission. In terms of radiative transport only, the emergence of thermal inversions can be understood as being controlled by the ratio of opacity at visible wavelengths (which controls the depth to which incident flux penetrates) to the opacity at thermal infrared wavelengths (which controls the cooling of the planetary atmosphere). Generally, if the optical opacity is high at low pressure, leading to absorption of stellar flux high in the atmosphere, and the corresponding thermal infrared opacity is low, the upper atmosphere will have less efficient cooling, leading to elevated temperatures at low pressure \citep{Madhusudhan1}. 

Early theoretical works suggested that ultrahot Jupiters have inverted T-P profiles in the presence of optical absorbers, such as TiO and VO \citep{Hubeny1, Fortney2}. H$^-$ is also considered as a potential absorber, capable of causing thermal inversion (see, e.g., \citealt{Lothringer1}). The strong opacity of TiO and VO could result in thermal inversions, that is, rising temperatures with higher altitudes. Conversely, several studies explored mechanisms explaining why TiO might not play a role in the upper atmosphere \citep{Spiegel1, Madhusudhan2, Parmentier2013}. Thermal inversions were found, for example, in the ultrahot Jupiters WASP-33b \citep{Haynes1}, WASP-121b \citep{Evans1}, WASP-103b \citep{Kreidberg2}, KELT-9b \citep{Pino1}, and KELT-7b \citep{Pluriel1}. The ultrahot Jupiter WASP-12b is an exception; it shows no signs of TiO absorption or temperature inversion \citep{Sing3, Akin1}. This is in disagreement with the predictions, which emphasize the importance of such detections. Observations of hot Jupiters with $T_\mathrm{eq} < 2000\,\mathrm{K}$, WASP-43b \citep{Stevenson1} and HD\,209458b \citep{Line1, Santos1}, detected non-inverted T-P profiles, in agreement with the predictions of \citet{Hubeny1} and \citet{Fortney2}. However, the diagnosis of temperature inversions has always been recognized to be a model-dependent process \citep{Madhusudhan1}. 

Furthermore, the observed occultation depth can be translated into a geometric albedo if only reflected starlight is measured \citep{Heng1}, or if thermal emission contribution is taken into account \citep{Wong1, Wong2}. The geometric albedo $A_\mathrm{g}$ is a wavelength-dependent quantity, defined as the albedo of the planet at full phase \citep{Russell1, Seager1}. It determines how much starlight enters the planet's atmosphere without being reflected at its top. The first confirmed secondary-eclipse detections of the planets HD\,209458b \citep{Deming1} and TrES-1b \citep{Charbonneau2} were reported with the Spitzer Space Telescope \citep{Werner1}. Dedicated exoplanet space-based optical telescopes, this means, the Kepler space telescope \citep{Borucki1}, the Transiting Exoplanet Survey Satellite \citep[TESS;][]{Ricker1}, and the Characterising Exoplanet Satellite \citep[CHEOPS;][]{Benz1, Fortier1} were also frequently used to measure the secondary eclipses and geometric albedos of transiting exoplanets (see, e.g., \citealt{Heng1, Angerhausen1} or \citealt{Esteves2} in the case of Kepler, \citealt{Wong1, Wong2} in the case of TESS, and \citealt{Lendl1, Brandeker1, Hooton1, Deline1, Scandariato1} or \citealt{Krenn1} in the case of CHEOPS). The unprecedented precision of these observations reveals that hot Jupiters have, in general, low geometric albedos, which means $A_\mathrm{g} < 0.3$ \citep{Cowan1, Esteves1}, in some cases, $A_\mathrm{g} < 0.1$ \citep{Angerhausen1, Esteves2}. 

In this work, we selected the KELT-7 system \citep{Bieryla1} as a subject of our follow-up. KELT-7b has a short orbital period of $\sim 2.734\,\mathrm{d}$, a mass of $\sim 1.28\,\mathrm{M_{Jup}}$, and a radius of $\sim 1.496\,\mathrm{R_{Jup}}$. This exoplanet is an ultrahot Jupiter with $T_\mathrm{eq} = 2028 \pm 17\,\mathrm{K}$ \citep{Tabernero1} transiting a bright ($V$ = 8.54 mag) F2V-type star (see Sect. \ref{host} for further details). When discovered, this host was the fifth most massive, fifth hottest, and the ninth brightest star known to host a transiting planet. Therefore, KELT-7b is an ideal target for atmospheric characterization. Recent works also focus on the atmosphere of KELT-7b. \citet{Pluriel1} detect strong absorption features in the transmission spectrum indicative of $\mathrm{H_2O}$ and $\mathrm{H^-}$. On the other hand, the emission spectrum lacks strong absorption features. The analysis reveals temperature inversion. Later, \citet{Stangret1} searched for absorption features of a broad range of atomic and molecular species in a sample of six hot Jupiters based on high-resolution transmission spectroscopy. The nondetection in the case of KELT-7b is explained by stellar pulsations and the Rossiter-McLaughlin effect. Similarly, \citet{Tabernero1} are only able to determine upper limits of $0.08 - 1.4$\% on the presence of H$\alpha$, Li\,I, Na\,I, Mg\,I, and Ca\,II.    

\begin{table*}
\centering
\caption{Log of CHEOPS photometric observations of KELT-7 used in this work.}
\label{cheopsobslog}
\begin{tabular}{ccccccc}
\hline
\hline
Visit  	& Start date 			& End date 			      & File  							  & Efficiency  		  & RMS                    & Number\\
No.    	& [UTC]      			& [UTC]    				  & key   							  & [\%]         		  & [ppm]                  & of frames\\   
\hline
\multicolumn{7}{c}{Occultations}\\
1 		& 2021-10-16 14:33 		& 2021-10-17 03:10 		& \texttt{CH\_PR100016\_TG013901}     & 54.0                  & 270 	               & 945\\
2       & 2021-10-27 12:14      & 2021-10-28 01:39      & \texttt{CH\_PR100016\_TG013902}     & 57.6                  & 262                    & 1071\\
3       & 2021-10-30 05:13      & 2021-10-30 18:28      & \texttt{CH\_PR100016\_TG013903}     & 51.3                  & 348                    & 941\\
4       & 2022-01-01 02:48      & 2022-01-01 14:25      & \texttt{CH\_PR100016\_TG013904}     & 58.5                  & 264                    & 941\\
5       & 2022-01-03 20:26      & 2022-01-04 08:03      & \texttt{CH\_PR100016\_TG013905}     & 56.4                  & 264                    & 907\\
6       & 2022-01-09 08:15      & 2022-01-09 18:54      & \texttt{CH\_PR100016\_TG013906}     & 59.8                  & 277                    & 882\\
7       & 2022-01-20 06:50      & 2022-01-20 18:27      & \texttt{CH\_PR100016\_TG013907}     & 55.8                  & 323                    & 897\\
8       & 2022-02-02 22:55      & 2022-02-03 12:24      & \texttt{CH\_PR100016\_TG013908}     & 54.6                  & 332                    & 1021\\
9       & 2022-12-03 13:02      & 2022-12-04 01:29      & \texttt{CH\_PR100016\_TG015801}     & 60.3                  & 264                    & 1039\\
10      & 2022-12-30 21:00      & 2022-12-31 07:50      & \texttt{CH\_PR100016\_TG016201}     & 61.2                  & 325                    & 918\\
11      & 2023-01-08 02:49      & 2023-01-08 13:37      & \texttt{CH\_PR100016\_TG016202}     & 61.3                  & 254                    & 918\\
12      & 2023-01-18 23:40      & 2023-01-19 12:27      & \texttt{CH\_PR100016\_TG016203}     & 54.4                  & 272                    & 963\\
\hline
\multicolumn{7}{c}{Transits}\\
13      & 2021-11-08 18:16      & 2021-11-09 05:42      & \texttt{CH\_PR110047\_TG000501}     & 56.8                  & 340                    & 837\\
14      & 2022-10-27 13:07      & 2022-10-28 02:23      & \texttt{CH\_PR110047\_TG000502}     & 56.8                  & 421                    & 970\\
\hline
\end{tabular}
\tablefoot{The table lists the time interval of individual observations (time notation follows the ISO-8601 convention), the file key, which supports fast identification of the observations in the CHEOPS archive, the efficiency, which is the ratio between the amount of science observing time available during a visit and the total amount of time in a visit, the point-to-point root mean square (RMS) of the \texttt{DRP}-processed light curves with an aperture radius of 24 pixels, and the number of obtained frames.}
\end{table*}

Furthermore, the rapid rotation of the host star, with $v\sin I_\mathrm{s} = 71.4 \pm 0.2\,\mathrm{km.s^{-1}}$ \citep{Tabernero1}, makes this system even more interesting. The rapid rotation at early-type stars leads to an oblate shape of the star and induces an equator-to-pole gradient in the effective temperature, called gravity darkening \citep{Zeipel1, Zeipel2}. The so-called von Zeipel theorem predicts that the flux emitted from the surface is proportional to the local effective gravity, and thus the effect induces cooler temperatures at a rapidly rotating star’s equator and hotter temperatures at the poles. If an exoplanet transits a rapidly rotating star, distorted transit light curves are expected, as was predicted by \citet{Barnes1}. If such asymmetries are measured (see, e.g., \citealt{Szabo1, Lendl1, Hooton1, Deline1} or \citealt{Jones1}), this can be used to determine the sky-projected angle $\lambda$ between the stellar rotational axis and the planet orbit normal; that is, we can detect the spin-orbit misalignment. In addition, the stellar inclination $I_\mathrm{s}$ can be derived, and thus the true misalignment $\Psi$ is possible to obtain. 

We observed KELT-7b photometrically using the CHEOPS space observatory. In addition, we used TESS photometric data and literature data (see Sect. \ref{obsdatared}). We aim to characterize the atmosphere of the planet mainly via emission spectroscopy, to provide a precise measurement of the albedo of KELT-7b, and to search for a possible distortion in the CHEOPS transit light curve of the planet. The paper is organized as follows. In Sect. \ref{obsdatared}, we provide a brief description of observations and data reduction. In Sect. \ref{host}, we summarize the most important stellar parameters based on \citet{Tabernero1}. The data analysis, including light-curve fitting, secondary eclipse detection, and search for transit asymmetry, is described in Sect. \ref{dataan}. Atmosphere modeling of KELT-7b is the subject of Sect. \ref{atmosection}. In Sect. \ref{discuss}, we discuss the results of the atmosphere modeling. We conclude with the results in Sect. \ref{concl}.       

\section{Observations and data reduction}
\label{obsdatared}

\subsection{CHEOPS data}
\label{cheopsdata}

CHEOPS performed a total of 14 observations (visits) of KELT-7 between October 2021 and January 2023 (see Table \ref{cheopsobslog} for further details). The secondary eclipse observations were performed within CHEOPS programme CH\_PR110016, while the transits of KELT-7b were observed under programme CH\_PR100047. The CHEOPS observations are available as subarray data products \citep{Benz1} at a cadence equal to the exposure time (26.0\,s for occultation and 28.0\,s for transit observations). The subarrays contain a circular region around the target with a radius of 100 pixels. Aperture photometry is available for the subarrays via the official CHEOPS \texttt{Data Reduction Pipeline} \citep[\texttt{DRP};][]{Hoyer2020}. It performs several image corrections, including bias, dark, and flat corrections, contamination estimation, and background-star correction. We processed all CHEOPS observations with the \texttt{DRP} version 14.1.2 using an aperture radius of 24 pixels. 

In general, CHEOPS observations are affected by instrumental noise such as stray light from the Earth and the Moon (Moon glint), smearing effects, or spacecraft jitter. The flux measurements usually show a particularly strong correlation with the spacecraft roll angle \citep[see, e.g.,][]{Lendl1, Bonfanti2021}. The spacecraft rotates around itself exactly once every orbit. Therefore, the roll-angle parameter is directly linked to the orbital position of the spacecraft. Instrumental noise must be accounted for during the data analysis to identify the transit and occultation signals of the planet (see Sect. \ref{dataan}). Before performing the data analysis, we removed all points flagged by the \texttt{DRP}; this includes those points contaminated, for example, by cosmic rays. We also removed points with peculiarly high backgrounds by removing any points with a background larger than four times the median background value, as well as points with unusually high pointing offsets by removing all points with a centroid offset of more than 1\,pixel. Finally, we also removed points with a smearing estimate larger than $3 \times 10^{-5}$. In the case of the occultation data, we also performed sigma clipping and removed all points with median absolute deviation (MAD) higher than four to discard outliers.

\subsection{TESS data}
\label{tessdata}

In this work, we used TESS photometric data of KELT-7 from Sectors 19, 43, 44, 45, 59, 71, and 73 at 2\,min cadence. The TESS photometric baseline, therefore, spans from November 2019 to January 2024 (see Table \ref{TESSobslog} for further details). In the case of Sectors 59, 71, and 73, there are also 20\,s cadence data available, which, however, were not used in this work. In our analysis, we used the Pre-search Data Conditioning Simple Aperture Photometry (PDCSAP) flux, provided by the Science Processing Operations Center (SPOC) pipeline \citep{Smith2012, Stumpe1, Stumpe2014, 2016SPIE.9913E..3EJ}. Contrary to the Simple Aperture Photometry (SAP) flux, the PDCSAP light curve product has long-term trends removed from the data using Co-trending Basis Vectors (CBVs). The pipeline attempts to remove systematic artifacts while keeping planetary transits intact. Therefore, PDCSAP flux has fewer systematic trends and is specifically intended for detecting exoplanets. In principle, the PDCSAP light curve product should also be corrected for light dilution. The data were downloaded from the Mikulski Archive for Space Telescopes\footnote{\url{https://mast.stsci.edu/portal/Mashup/Clients/Mast/Portal.html}} (MAST). The average uncertainty of the $109\,739$ data points is 400\,ppm. We did not apply outlier removal in the TESS dataset.  

\begin{table}
\centering
\caption{Log of TESS photometric observations of KELT-7 used in this work.}
\label{TESSobslog}
\begin{tabular}{cccc}
\hline
\hline
Sector & Start date & End date     & Number of PDCSAP\\
No.    & [UTC]      & [UTC]        & data points\\
\hline
19 & 2019-11-27 & 2019-12-24 & $16\,741$\\
43 & 2021-09-16 & 2021-10-12 & $16\,268$\\
44 & 2021-10-12 & 2021-11-06 & $16\,244$\\
45 & 2021-11-06 & 2021-12-02 & $16\,149$\\
59 & 2022-11-26 & 2022-12-23 & $16\,734$\\
71 & 2023-10-16 & 2023-11-11 & $15\,114$ \\
73 & 2023-12-07 & 2024-01-03 & $12\,489$\\
\hline
Total & -- & -- & $109\,739$\\
\hline
\end{tabular}
\end{table}

\subsection{Additional data}
\label{archivedata}

During the analysis, we also used the following literature data. \citet{Martioli1} present near-infrared high-precision photometric observations of secondary eclipses for eight transiting hot Jupiters, including KELT-7b. The observations were carried out using the Wide-field InfraRed Camera (WIRCam) instrument \citep{Puget1} installed on the Canada-France-Hawaii Telescope (CFHT). KELT-7b was observed using the $K_\mathrm{cont}$ filter (at a wavelength of $\lambda \sim 2.2\,\mu\mathrm{m}$). The observations reveal an occultation depth of $D_\mathrm{occ, CFHT} = 400 \pm 120\,\mathrm{ppm}$. However, we later discarded this data point from the analysis, because it was very probably inconsistent with any tested model (see Fig. \ref{3DEmisison}). Unfortunately, ground-based observations are often affected by strong correlated noise (see, e.g., \citealt{Hooton2}). \citet{Garhart1} report transit depth and occultation depth measurements for a sample of 36 transiting hot Jupiters observed at $\lambda \sim 3.6\,\mu\mathrm{m}$ and $\lambda \sim 4.5\,\mu\mathrm{m}$ using the Spitzer space telescope \citep{Werner1} and its InfraRed Array Camera (IRAC) instrument \citep{Fazio1}. For KELT-7b, the authors report transit depths of $D_\mathrm{tra, Spitzer, 3.6} = 7925 \pm 62\,\mathrm{ppm}$ and $D_\mathrm{tra, Spitzer, 4.5} = 8092 \pm 36\,\mathrm{ppm}$. The observed occultation depths of KELT-7b are $D_\mathrm{occ, Spitzer, 3.6} = 1688 \pm 46\,\mathrm{ppm}$ and $D_\mathrm{occ, Spitzer, 4.5} = 1896 \pm 57\,\mathrm{ppm}$. Finally, we also used literature transit depth and occultation depth data obtained from 25 spectral bins ($\lambda \sim 1.12 - 1.63\,\mu\mathrm{m}$) using the Wide Field Camera 3 (WFC3) instrument \citep{Leckrone1} installed on the Hubble Space Telescope (HST), which were published by \citet{Pluriel1}.

\section{The planet's host star}
\label{host}

Properties of the planet's host star, KELT-7, were obtained by \citet{Tabernero1} only recently. The authors employed the \texttt{SteParSyn} code\footnote{\url{https://github.com/hmtabernero/SteParSyn}} \citep{Tabernero2} to retrieve the stellar atmospheric parameters and their associated uncertainties. Using these stellar parameters, they calculated the age of the star, its mass, and its radius with the PARAM web interface\footnote{\url{http://stev.oapd.inaf.it/cgi-bin/param}} \citep{daSilva1}, and the PARSEC stellar evolutionary tracks and isochrones \citep{Bressan1}. During the joint fit and eclipse detection (see Sect. \ref{sececl}), we used stellar parameters derived by these authors, including the stellar radius $R_\mathrm{s} = 1.712 \pm 0.037\,\mathrm{R_\odot}$, the stellar mass $M_\mathrm{s} = 1.517 \pm 0.022\,\mathrm{M_\odot}$, the effective temperature $T_\mathrm{eff} = 6699 \pm 24\,\mathrm{K}$, the surface gravity $\log\,g = 4.15 \pm 0.09\,\mathrm{dex}$, and metallicity [Fe/H] = $0.24 \pm 0.02\,\mathrm{dex}$. The age of the star is estimated to be $1.2 \pm 0.7\,\mathrm{Gyr}$. Further stellar parameters are listed in Table\,1 in \citet{Tabernero1}. 

\section{Data analysis}
\label{dataan}

\subsection{CHEOPS instrumental noise}
\label{secinstr}
CHEOPS flux measurements are known to often show a strong correlation with the spacecraft roll angle (see Sect. \ref{cheopsdata}). However, in the case of the KELT-7 CHEOPS observations, we do not find such a strong correlation. For this reason, we refrained from using a more elaborate roll-angle correction and only used a first-order linear model of the sine and cosine of the roll-angle parameter to remove roll-angle-related trends. We also added a first-order linear detrending model on the offsets of the x- and y-centroid positions relative to their mean values. All of the linear detrending vectors were fitted simultaneously with the astrophysical model. 

\subsection{Stellar activity}
\label{stellaractivity}

\begin{figure*}
\centering
\centerline{
\includegraphics[width=0.66\columnwidth]{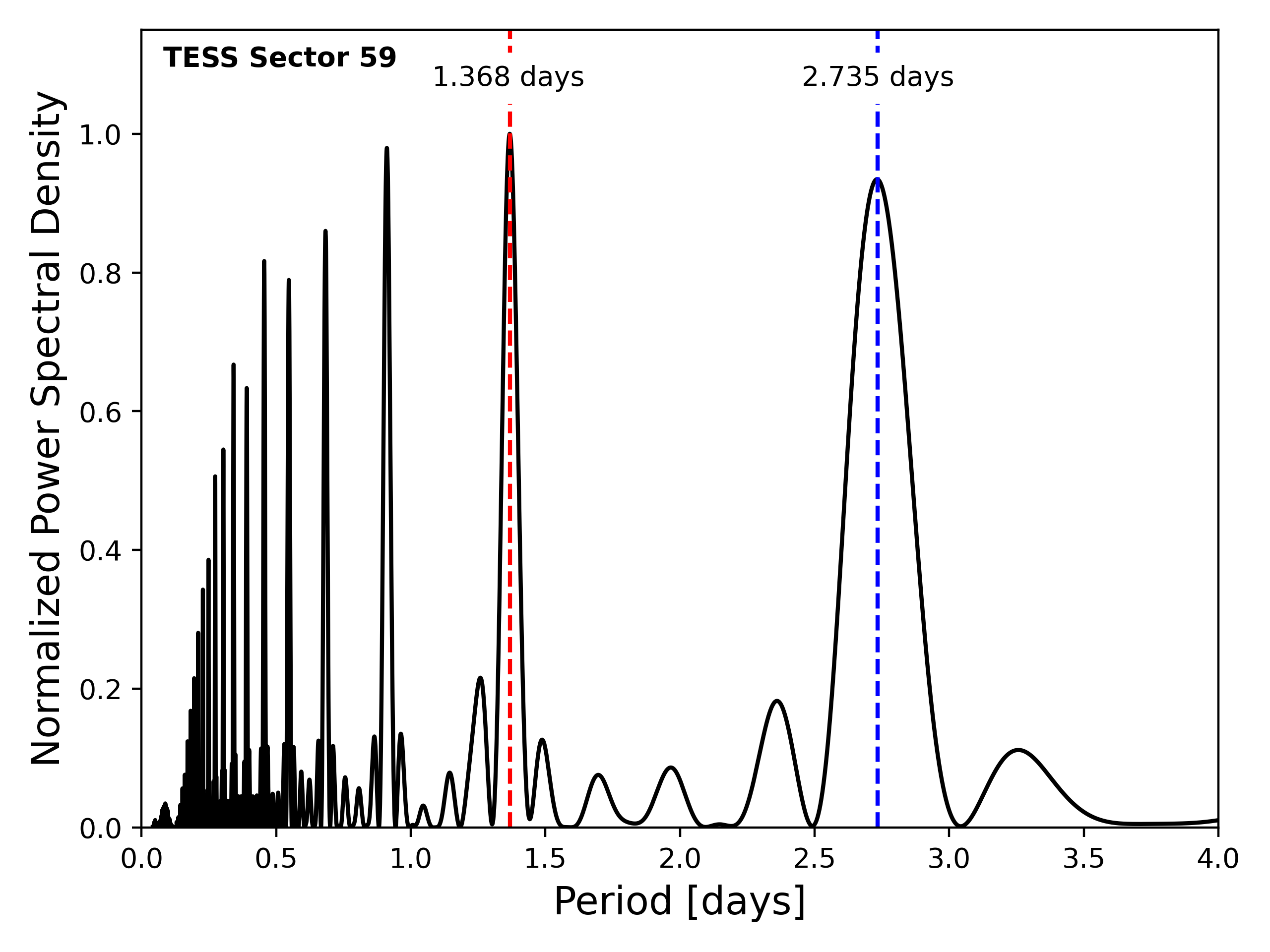}
\includegraphics[width=0.66\columnwidth]{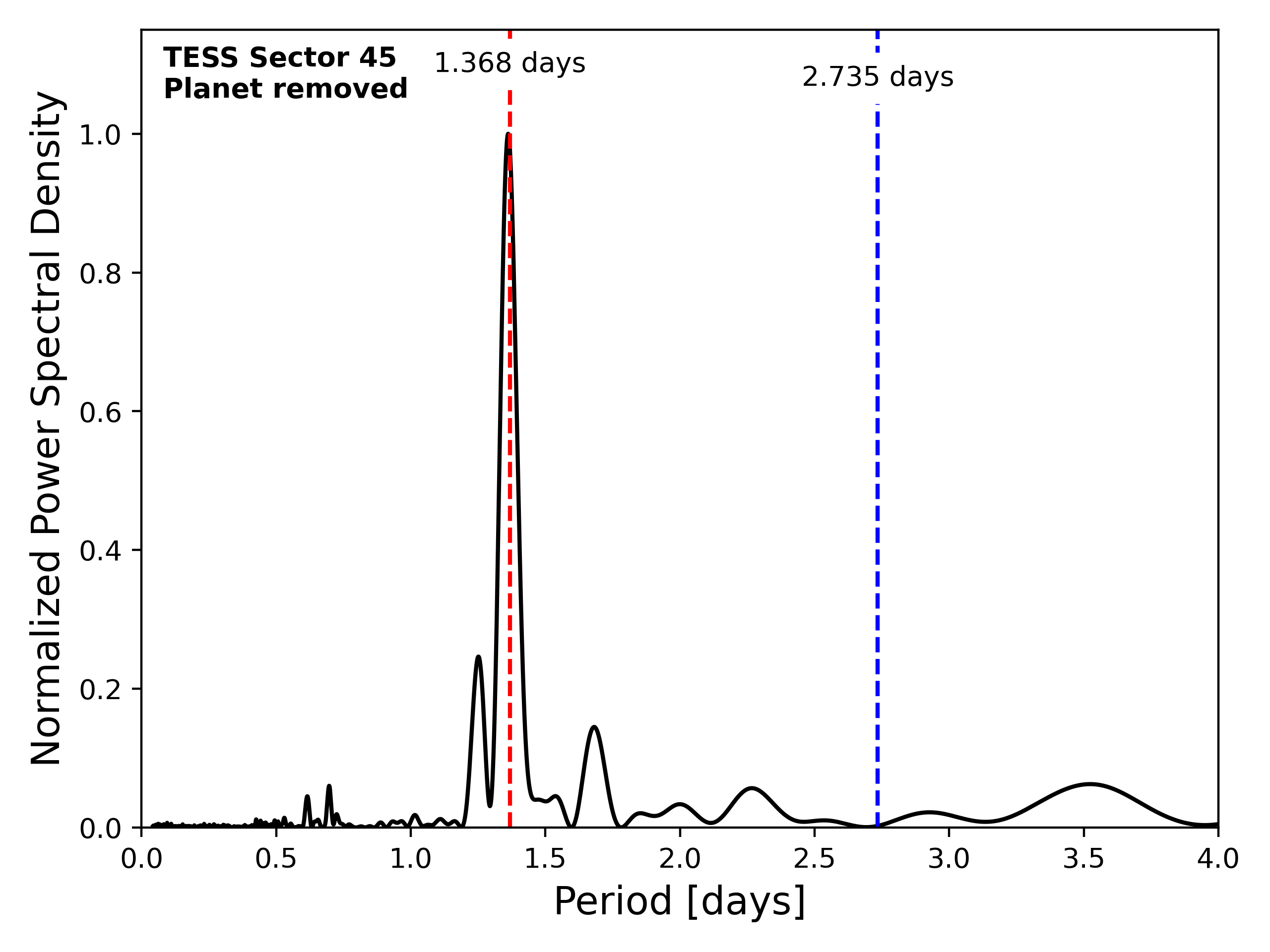}
\includegraphics[width=0.66\columnwidth]{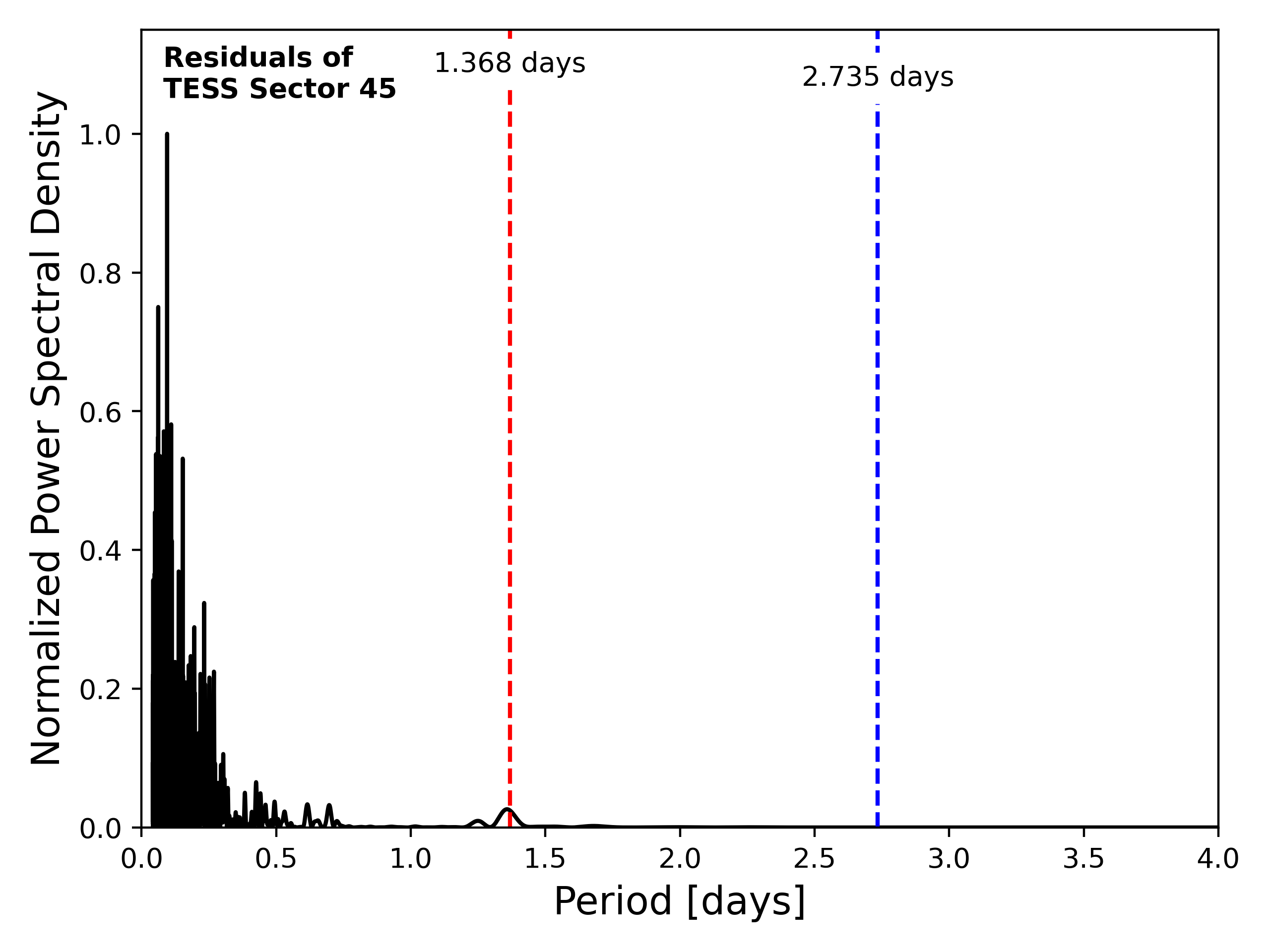}}
\caption{Periodograms of KELT-7 TESS PDCSAP observations. The blue dashed line represents the fitted orbital period of the planet. The red dashed line represents the median value of the prior imposed on the stellar rotational period. Left panel: Periodogram of TESS Sector 59 raw data, which was used to determine the prior. Middle panel: Periodogram of the residuals of TESS Sector 45 data after removal of the transits and occultations. Right panel: Periodogram of the residuals of TESS Sector 45 data.}
\label{periodogram} 
\end{figure*}

\citet{Tabernero1} show that KELT-7 TESS light curves are heavily affected by stellar activity due to the rapid rotation of the star. Accordingly, the authors obtain a stellar rotation period of $P_\mathrm{rot, s} = 1.38 \pm 0.05\,\mathrm{d}$. Following their findings, we also observe periodic flux changes in the TESS data that most likely are caused by stellar activity (see Figs.~\ref{tessallsec1} and \ref{tessallsec2}; left panels). We note that the same variability signal can be seen in both the PDCSAP and SAP flux. The effects of stellar activity must be accounted for in both TESS and CHEOPS observations when deriving the transit and eclipse parameters. To properly model the stellar noise, we employed a Gaussian process (GP) model. It was built using a Simple Harmonic Oscillator (SHO) kernel based on \texttt{celerite}\footnote{\url{https://github.com/dfm/celerite}} models \citep{celerite1}. The kernel is parameterized with an undamped angular frequency of the oscillator $\omega_0$, a quality factor $Q$, and the amplitude of the power spectral density $S_0$ at $\omega = \omega_\mathrm{0}$ \citep{celerite1,celerite2}:

\begin{equation}
    S(\omega) = \sqrt{\frac{2}{\pi}} \frac{S_0 \omega_0^4}{(\omega^2 - \omega_0^2)^2 + \omega_0^2 \omega^2 / Q^2}.
\end{equation}

\noindent The parameters $\omega_0$ and $Q$ can also be expressed via the undamped period of the oscillator $\rho_\mathrm{SHO}$ and the damping timescale of the process $\tau_\mathrm{SHO}$:

\begin{equation}
\label{eq_rho}
    \rho_\mathrm{SHO} = \frac{2 \pi}{\omega_0},
\end{equation}

\begin{equation}
    \tau_\mathrm{SHO} = \frac{2 Q}{\omega_0}.
\end{equation}

To model stellar activity, the parameters $\rho_\mathrm{SHO}$, $\tau_\mathrm{SHO}$, and $S_0$ can be interpreted as the stellar rotation period $P_{\mathrm{rot, s}}$, the characteristic damping timescale of star spot dissipation $\tau_{\mathrm{dis}}$, and the amplitude of the activity-induced variations $S_0$, respectively. Both $P_{\mathrm{rot, s}}$ and $\tau_{\mathrm{dis}}$ are independent of time of observation and observed filter, while $S_0$ depends on the observed wavelength range. Therefore, we assumed a common $P_{\mathrm{rot, s}}$ and $\tau_{\mathrm{dis}}$ across all observations while accounting for independent $S_0$ parameters for TESS and CHEOPS observations. We also added a white-noise term $\sigma$ per instrument. To constrain the GP parameter $\rho_\mathrm{SHO}$ (i.e., the proxy of the stellar rotation period), we analyzed the periodogram of TESS Sector 59 (see left panel of Fig. \ref{periodogram}). We find the maximum peak of the periodogram at $1.368\,\mathrm{d}$. Assuming a Gaussian distribution to fit for the width of the peak we defined a prior on $P_{\rm rot, s}$ of $1.368 \pm 0.030\,\mathrm{d}$ and translated it to a prior on $\omega_0$ of $4.59 \pm 0.10$ following Eq. (\ref{eq_rho}). A similar peak can be found when analyzing the periodograms of all the other TESS sectors. We note that this prior on the stellar rotation period is very similar to half of the orbital period of the planet, which would be $1.3674\,\mathrm{d}$. To ensure that this peak in the periodograms is not caused by the planet, but due to the stellar variability, we also checked the periodograms of the TESS sectors after removing the transits and occultations, which still contain the identical peak (see middle panel of Fig. \ref{periodogram}). Additionally, we also note that the final fitted rotational period of the star is $1.320 \pm 0.020\,\mathrm{d}$ (see Sect. \ref{sececl}), which is more than $3\sigma$ different from half of the orbital period of the planet.

\begin{table*}
\footnotesize
\centering
\caption{Fitted \texttt{CONAN3} parameters of the KELT-7 planetary system.}
\label{finalparams}
\begin{tabular}{cccc}
\hline
\hline
Parameter [unit] 									          & Description 						    & Prior		                                                & Value\\
\hline
$T_\mathrm{c}$ [$\mathrm{BJD}_\mathrm{TDB} - 2457000$] 	      & reference mid-transit time              & $\mathcal{U}(2880.34,2880.35)$                            & $2880.342376 \pm 0.000023$\\
$R_\mathrm{p,CHEOPS}/R_\mathrm{s}$                            & CHEOPS planet-to-star radius ratio      & $\mathcal{U}(-1,1)$                                       & $0.08998 \pm 0.00020$\\
$R_\mathrm{p,TESS}/R_\mathrm{s}$                              & TESS planet-to-star radius ratio        & $\mathcal{U}(-1,1)$                                       & $0.08956 \pm 0.00009$\\
$b$                                                           & impact parameter                        & $\mathcal{U}(0,1)$                                        & $0.607 \pm 0.004$\\
$W_\mathrm{tra}$ [d]                                          & total transit duration                  & $\mathcal{U}(0,0.3)$                                      & $0.14449 \pm 0.00011$\\
$P_\mathrm{orb}$ [d] 						         	      & orbital period                          & $\mathcal{U}(2.73,2.74)$                                  & $2.73476613 \pm 0.00000013$\\
$D_\mathrm{occ,CHEOPS}$ [ppm]                                 & CHEOPS occultation depth                & $\mathcal{U}(0,500)$                                      & $36 \pm 11$\\
$D_\mathrm{occ,TESS}$ [ppm]                                   & TESS occultation depth                  & $\mathcal{U}(0,500)$                                      & $69 \pm 9$\\
$u_\mathrm{1,CHEOPS}$                                         & CHEOPS quadratic LD coefficient         & $\mathcal{N}(0.35,0.03)$                                  & $0.319 \pm 0.016$\\
$u_\mathrm{2,CHEOPS}$                                         & CHEOPS quadratic LD coefficient         & $\mathcal{N}(0.31,0.04)$                                  & $0.242 \pm 0.021$\\
$u_\mathrm{1,TESS}$                                           & TESS quadratic LD coefficient           & $\mathcal{N}(0.21,0.04)$                                  & $0.220 \pm 0.016$\\
$u_\mathrm{2,TESS}$                                           & TESS quadratic LD coefficient           & $\mathcal{N}(0.33,0.05)$                                  & $0.212 \pm 0.022$\\
$\log \sigma_\mathrm{CHEOPS}$ [log relative flux]             & CHEOPS white-noise term in the GP       & $\mathcal{U}(-20,-1)$                                     & $-9.17 \pm 0.03$\\
$\log \sigma_\mathrm{TESS}$ [log relative flux]               & TESS white-noise term in the GP         & $\mathcal{U}(-20,-1)$                                     & $-8.540\pm 0.008$\\
$\log S_\mathrm{0,CHEOPS}$                                    & CHEOPS scaled amplitude in the SHO GP   & $\mathcal{U}(-30,-1)$                                     & $-15.85 \pm 0.22$\\
$\log S_\mathrm{0,TESS}$                                      & TESS scaled amplitude in the SHO GP     & $\mathcal{U}(-30,-1)$                                     & $-16.28 \pm 0.19$\\
$\log Q$                                                      & quality factor in the SHO GP            & $\mathcal{U}(-10,5)$                                      & $-1.15 \pm 0.12$\\
$\log \omega_0$ [$\log \mathrm{d}^{-1}$]                      & frequency in the SHO GP                 & $\mathcal{N}(1.52,0.02)$                                  & $1.560 \pm 0.015$\\
\hline
\end{tabular}
\end{table*}

\begin{table}
\centering
\caption{Derived \texttt{CONAN3} parameters of the KELT-7 planetary system.}
\label{derivparams}
\begin{tabular}{ccc}
\hline
\hline
Parameter [unit] & Value\\
\hline
$R_\mathrm{p,CHEOPS}^{\star}$ [$\mathrm{R_{Jup}}$]      & $1.499 \pm 0.036$\\
$R_\mathrm{p,TESS}^{\star}$ [$\mathrm{R_{Jup}}$]        & $1.492 \pm 0.034$\\
$D_\mathrm{tra,CHEOPS}$ [ppm]                           & $8097 \pm 36$\\
$D_\mathrm{tra,TESS}$ [ppm]                             & $8022 \pm 17$\\
$a/R_\mathrm{s}$                                        & $5.510 \pm 0.016$\\
$a$ [au]                                                & $0.0439 \pm 0.0010$\\
$i$ [deg]                                               & $83.68 \pm 0.06$\\
$W_\mathrm{occ}$ [d]                                    & $0.158\pm 0.004$\\
$\rho_\mathrm{s}^{\star\star}$ [$\rho_\odot$]           & $0.302\pm 0.002$\\
$P_\mathrm{rot, s}$ [d]                                 & $1.320 \pm 0.020$\\
$\tau_{\rm dis}$ [d]                                    & $0.133 \pm 0.018$\\
\hline
\end{tabular}
\tablefoot{$^{\star}$Derived based on a stellar radius of $R_\mathrm{s} = 1.712 \pm 0.037~\mathrm{R}_\odot$ \citep{Tabernero1}. $^{\star\star}$Derived based on a stellar radius of $R_\mathrm{s} = 1.712 \pm 0.037~\mathrm{R}_\odot$ and a stellar mass of $M_\mathrm{s} = 1.517 \pm 0.022~\mathrm{M}_\odot$ \citep{Tabernero1}.}
\end{table}

\subsection{Planetary model and limb darkening law}

To fit the planetary model, we adopted wide uniform priors on the impact parameter $b$, the total transit duration $W_\mathrm{tra}$, the reference mid-transit time $T_\mathrm{c}$, and the orbital period $P_\mathrm{orb}$. The values of the adopted priors are listed in Table \ref{finalparams}. We also fitted individual planet-to-star radius ratios and occultation depths for each of the observed filters (TESS and CHEOPS), which were also subject to uniform priors. We did not incorporate ellipsoidal variation and Doppler beaming into our planetary model, as their theoretical estimates are only a few ppm, and therefore do not significantly impact our model fitting. Additionally, based on the TESS phase-curve analysis, we found insignificant nightside flux estimates, which means that we could not account for any meaningful heat-distribution efficiency. To account for limb darkening (LD), we adopted a quadratic LD law. We computed theoretical LD coefficients, including their uncertainties for KELT-7 using the \texttt{LDCU} Python package (see Table \ref{finalparams}). \texttt{LDCU}\footnote{\url{https://github.com/delinea/LDCU}} is a modified version of the Python routine implemented by \citet{espinoza2015} that computes the LD coefficients and their corresponding uncertainties using a set of stellar intensity profiles that account for the uncertainties in the stellar parameters. The stellar intensity profiles are generated based on two libraries of synthetic stellar spectra: ATLAS \citep{Kurucz1979} and PHOENIX \citep{husser2013}.

\subsection{Joint fit and eclipse detection}
\label{sececl}

To retrieve the secondary eclipse depths, we proceeded by jointly fitting all planetary parameters, the basis vectors of the linear-detrending models for CHEOPS, and the GP parameters within a Markov chain Monte Carlo (MCMC) framework, using the \texttt{COde for transiting exoplaNet ANalysis 3} (\texttt{CONAN3}) Python package \citep{Lendl2017, Lendl2020b}. We used 40 chains, with each chain performing $30\,000$ steps, of which the first $10\,000$ steps were discarded. We used the Gelman-Rubin test \citep{Gelman1992} to check the convergence of our fit. We also analyzed the periodograms of the residuals to ensure that the combined variability and planetary model can account for all periodic variations in the data (see right panel of Fig. \ref{periodogram}). The medians and $1\sigma$ confidence intervals of the fitted parameters, including the secondary eclipse depths, are listed in Table \ref{finalparams}. The derived parameters are presented in Table \ref{derivparams}. The phase-folded and fitted CHEOPS and TESS light curves are depicted in Fig. \ref{cheopstessobsall}. Supplementary figures can be found in Appendix \ref{addfigs}.

Finally, we performed a test fit -- reran the described joint-fit procedure, keeping everything identical except for the stellar parameters. In this case, we applied the stellar parameters published by \citet{Bieryla1} -- and used previously by \citet{Pluriel1} -- instead of the stellar parameters presented by \citet{Tabernero1}. We tested the impact of the change on the fitted parameters. We do not detect any significant difference. The fitted parameters are well within the $1\sigma$ uncertainties, similarly to the two sets of stellar parameters, which are also consistent with each other within $1\sigma$ of their uncertainties.

\subsection{Search for transit asymmetry}
\label{transitasymmetry}

Based on spectroscopy, \citet{Bieryla1} find that the normal of the orbit of the planet is likely to be well-aligned with the stellar spin axis, with a projected spin-orbit angle of $\lambda = 9.7 \pm 5.2\,\mathrm{deg}$. Later, \citet{Zhou1} confirm the spin-orbit alignment of the system with an improved value of $\lambda = 2.7 \pm 0.6\,\mathrm{deg}$. \citet{Tabernero1} obtain a projected spin-orbit angle of $\lambda = -10.55 \pm 0.27\,\mathrm{deg}$, and a 3D spin-orbit angle of $\Psi = 12.4 \pm 11.7\,\mathrm{deg}$. 

Given the rapid rotation of KELT-7 (see Sect. \ref{intro}), we tried fitting the transit photometric data from CHEOPS with a model that includes the gravity-darkening effect. We used the \texttt{Transit and Light Curve Modeller} code (\texttt{TLCM}; \citealt{tlcm,power_wavelets_1}), which was previously used to model gravity-darkened transits in the WASP-189 \citep{Lendl1}, MASCARA-1 \citep{Hooton1}, WASP-33 \citep{Kalman_W33}, and KELT-20 \citep{Singh_K20} systems. Alongside the usual parameters to describe the transit ($P_\mathrm{orb}$, $R_\mathrm{p, CHEOPS}/R_\mathrm{s}$, $a/R_\mathrm{s}$, $b$, and the LD coefficients), we fitted for the stellar inclination $I_\mathrm{s}$, and longitude of the node $\Omega_\mathrm{s}$ of the stellar rotation axis. To account for the systematic noise in the light curve, including the roll-angle effect, we tried three different approaches, each of which introduces additional fitting parameters. The first approach was to fit the roll-angle effect with two sine and two cosine terms per transit. The second was to rely solely on the wavelet method of red-noise fitting \citep{power_wavelets_1, carter_winn}, which involves fitting for the red- and white-noise levels of the light curve, $\sigma_\mathrm{r}$ and $\sigma_\mathrm{w}$. Finally, we tried a combination of the previous two approaches, employing both trigonometric terms and wavelets.

We obtain consistent results from all three methods of dealing with correlated noise in the light curves. The transit parameters resulting from these fits are also in good agreement (within $1\sigma$) with those presented in Table \ref{finalparams}. We find no significant difference in the quality of the fits obtained from \texttt{TLCM} with and without gravity darkening. In other words, we find no evidence of significant transit asymmetry. We find $I_\mathrm{s} = 86 \pm 25\,\mathrm{deg}$, and we are unable to place any meaningful constraint on the sky-projected orbital obliquity ($\lambda = 8 \pm 105\,\mathrm{deg}$).\footnote{In general there is a four-way degeneracy in $I_\mathrm{s}$ and $\lambda$ deduced from gravity darkening, such that we are unable to distinguish between $\lambda$, $-\lambda$, $180\degr - \lambda$, and $180\degr + \lambda$ (see Smith et al. in prep. for more details). Here, $\lambda$ is so poorly constrained that we report only one of these solutions, and draw no conclusions from the result.}

\begin{figure*}
\centering
\centerline{
\includegraphics[width=\columnwidth]{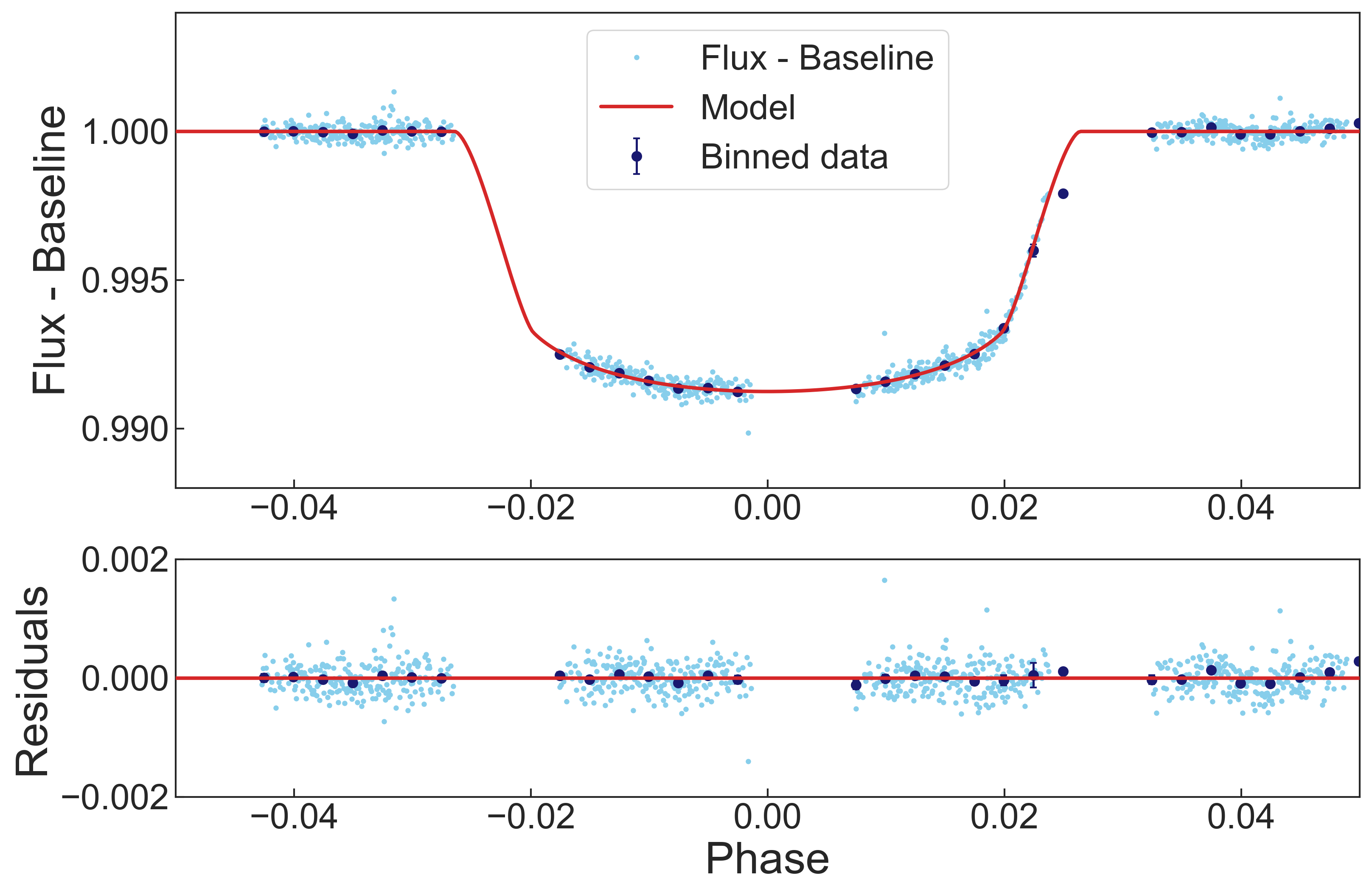}
\includegraphics[width=\columnwidth]{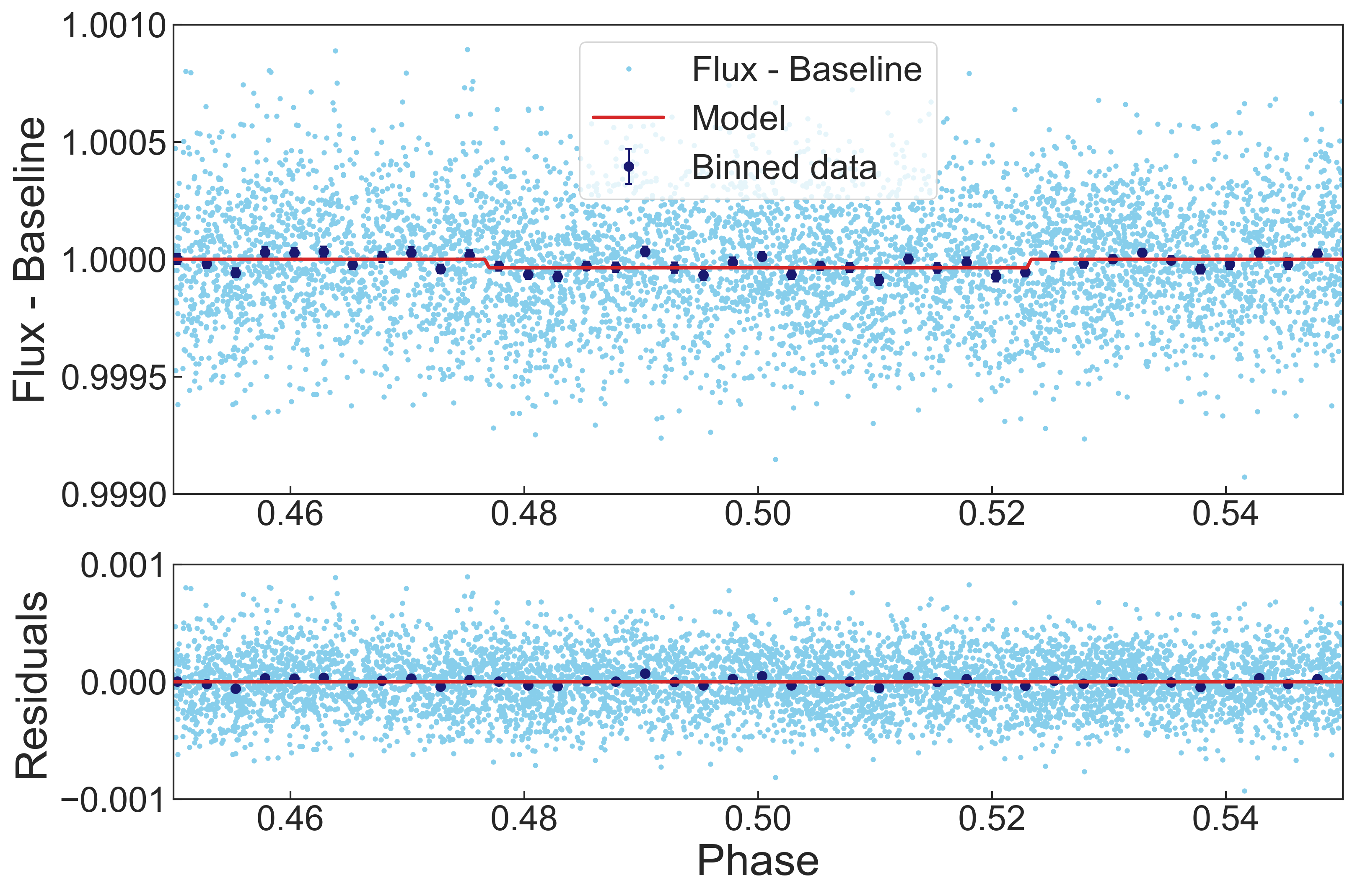}}
\centerline{
\includegraphics[width=\columnwidth]{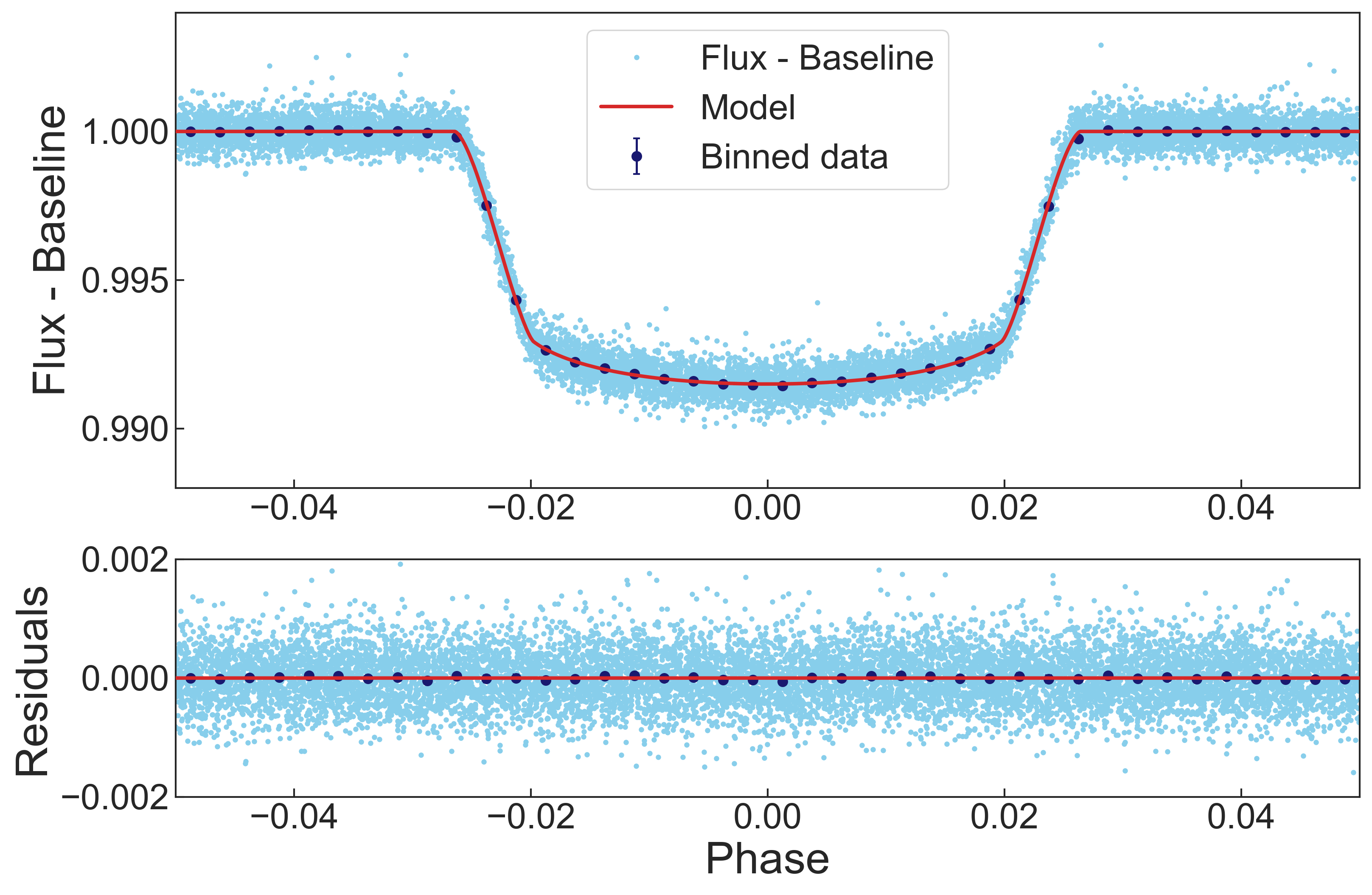}
\includegraphics[width=\columnwidth]{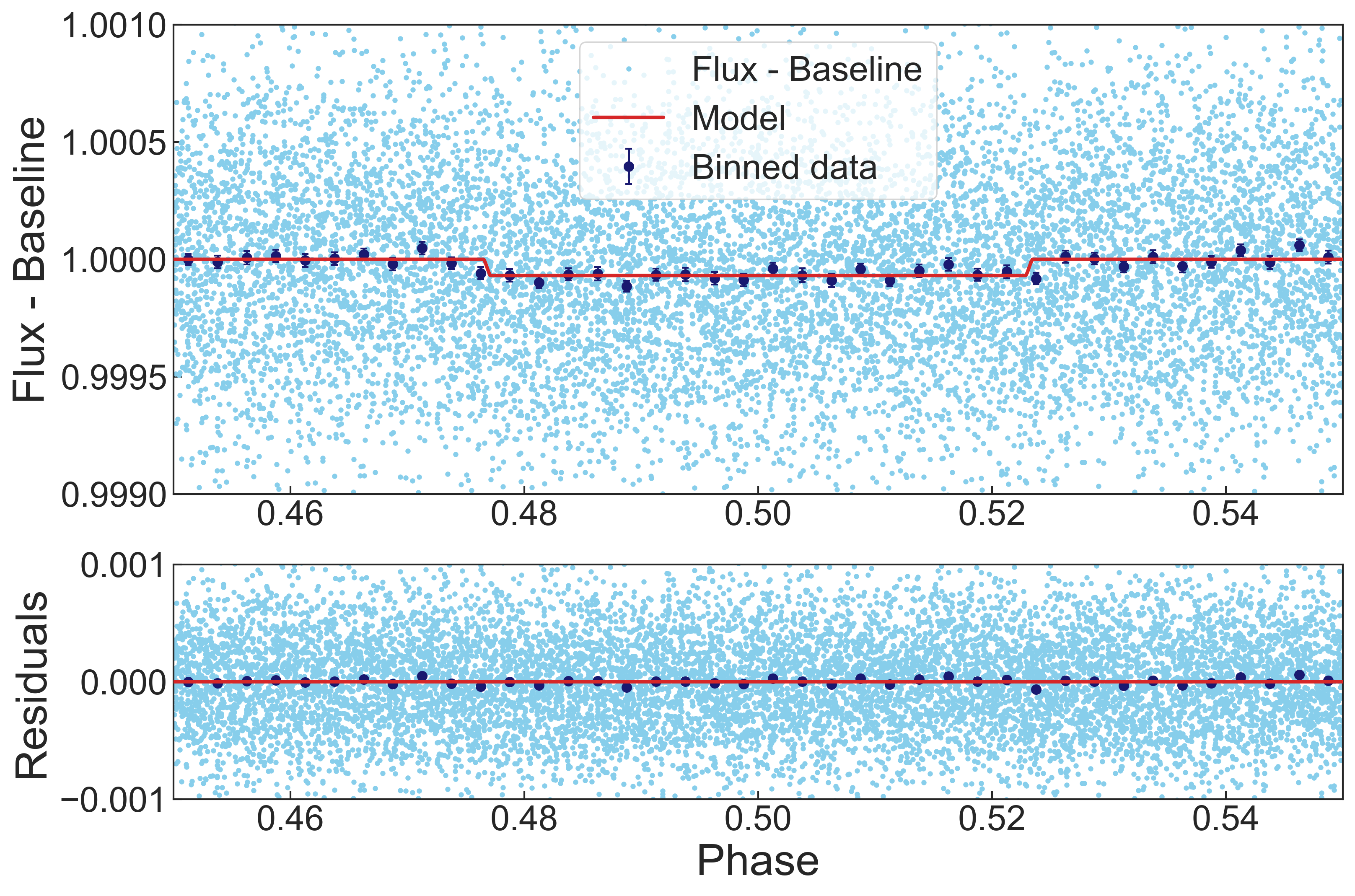}}
\caption{Phase-folded, detrended, and binned CHEOPS (top panels) and TESS (bottom panels) transit (left panels) and occultation (right panels) light curves of KELT-7b, overplotted with the best-fitting \texttt{CONAN3} model. Residuals are also shown.}
\label{cheopstessobsall} 
\end{figure*}

\section{Atmosphere modeling}
\label{atmosection}

The observed emission of KELT-7b, as measured by CHEOPS and TESS, may include contributions from both reflected stellar light (characterized by its albedo properties) and thermal emission from the planet (characterized by its thermal structure and composition). To distinguish between reflected and thermal signals, we focused on the thermal emission of KELT-7b using atmospheric retrievals from its infrared emission observations (HST and Spitzer, see Sect. \ref{archivedata}). To avoid biasing the results with data points affected by reflected light, which is not modeled in this case, we excluded the CHEOPS and TESS occultation observations from the retrieval (see Table \ref{finalparams}). To perform a robust assessment of our atmospheric analysis, we conducted Bayesian retrievals using two independent frameworks. To evaluate the impact of our model assumptions and to compare our findings with those in the literature \citep{Pluriel1, ChangeatEtal2022apjsHSTeclipseHotJupiters}, we systematically tested multiple model assumptions by applying equivalent assumptions across both frameworks, adapted where necessary to their respective implementations. Throughout our retrieval analyses, we assumed the system parameters reported by \citet{Bieryla1} and used by \citet{Pluriel1}, unless otherwise specified.

\subsection{Atmospheric retrieval with {\pyratbay}}
\label{Retrieval1}

To characterize the atmospheric properties of KELT-7b, we employed the open-source {\pyratbay} modeling framework \citep{CubillosBlecic2021mnrasPyratBay}. The {\pyratbay} package combines parameterized atmospheric modeling, spectral synthesis, and Bayesian posterior sampling, which together constrain the planetary atmospheric profiles based on the occultation observations. In this work, we modeled the dayside KELT-7b atmosphere as a one-dimensional (1D) profile as a function of pressure, adopting a temperature profile \citep{MadhusudhanSeager2009apjRetrieval}. For composition, we tested two alternatives: one following the free-chemistry parameterization (i.e., modeling the abundance of each absorber as a constant-with-altitude free parameter), and another one assuming thermochemical equilibrium consistent with the temperature profile (Cubillos et al., in prep.). We parameterized the composition with two free parameters that determine the atmospheric metallicity (abundance of all metal elements, relative to solar [M/H]) and the carbon elemental abundance (relative to oxygen, C/O).

The radiative-transfer calculation considered opacities from the main molecular species expected for hot Jupiters, including CO \citep{LiEtal2015apjsCOlineList}, \ch{CO2} \citep{RothmanEtal2010jqsrtHITEMP}, \ch{CH4} \citep{HargreavesEtal2015apjHotCH4}, \ch{H2O} \citep{PolyanskyEtal2018mnrasPOKAZATELexomolH2O}, HCN \citep{HarrisEtal2006mnrasHCNlineList, HarrisEtal2008mnrasExomolHCN}, \ch{NH3} \citep{Yurchenko2015jqsrtBYTe15exomolNH3, ColesEtal2019mnrasNH3coyuteExomol}, FeH \citep{Bernath2020jqsrtMoLLIST}, TiO \citep{McKemmishEtal2019mnrasTOTOexomolTiO}, and VO \citep{McKemmishEtal2016mnrasVOMYTexomolVO}. We pre-processed the larger ExoMol line lists with the \textsc{repack} algorithm \citep{Cubillos2017apjRepack} to extract the dominant transitions before sampling them to a fixed grid at a resolving power of 15\,000. The code also included Rayleigh scattering by H, H$_2$, and He \citep{Kurucz1970saorsAtlas}; H$_2$--H$_2$ and H$_2$--He collision-induced absorption \citep{BorysowEtal1988apjH2HeRT, BorysowFrommhold1989apjH2HeOvertones, BorysowEtal1989apjH2HeRVRT, BorysowEtal2001jqsrtH2H2highT, Borysow2002jqsrtH2H2lowT, JorgensenEtal2000aaCIAH2He}; and \ch{H-} continuous absorption \citep{John1988aaHydrogenIonOpacity}. For transmission geometry, we also included the opacity from the Na and K resonant lines \citep{BurrowsEtal2000apjBDspectra}. The posterior sampling was handled by the \textsc{mc3} package \citep{CubillosEtal2017apjRednoise}, using the Nested-sampling algorithm \citep[via \textsc{pymultinest},][]{FerozEtal2009mnrasMultiNest, BuchnerEtal2014aaBayesianXrayAGN} with 2000 live points.

\begin{figure*}[ht]
\centering
\includegraphics[width=\linewidth, clip]{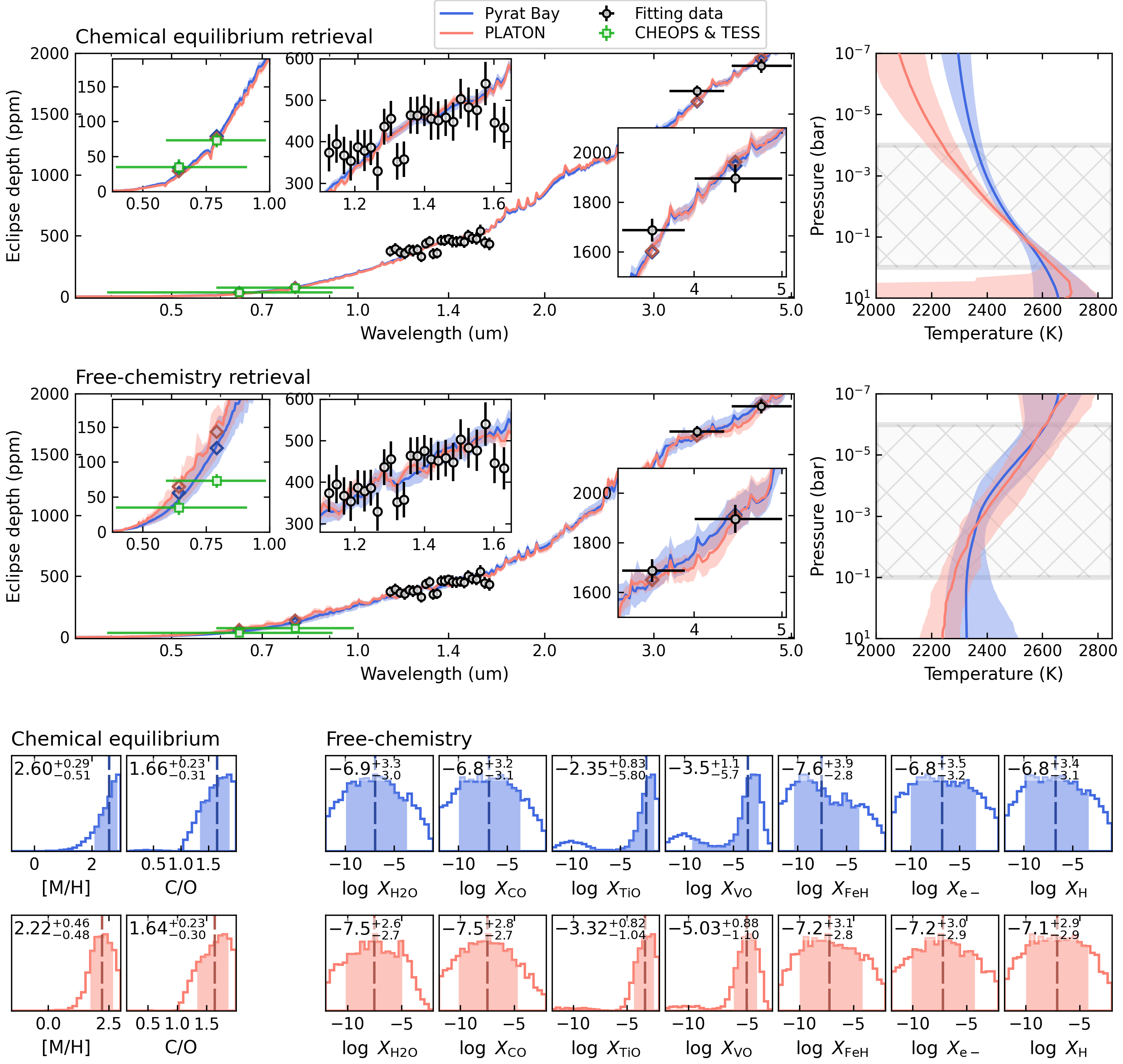}
\caption{KELT-7b atmospheric retrieval of the infrared occultations. Top left panel: Retrievals assuming thermochemical equilibrium with {\pyratbay} (blue) and {\platon} (pink). The solid curves with shaded areas show the median and $1\sigma$ span of the posterior model spectra, displayed at a resolution of $R=150$. The black circle markers with error bars show observations used to constrain the models (HST and Spitzer). The green square markers show the CHEOPS and TESS occultation depths (not used as retrieval constraints). The diamond markers show the model spectra integrated over the observing bands. The insets zoom in on the regions probed by the observations. Top right panel: Retrieved T-P profiles for each retrieval code (median and 1$\sigma$ span from the posterior distribution, same color coding as previous panel). The gray hatched area denotes the range of pressures probed by the observations. Middle panels: Same as above, but for the free-chemistry retrievals. Bottom panels: Posterior distribution of the atmospheric composition parameters (same color coding as above). The labels on top of each posterior show the mean and $1\sigma$ uncertainties for each parameter posterior (denoted with a dashed line and shaded area, respectively). Some parameters have been omitted from this figure (see Table \ref{table:retrievals} for the full list of free parameters).}
\label{fig:emission_retrieval}
\end{figure*}

\subsection{Atmospheric retrieval with \textsc{platon}}
\label{Retrieval2}

\begin{figure*}
\centering
\includegraphics[width=\linewidth, clip]{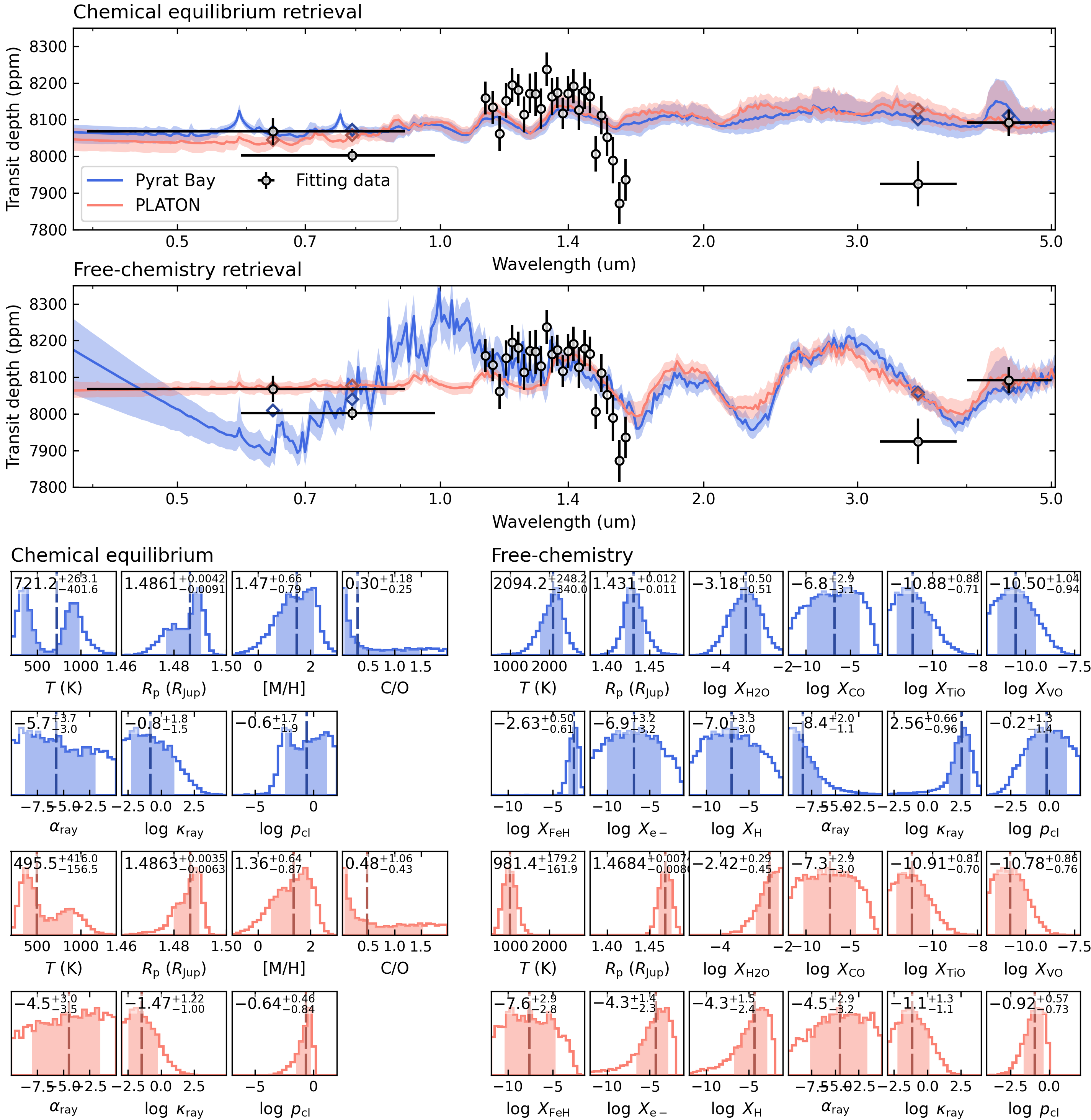}
\caption{KELT-7b atmospheric retrieval of the transmission observations assuming thermochemical equilibrium (top panel) and free-chemistry (middle panel). The solid curves with shaded areas show the median and $1\sigma$ span of the posterior model spectra for {\pyratbay} (blue) and {\platon} (pink), displayed at a resolution of $R=150$. The black circle markers with error bars show observations used to constrain the models. The diamond markers show the model spectra integrated over the observing bands. Bottom panels: Posterior distribution of the model parameters (same color coding as above). The labels on top of each posterior show the mean and $1\sigma$ uncertainties for each parameter posterior (denoted with a dashed line and shaded area, respectively). Some parameters have been omitted from this figure (see Table \ref{table:retrievals} for the full list of free parameters).}
\label{fig:transmission_retrieval} 
\end{figure*}

To test the robustness of our atmospheric characterization and its sensitivity to different modeling assumptions, we employed a second atmospheric retrieval tool. We adopted the open-source \textsc{platon} software, version 6.2 \citep{zhang2025}. For the eclipse retrieval, we assumed a cloud-free atmosphere using the default opacities provided by \textsc{platon} at a resolution of $R=20\,000$, which include gas and collision-induced absorption from H$_2$O, CO, CO$_2$, CH$_4$, TiO, VO, Na, K, and FeH (see the aforementioned release paper for further details). We also included H$^{-}$ bound-free and free-free continuous absorption given the high expected temperature of the planetary atmosphere \citep[see, e.g.,][]{John1988aaHydrogenIonOpacity,arcangeli2018}.

The emission spectroscopy retrievals were carried out for both the thermochemical-equilibrium (where abundances are described through metallicity [M/H] and the C/O ratio parameters) and the free-chemistry scenarios (where species abundances are directly fit as constant-with-altitude volume mixing ratios). We adopted \citet{MadhusudhanSeager2009apjRetrieval}'s parameterization for the temperature profile. To sample the parameter posterior distributions, we used \textsc{pymultinest} \citep{feroz2008, FerozEtal2009mnrasMultiNest,feroz2019, BuchnerEtal2014aaBayesianXrayAGN} with the uniform priors shown in Table \ref{table:retrievals}, and 1000 live points. For the thermochemical equilibrium scenario, we explored the full metallicity and C/O ratio ranges allowed by \textsc{platon}'s model grid, while in the free chemistry scenario, all absorbers were assigned a prior on the volume mixing ratio ranging from 10$^{-12}$ to 10$^{-2}$.

\subsection{Occultation retrieval results}
\label{sec:retrieval_eclipse}

Figure \ref{fig:emission_retrieval} shows the {\pyratbay} and {\platon} retrieved occultation spectra, T-P profiles, and parameter posteriors (see Table \ref{table:retrievals} as well). We find that both retrieval codes produce consistent results when subjected to the same set of assumptions. When assuming a thermochemical-equilibrium atmospheric composition, the retrievals return a non-inverted T-P profile (probing mainly the $1-10^{-4}\,\mathrm{bar}$ range) with a composition characterized by a C/O ratio greater than one (C/O > 1.1, at the 3$\sigma$ lower boundary) and a super-solar metallicity in the $170 - 400 \times$ solar range ([M/H] = $2.6_{-0.5}^{+0.3}$ for {\pyratbay}, [M/H] = $2.2\pm0.5$ for {\platon}). The main driver for this behavior is the relatively weak \ch{H2O} absorption feature at 1.4\,{\microns}, since a $\mathrm{C/O} > 1$ scenario leads to a depletion of \ch{H2O} abundance. In contrast, when adopting a free-chemistry atmospheric parameterization, the retrievals return an inverted T-P profile between $0.1-10^{-6}\,\mathrm{bar}$. In this case, the strong optical absorbers TiO and VO are the only species with well-constrained abundances, albeit at -- likely unphysically -- high concentrations. The retrievals constrain the water abundance (volume mixing ratio) to less than $\sim 10 - 100\,\mathrm{ppm}$. Once again, this may be due to the absence of a clear \ch{H2O} absorption feature at 1.4\,{\microns}.

We tested a range of configurations to explore the dependence of the retrieval results on the model assumptions. We performed three additional comparisons: (1) retrievals adopting the system parameters from this work and those used in \citet{Pluriel1}, (2) retrievals employing a different thermal profile parameterization \citep{Guillot2010aaRadiativeEquilibrium}, and (3) retrievals including only the molecular absorbers of \citet{Pluriel1} versus a larger set of absorbers (see Sect. \ref{Retrieval1}). None of these tests led to qualitatively different retrieval results.

\citet{Pluriel1} and \citet{ChangeatEtal2022apjsHSTeclipseHotJupiters} have previously presented atmospheric retrieval analyses of KELT-7b based on the HST/WFC3 and Spitzer occultations. Adopting a free-chemistry parameterization, they both find an inverted T-P profile, a nondetection of \ch{H2O}, and a detection of \ch{H-} absorption. \citet{ChangeatEtal2022apjsHSTeclipseHotJupiters} further performed retrievals assuming equilibrium chemistry, finding solar to super-solar metallicities, C/O ratios greater than one, and a different thermal structure. These findings are well in agreement with our results, with the main difference being the optical absorber found for the free-chemistry parameterization. This is not unexpected -- as both TiO and VO or \ch{H-} optical absorbers are not strongly constrained by the near-infrared observations, in both scenarios, they contribute to a higher brightness temperature at the blue end of the WFC3 band. Thus, our comparison tests and the agreement with previous analyses from the literature lead us to conclude that the choice of free or thermochemical-equilibrium chemistry is the main factor driving the retrievals to different atmospheric scenarios.

We note that the free-chemistry retrievals generally yield better fits to the observations than the thermochemical-equilibrium retrievals. However, assuming free constant-with-altitude abundances risks adopting scenarios at odds with plausible physical conditions. The dayside atmosphere of ultrahot Jupiters similar to KELT-7b is expected to reach temperatures exceeding 2000\,K. Since disequilibrium-chemistry processes, such as photochemistry and transport-induced quenching, become less and less important with increasing effective temperature, at these extreme temperatures the chemical reaction rates are fast enough to overcome the effect of disequilibrium chemistry \citep{KopparapuEtal2012apjWASP12bChemistry, Moses2014rsptaChemicalKinetics, MadhusudhanEtal2016ssrChemistryFormation, VenotEtal2018exaChemicalCompositionAriel}. Thermochemical equilibrium is therefore the expected assumption to model the dayside atmospheric composition of planets similar to KELT-7b and their resulting emission spectra. If disequilibrium chemistry occurs at all (e.g., photochemistry), it would occur at high altitudes above the pressures probed by the observations presented in this work, and thus the modeled emission spectra of planets similar to KELT-7b would not be significantly impacted \citep{ShulyakEtal2020aaHotJupiterChemistry}. In contrast, thermochemical-equilibrium calculations indicate that we expect a strong variation in abundances with altitude at the pressures where \ch{H2} dissociates into H, which can occur precisely at the pressures probed by near-infrared observations. In the specific case of KELT-7b, the free-chemistry retrievals point to high abundances of TiO (this work) or \ch{e-} \citep{Pluriel1, ChangeatEtal2022apjsHSTeclipseHotJupiters}, which are orders of magnitude above expected values from self-consistent chemical models. The suboptimal thermochemical-equilibrium fit suggests that there may be missing physics in these retrieval models, which may be resolved with the availability of improved data.

Lastly, we must consider that combining multi-epoch observations can lead to biases in the atmospheric interpretation due to stellar activity, instrumental systematics, or different assumptions made for each data reduction \citep{EdwardsEtal2024rastiAbsoluteDepthsKELT11b}. Observations with the James Webb Space Telescope (JWST) have demonstrated that not only can there be transit- or eclipse-depth offsets between different observations \citep[see, e.g.,][]{Fu1, Madhusudhan3, LouieEtal2025ajWASP17bTransitJWST, MayoEtal2025arxivWASP166bTransitJWST}, but also between different detectors in the same observation \citep[see, e.g.,][]{CarterEtal2024natasDataSynthesisWASP39b, GressierEtal2024apjL98-59dTransitJWSTsulphur, FournierTondreauEtal2024arxivWASP52bTransitJWST}. JWST depth offsets can be effectively detrended in retrievals by applying ad-hoc, offset-free parameters; however, for the sparse, low-resolution wavelength coverage of HST and Spitzer, the addition of offset parameters will likely lead to a strongly degenerate solution with the astrophysical signal.

Considering the model-dependent outcome of the KELT-7b atmospheric retrievals and the discussion above, we should be cautious when interpreting the retrieval results. Qualitatively speaking, extending the retrieved emission models over the CHEOPS and TESS optical bands suggests that the planet's thermal emission is consistent (equilibrium chemistry) or larger (free chemistry) than the observed occultation depths, which suggests that the planet has a low albedo, producing little reflected light. Section \ref{sec:albedo} presents an alternative analysis of the albedo properties of KELT-7b that relies on the observed brightness temperatures.

\subsection{Transmission retrieval results}
\label{sec:retrieval_transit}

{\renewcommand{\arraystretch}{1.15}
\begin{table*}
\centering
{
\caption{Priors and posterior parameter estimations (median and 68\% credible intervals) for the atmospheric retrievals.}
\label{table:retrievals}
\begin{tabular}{l|ccc|ccc}
\hline
\hline
& \multicolumn{3}{c|}{\textsc{Pyrat Bay} retrievals} & \multicolumn{3}{c}{\textsc{platon} retrievals} \\
Parameter & Prior & Equilibrium  & Free-chemistry & Prior & Equilibrium  & Free-chemistry  \\
\hline
Occultation\\

$\log p_1$  & $\mathcal{U}(-9, 2)$      &  $-2.2^{+1.9}_{-3.9}$     &  $-6.2^{+1.4}_{-1.3}$      & $\mathcal{U}(-9, 1)$       &  $5.4^{+1.7}_{-6.3}$              &  $-1.5^{+1.9}_{-1.4}$        \\                            
$\log p_2$  & $\mathcal{U}(-9, 2)$      &  $-5.9^{+2.9}_{-2.0}$     &  $-0.2^{+1.4}_{-2.5}$      & $\cdots$                   & $\cdots$                          &  $\cdots$                     \\                                      
$\log p_3$  & $\mathcal{U}(-9, 2)$      &  $0.65^{+0.89}_{-1.14}$   &  $-0.5^{+1.7}_{-1.9}$      & $\mathcal{U}(-9, 1)$       &  $3.5^{+2.8}_{-4.7}$              &  $5.0^{+1.7}_{-1.8}$         \\                            
$a_1$        & $\mathcal{U}(0.2, 2.0)$   &  $1.27^{+0.48}_{-0.31}$   &  $1.33^{+0.41}_{-0.48}$   & $\mathcal{U}(0.2, 2)$      &  $0.871^{+0.292}_{-0.091}$        &  $1.19^{+0.47}_{-0.41}$      \\      
$a_2$        & $\mathcal{U}(0.2, 2.0)$   &  $0.78^{+0.41}_{-0.27}$   &  $0.69^{+0.18}_{-0.17}$   & $\mathcal{U}(0.2, 2)$      &  $1.16^{+0.54}_{-0.56}$           &  $1.32^{+0.39}_{-0.42}$      \\                            
$T_0$ (K)    & $\mathcal{U}(500, 5000)$  &  $2284^{+50}_{-73}$ &  $2650^{+130}_{-150}$           & $\mathcal{U}(X, X)$        &  $2059^{+97}_{-102}$        &  $2695^{+157}_{-185}$  \\
$T_3$ (K)    & $\cdots$ & $\cdots$  & $\cdots$                                                   & $\mathcal{U}(500, 3000)$   &  $2232^{+543}_{-1114}$      &  $2218^{+76}_{-110}$   \\                            
{[M/H]}      & $\mathcal{U}(-1, 3)$      &  $2.60^{+0.29}_{-0.51}$  &   $\cdots$                 & $\mathcal{U}(-2, 3)$     & $2.22^{+0.46}_{-0.48}$ & $\cdots$ \\
C/O          & $\mathcal{U}(0.01, 2.0)$  &  $1.66^{+0.23}_{-0.31}$  &   $\cdots$                 & $\mathcal{U}(0.001, 2)$  & $1.64^{+0.23}_{-0.30}$ & $\cdots$ \\
$\log X_{\rm H2O}$  & $\mathcal{U}(-12, -1)$   &  $\cdots$ &  $-6.9^{+3.3}_{-3.0}$               & $\mathcal{U}(-12, -2)$   & $\cdots$ & $-7.5^{+2.6}_{-2.7}$  \\
$\log X_{\rm CO}$   & $\mathcal{U}(-12, -1)$   &  $\cdots$ &  $-6.8^{+3.2}_{-3.1}$               & $\mathcal{U}(-12, -2)$   & $\cdots$ & $-7.5^{+2.8}_{-2.7}$  \\
$\log X_{\rm CO2}$ & $\mathcal{U}(-12, -1)$  &  $\cdots$ &  $-7.6^{+3.6}_{-2.6}$                 & $\mathcal{U}(-12, -2)$   &  $\cdots$ &  $-8.5^{+2.1}_{-2.0}$   \\
$\log X_{\rm CH4}$ & $\mathcal{U}(-12, -1)$  &  $\cdots$ &  $-4.4^{+2.3}_{-4.8}$                 & $\mathcal{U}(-12, -2)$   &  $\cdots$ &   $-7.2^{+2.4}_{-2.9}$   \\
$\log X_{\rm TiO}$  & $\mathcal{U}(-12, -1)$   &  $\cdots$ &  $-2.4^{+0.8}_{-5.8}$               & $\mathcal{U}(-12, -2)$   &  $\cdots$ & $-3.3^{+0.8}_{-1.0}$  \\
$\log X_{\rm VO}$   & $\mathcal{U}(-12, -1)$   &  $\cdots$ &  $-3.5^{+1.1}_{-5.7}$               & $\mathcal{U}(-12, -2)$   & $\cdots$ & $-5.0^{+0.9}_{-1.1}$  \\
$\log X_{\rm FeH}$  & $\mathcal{U}(-12, -1)$   &  $\cdots$ &  $-7.6^{+3.9}_{-2.8}$               & $\mathcal{U}(-12, -2)$   & $\cdots$ & $-7.2^{+3.1}_{-2.8}$  \\
$\log X_{\rm e-}$   & $\mathcal{U}(-12, -1)$   &  $\cdots$ &  $-6.8^{+3.5}_{-3.2}$               & $\mathcal{U}(-12, -2)$   & $\cdots$ & $-7.2^{+3.0}_{-2.9}$  \\
$\log X_{\rm H}$    & $\mathcal{U}(-12, -1)$   &  $\cdots$ &  $-6.8^{+3.4}_{-3.1}$               & $\mathcal{U}(-12, -2)$   & $\cdots$ & $-7.1^{+2.9}_{-2.9}$  \\
\\
Transmission\\
$T_{\rm iso} ({\rm K})$     & $\mathcal{U}(200, 3500)$   &  $721^{+263}_{-402}$    &  $2094^{+248}_{-340}$ & $\mathcal{U}(200, 3000)$   & $496^{+416}_{-157}$       & $981^{+179}_{-162}$ \\ 
$R_{\rm p}$ ($R_{\rm Jup}$)  & $\mathcal{U}(0.65, 1.94)$  &  $1.4861^{+0.0042}_{-0.0091}$ &  $1.431^{+0.012}_{-0.011}$  & $\mathcal{U}(0.65, 1.94)$  & $1.486^{+0.004}_{-0.006}$ & $1.486^{+0.007}_{-0.008}$ \\ 
{[M/H]}                      & $\mathcal{U}(-1, 3)$       &  $1.47^{+0.66}_{-0.79}$       &  $\cdots$                   & $\mathcal{U}(-2, 3)$       & $1.36^{+0.64}_{-0.87}$    & $\cdots$ \\
C/O                          & $\mathcal{U}(0.01, 2.0)$   &  $0.30^{+1.18}_{-0.25}$       &  $\cdots$                   & $\mathcal{U}(0.001, 2)$    & $0.48^{+1.06}_{-0.43}$    & $\cdots$ \\  
$\log\ X_{\rm H2O}$          & $\mathcal{U}(-12, -1)$     &  $\cdots$                     &  $-3.18^{+0.50}_{-0.51}$    & $\mathcal{U}(-12, -2)$     & $\cdots$ & $-2.4^{+0.3}_{-0.5}$   \\  
$\log\ X_{\rm CO}$           & $\mathcal{U}(-12, -1)$     &  $\cdots$                     &  $-6.8^{+2.9}_{-3.1}$       & $\mathcal{U}(-12, -2)$     & $\cdots$ & $-7.3^{+2.9}_{-3.0}$   \\                          
$\log\ X_{\rm CO2}$          & $\mathcal{U}(-12, -1)$     &  $\cdots$                     &  $-8.2^{+2.0}_{-2.3}$       & $\mathcal{U}(-12, -2)$     & $\cdots$ & $-7.8^{+2.4}_{-2.5}$   \\       
$\log\ X_{\rm CH4}$          & $\mathcal{U}(-12, -1)$     &  $\cdots$                     &  $-8.5^{+2.1}_{-2.1}$       & $\mathcal{U}(-12, -2)$     & $\cdots$ &  $-9.0^{+1.9}_{-1.9}$  \\
$\log\ X_{\rm TiO}$          & $\mathcal{U}(-12, -1)$     &  $\cdots$                     &  $-10.88^{+0.88}_{-0.71}$   & $\mathcal{U}(-12, -2)$     & $\cdots$ & $-10.9^{+0.8}_{-0.7}$  \\
$\log\ X_{\rm VO}$           & $\mathcal{U}(-12, -1)$     &  $\cdots$                     &  $-10.50^{+1.04}_{-0.94}$   & $\mathcal{U}(-12, -2)$     & $\cdots$ & $-10.8^{+0.9}_{-0.8}$  \\
$\log\ X_{\rm FeH}$          & $\mathcal{U}(-12, -1)$     &  $\cdots$                     &  $-2.6^{+0.5}_{-0.6}$    & $\mathcal{U}(-12, -2)$     & $\cdots$ & $-7.6^{+2.9}_{-2.8}$   \\   
$\log\ X_{\rm e-}$           & $\mathcal{U}(-12, -1)$     &  $\cdots$                     &  $-6.9^{+3.2}_{-3.2}$       & $\mathcal{U}(-12, -2)$     & $\cdots$ & $-4.3^{+1.4}_{-2.3}$   \\
$\log\ X_{\rm H}$            & $\mathcal{U}(-12, -1)$     &  $\cdots$                     &  $-7.0^{+3.3}_{-3.0}$       & $\mathcal{U}(-12, -2)$     & $\cdots$ & $-4.3^{+1.5}_{-2.4}$   \\
$\alpha_{\rm ray}$           & $\mathcal{U}(-10, 0)$      &  $-5.7^{+3.7}_{-3.0}$         &  $-8.4^{+2.0}_{-1.1}$       & $\mathcal{U}(-10, 0)$      & $-4.5^{+3.0}_{-4.5}$      & $496^{+2.9}_{-3.2}$ \\ 
$\log\ \kappa_{\rm ray}$     & $\mathcal{U}(-3, 5)$       &  $-0.8^{+1.8}_{-1.5}$         &  $2.56^{+0.66}_{-0.96}$     & $\mathcal{U}(-3, 5)$       & $-1.47^{+1.22}_{-1.00}$   & $-1.1^{+1.3}_{-1.1}$\\      
$\log\ p_{\rm cloud}$        & $\mathcal{U}(-7, 2)$       &  $-0.6^{+1.7}_{-1.9}$         &  $-0.2^{+1.3}_{-1.4}$       & $\mathcal{U}(-5.99, 0)$    & $-0.64^{+0.46}_{-0.84}$   & $-0.92^{+0.57}_{-0.73}$\\         
\hline
\end{tabular}
}
\tablefoot{Using \textsc{pyrat bay} and \textsc{platon}, assuming thermochemical-equilibrium and free-chemistry abundances. $X_{\rm s}$ denotes the volume mixing ratio for the given species. All pressure values are in bar units.}
\end{table*}
}

In addition, we also retrieved the atmospheric properties of KELT-7b from the HST and Spitzer transmission observations using {\pyratbay} and {\platon}. Since, in transmission geometry, stellar reflected light is negligible, we also included the CHEOPS and TESS measurements as retrieval constraints.

Our transmission retrievals also considered the impact of clouds and hazes through an opaque cloud deck, parameterized by a cloud-top pressure $p_\mathrm{cloud}$, and Rayleigh scattering, parameterized by the opacity slope $\alpha_{\rm ray}$ and strength $\kappa_{\rm ray}$ \citep{LecavelierEtal2008aaRayleighHD189733b}. We adopted an isothermal temperature profile at $T_{\rm iso}$, given the weaker sensitivity of transmission to the thermal structure, and fitted the planet radius at a reference pressure of $10\,\mathrm{bar}$ to solve the hydrostatic equilibrium equation. For the composition parameterization, we also tested both equilibrium and free chemistry.

Figure \ref{fig:transmission_retrieval} and Table \ref{table:retrievals} present our transmission retrieval results. Both of our retrieval tools yield consistent results. In the thermochemical-equilibrium case, they favor a $10 - 100 \times \mathrm{solar}$ metallicity, with $\lesssim 1 \times \mathrm{solar}$ C/O and no clouds ($p_\mathrm{cloud} \gtrsim 1\,\mathrm{bar}$). An upper limit is recovered on the strength of scattering ($\simeq~2.5 \times \mathrm{Rayleigh}$), and the scattering slope remains unconstrained. For the free-chemistry case, both of our retrievals detect H$_2$O at a temperature range where this molecule does not dissociate \citep{Parmentier2}, and require the presence of an additional optical absorber. {\pyratbay} finds absorption from FeH and hazes, whereas {\platon} finds \ch{H-} absorption. This discrepancy is not unexpected, given the limited spectral resolution and coverage in the optical range, which preclude unambiguous identification of the source of the optical opacity \citep[see, e.g.,][]{kesseli2020}. As in the case of the occultation retrievals, the transmission free-chemistry retrievals in general yield a better fit than the equilibrium retrievals.

Our results also qualitatively agree with the previous analyses of the transmission observations. When adopting a free-chemistry parameterization, \citet{Pluriel1} and \citet{ChangeatEtal2022apjsHSTeclipseHotJupiters} find a cloud-free atmosphere with absorption from \ch{H2O} and an optical absorber (\ch{H-}), whereas their equilibrium-chemistry retrievals do not fit the transit data well.

\subsection{Independent reduction of the HST/WFC3 transmission spectrum} 
\label{sec:wfc3_reduction}

To test whether any instrumental artifact could explain the poor fit of the transmission spectrum thermochemical-equilibrium retrievals in the near-infrared band, we independently reduced the two HST/WFC3 transits obtained for Hubble proposal 14767 (PI D. Sing), publicly available on MAST. We used the Intermediate MultiAccum (IMA) files and both scanning directions, and adopted the method described in \citet{bruno2018}. Each scanning direction was analyzed individually, resulting in two separate transit datasets. In particular, because of the brightness of the star, its two-dimensional spectral width in the detector scanning direction required an extraction window as large as 50 rows per nondestructive read. We did not observe any problematic features in the spectra (see Fig. \ref{fig:spectra_top}).

We retained the spectra in the $1.115 - 1.617\,\mu\mathrm{m}$ range and binned them using 6-pixel-wide bins to obtain the spectrophotometric transits. The full-range light curves were fitted with a least-squares minimization algorithm implementing the \textsc{batman} transit model \citep{kreidberg2015}. We assumed a circular orbit and relied on the scaled semi-major axis and orbital inclination reported by \citet{Bieryla1}, while fitting for the transit depth and mid-transit time. The stellar parameters obtained by the same authors were used to compute quadratic LD coefficients using the \textsc{exoctk} package \citep{stevenson2018,fowler2018,bourque2021}.

The transit model was multiplied by an exponential function to include the HST ramp, a second-degree polynomial to model stellar flux variations around the transits, and a scaling constant $C$, following standard practice \citep[see, e.g.,][]{stevenson2014}:

\begin{equation}
S(t) = C(1 + r_0\theta + r_1\theta^2 )(1 - e^{r_2\phi + r_3} + r_4\phi).
\end{equation}

\noindent Here, $\theta$ is the planetary orbital phase, $\phi$ is the HST orbit phase (with an additional phase offset fixed at 0.15, determined through trial and error), and $r_\mathrm{0\dotsc4}$ are parameters to fit. Once the parameters for the systematic noise were determined, all but the scaling constant were fixed to their best-fit value, and \textsc{emcee}, version 3.1.6 \citep{foreman-mackey2013}, was used to sample the posterior distributions of $C$ and the transit parameters. Setting 200 walkers and 2500 steps for the MCMC chains, and discarding the first 500 iterations as burn-in, was enough for each chain to be longer than 50 times its integrated autocorrelation time \citep{goodman-weare2010}.

\begin{figure}
\centering
\includegraphics[width=\columnwidth]{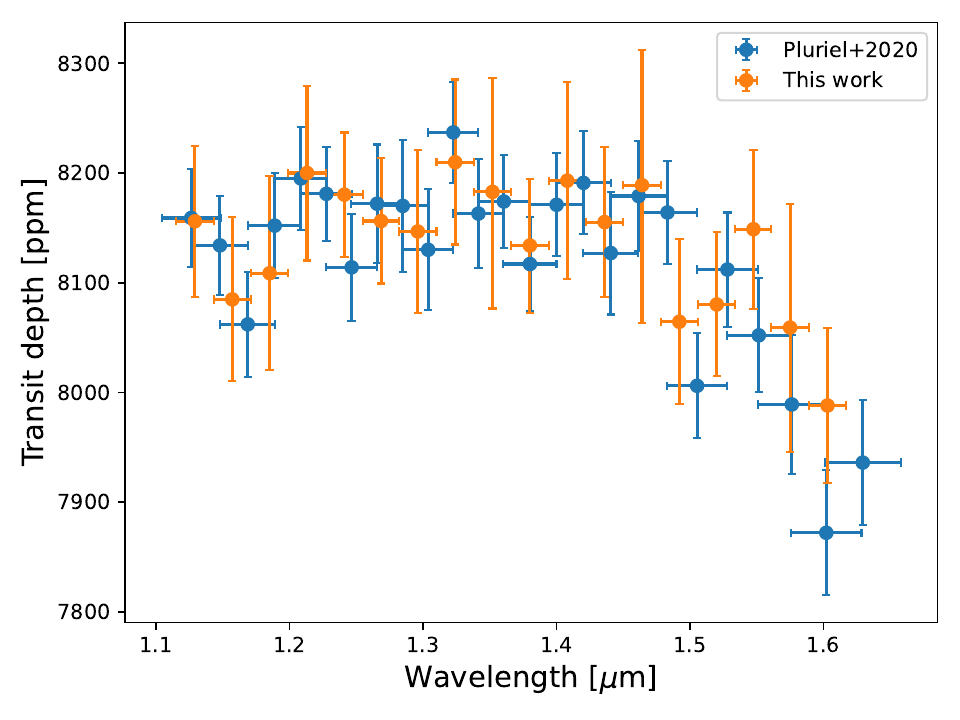}
\caption{Comparison of our independent extraction of the HST/WFC3 transmission spectrum and the one published by \citet{Pluriel1}.}
\label{transmission_spectra_compared}
\end{figure}

For all spectroscopic channels, the transit depth posterior distributions of the two transits were merged, and the final transmission spectrum was derived from the 16th, 50th, and 84th percentiles of the combined posterior distribution. Our output is compared to \cite{Pluriel1}'s in Fig. \ref{transmission_spectra_compared}, and confirms the transmission spectrum trend in the WFC3 band.

\subsection{Albedo}
\label{sec:albedo}

\begin{figure}
\centering
\includegraphics[width=\columnwidth]{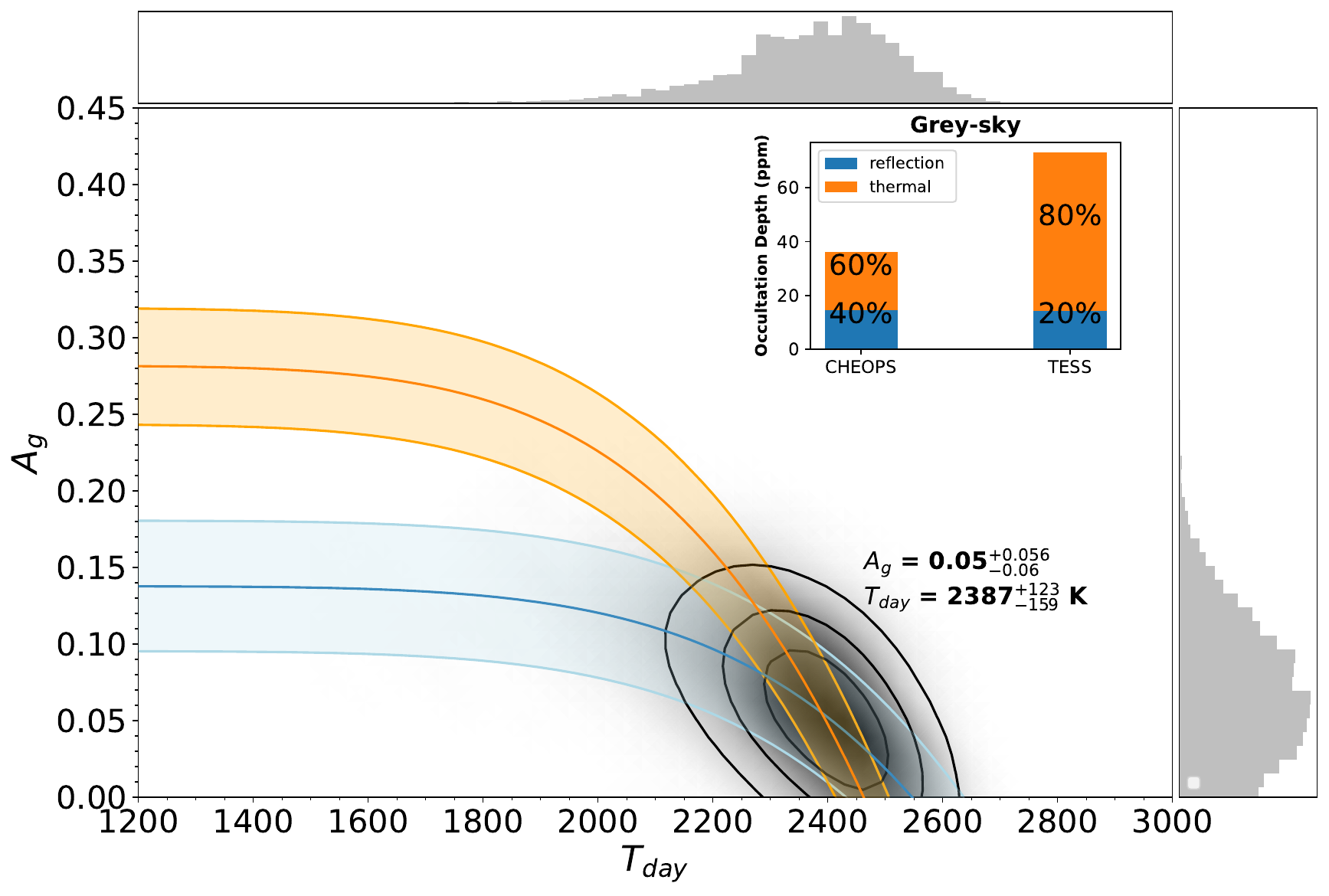}
\caption{Geometric albedo ($A_\mathrm{g}$) as a function of the dayside brightness temperature for the estimated occultation depths in CHEOPS (blue) and TESS (red) passbands. The plot shows where the two curves intersect as well as the values of the two parameters. The black concentric curves depict 1, 2, and $3\sigma$ distributions. The inset shows the reflection and emission contributions to the occultation depth for a gray-sky atmosphere.} 
\label{albedo} 
\end{figure}

A planetary brightness at occultation -- that is, the brightness of the dayside hemisphere -- can be expressed as the sum of the thermal emission ($\mathbb{E}$), which is a function of the brightness temperature ($T_\mathrm{day}$), and reflection ($\mathbb{R}$) of the incident stellar light, which depends on the geometric albedo ($A_\mathrm{g}$). The resultant brightness can be expressed as follows:

\begin{equation}
\frac{F_\mathrm{p}}{F_\mathrm{s}} = D_\mathrm{occ} = \mathbb{R}\left( A_\mathrm{g} \right) + \mathbb{E}\left( T_\mathrm{day} \right).
\end{equation}

Following the methodology described in \citet{Singh_K20}, we estimated the respective thermal emission and reflection/scattering contributions to the planetary brightness by assuming a gray-sky atmosphere, such that the geometric albedo and the brightness temperature are identical in the CHEOPS and TESS passbands. We obtain a very low geometric albedo of $A_\mathrm{g} = 0.05 \pm 0.06$ (<$1\sigma$). This corresponds to a brightness temperature of $T_\mathrm{day} = 2387^{+123}_{-159}\,\mathrm{K}$ (see Fig. \ref{albedo}). These numbers indicate the $1\sigma$ upper limit of approximately $40\%$ and $20\%$ reflection contamination in the CHEOPS and TESS passbands, respectively. For comparison, the brightness temperature retrieved at $10^{-1}\,\mathrm{bar}$ -- where the atmosphere is most sensitive to optical wavelengths -- is $\sim 2500\,\mathrm{K}$ (see Fig. \ref{fig:emission_retrieval}). Following this analytical approach, the brightness temperatures at $A_\mathrm{g} = 0$ (no reflection) in TESS and CHEOPS are $2462^{+45}_{-48}\,\mathrm{K}$ and $2547^{+90}_{-115}\,\mathrm{K}$, respectively. 

Utilizing the eclipse (see Fig. \ref{fig:emission_retrieval}) and transit spectra (see Fig. \ref{fig:transmission_retrieval}), we determined the planetary bolometric temperature. The resulting temperatures are approximately 2470\,K and 2415\,K, corresponding to the cases of equilibrium chemistry and free $\mathrm{H}^{-}$ chemistry, respectively. Based on these temperature estimates, we derived upper limits for the Bond albedo \citep[$\epsilon$ = 0,][]{Cowan1}: $0.0 \pm 0.2$ and $0.1 \pm 0.2$, respectively. In both cases, the Bond albedo remains consistent with zero within the given uncertainties. This indicates that the planet effectively absorbs nearly all incoming stellar irradiation to heat its dayside atmosphere.

\begin{figure*}[ht]
\centering
\includegraphics[width=.62\textwidth]{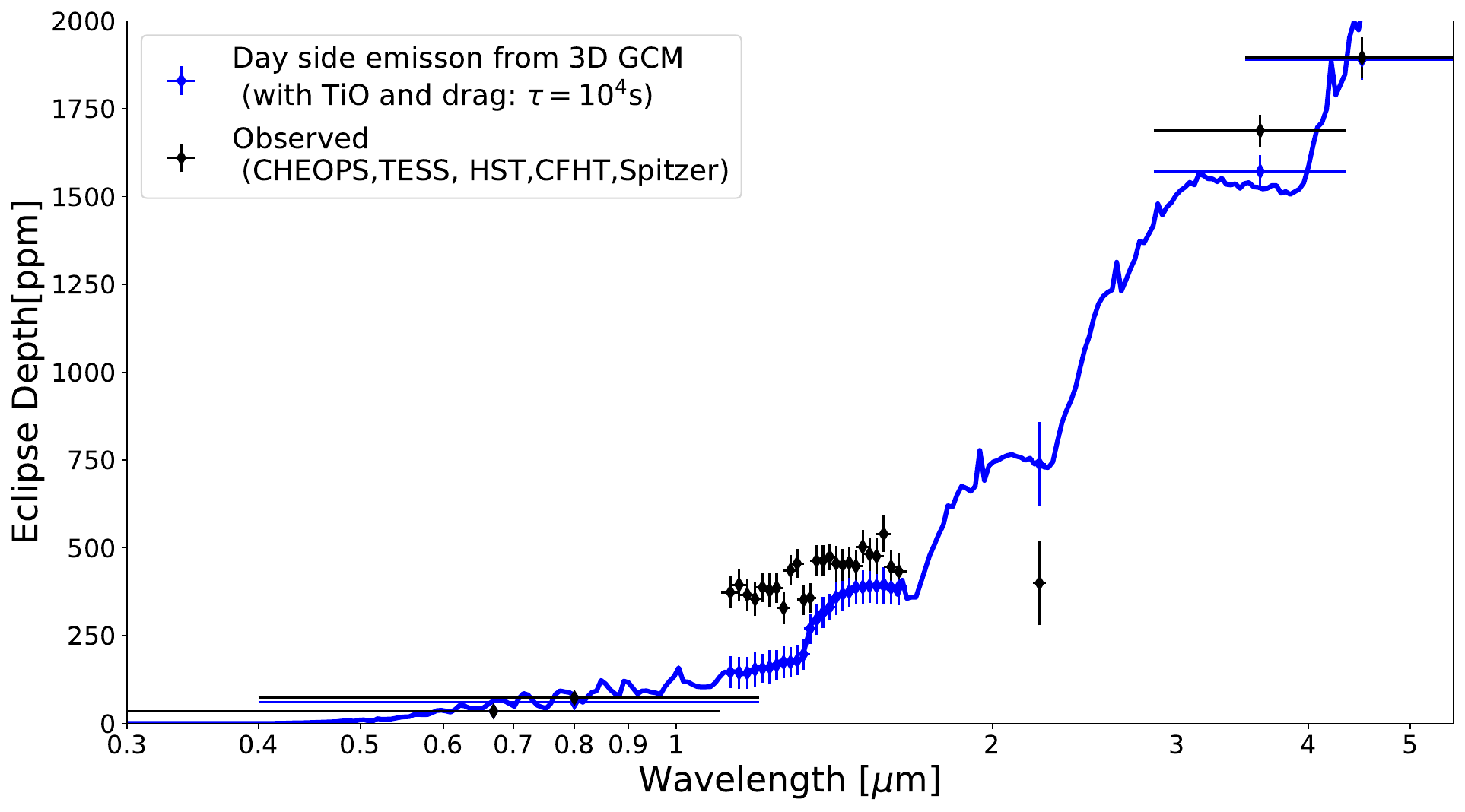}
\includegraphics[width=.3\textwidth]{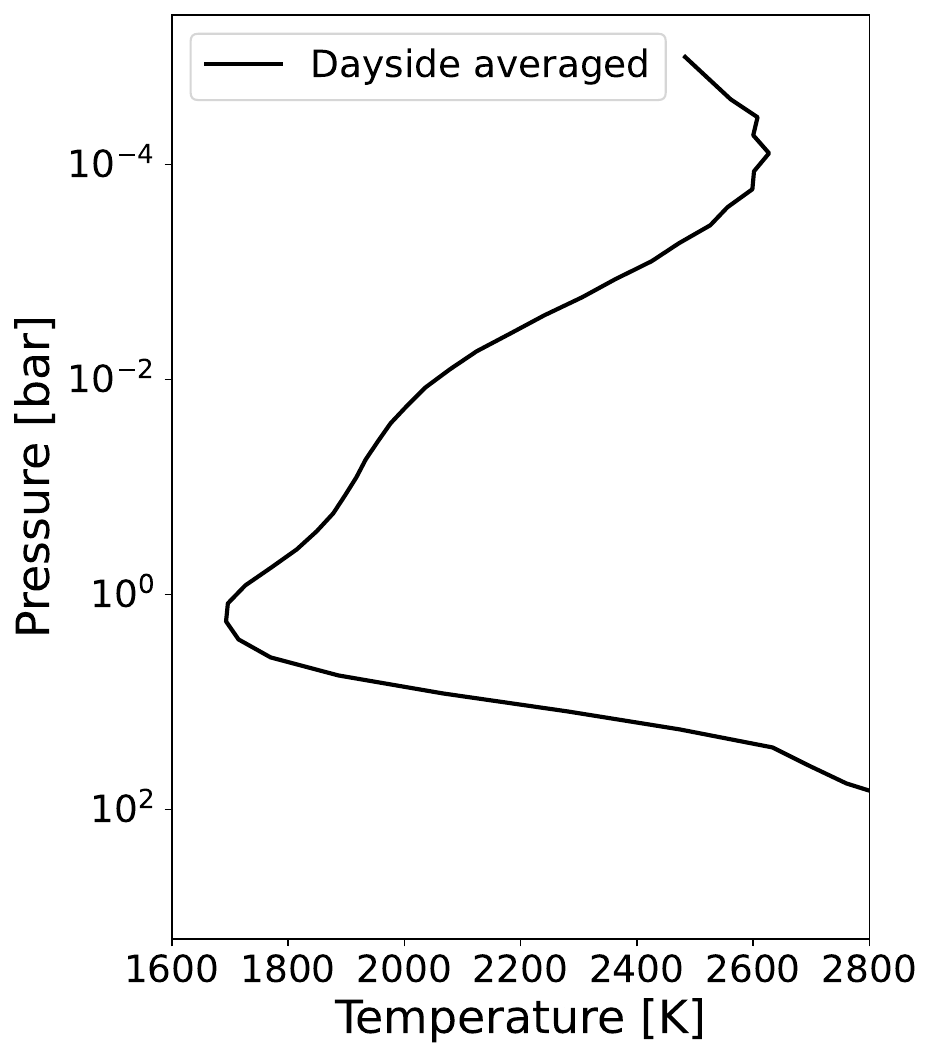}
\caption{Left panel: Theoretically calculated dayside emission from the 3D GCM \texttt{expeRT/MITgcm} for KELT-7b including TiO and VO as well as high magnetic drag (blue line) compared to the observational data (black dots). The model data are binned down for better comparison with the observational data (blue dots, assuming the same uncertainty as the observational data). The 3D GCM model agrees within $1\sigma$ with the CHEOPS and TESS observations. Right panel: Associated dayside averaged T-P profile.}
\label{3DEmisison} 
\end{figure*}

\subsection{3D climate atmosphere modeling}
\label{3dgcm}

As a sanity check for the retrieval models, we also simulated the dayside emission of the planet with the 3D general circulation model (GCM) \texttt{expeRT/MITgcm} \citep{Schneider2022} as part of the \texttt{ExoRad} climate framework \citep{Carone2020}. We used the planetary parameters from Table \ref{derivparams} and the stellar parameters adopted from \citet{Tabernero1}, described in Sect. \ref{host}, assuming solar metallicity and equilibrium chemistry for the planetary atmospheric composition.  We further employed tabulated opacities for the following species: H$_2$O from ExoMol \citep{TennysonEtal2016jmsExomol, Tennyson2020}, Na \citep{Allard2019Na_K}, K \citep{Allard2019Na_K}, CO$_2$, CH$_4$, NH$_3$, CO, H$_2$S, HCN, SiO, PH$_3$, and FeH, as well as H$^-$ absorption and electron scattering, suitable for an ionized atmosphere \citep[see, e.g.,][]{Helling2023}. Appendix \ref{Detail_GCM} contains a more detailed description of the model setup. 

We find that to match the eclipse depths of KELT-7b in the CHEOPS and TESS bands simultaneously with those observed with Spitzer in the IRAC 1 and 2 bands, TiO and VO opacities and strong magnetic drag with $\tau_{\rm drag}=10^4\,\mathrm{s}$ are needed. Ultrahot Jupiters similar to KELT-7b exhibit a strong horizontal gradient in ionization, as the dayside is thermally ionized, in contrast to the nightside \citep{Helling2019, Helling2021}. The degree of ionization also decreases with depth on the dayside, enabling magnetic coupling of the atmosphere to a global magnetic field \citep{Rauscher2013, Helling2023, Beltz2022}. The inclusion of magnetic drag, which mimics the coupling of magnetic fields with the partially ionized flow, has become state-of-the-art for ultrahot Jupiters \citep[see, e.g.,][]{Wardenier2023, Demangeon2024}. The exact choice of $\tau_{\rm drag}$ is still debated, especially in the context of the uniform-drag assumption implemented here \citep{Tan2019, Coulombe2023, Beltz2022}. In this work, we used the smallest $\tau_{\rm drag}$ that effectively disrupts superrotation on the dayside in our GCM and shifts the onset of the dayside temperature inversion to deeper, higher-pressure layers compared to a simulation without drag, which retains efficient superrotation and horizontal wind transport (see Appendix \ref{Detail_GCM}). Given the limited data, we only tested a few scenarios to constrain the range of the problem. One model without TiO and no drag, and two models with TiO -- with and without strong drag -- were explored. The latter two are shown in Appendix \ref{Detail_GCM}. We find that, of these models, the one using strong drag with TiO best matched the data.

We further used an interface between \texttt{expeRT/MITgcm} and \texttt{petitRADTRANS} \citep{Molliere2019} to generate dayside emission spectra. While the 3D climate model matches the TESS, CHEOPS, and Spitzer $4.5\,\mu\mathrm{m}$ observations well, it significantly underestimates the flux measured by HST/WFC3. At $3.6\,\mu\mathrm{m}$, the predicted flux appears to be lower by about $2\sigma$ compared to Spitzer observations. The CFHT data point could not be reconciled with any tested atmospheric scenario; therefore, we discarded it from the analysis (see Fig. \ref{3DEmisison}). Notably, the 3D climate model yields a temperature inversion that was not recovered by either retrieval model under the thermochemical-equilibrium assumption, only when adopting a free-chemistry parameterization. These models, however, use the HST/WFC3 data, which disagree with the predictions of the 3D GCM model. On the other hand, the choice of high magnetic drag in our 3D climate model yields inefficient horizontal heat distribution, which is consistent with the albedo results. 

\section{Discussion}
\label{discuss}

Our attempt to understand the atmosphere of KELT-7b yields disparate results, depending on the approach and data used. Assuming the same geometric albedo and dayside temperature in the CHEOPS and TESS passbands, we find a low geometric albedo ($A_\mathrm{g} = 0.05 \pm 0.06$) and high dayside temperatures, consistent with inefficient horizontal heat distribution $\epsilon$ close to zero. Likewise, a 3D GCM simulation yields inefficient heat transfer, a TiO-induced temperature inversion, and a hot dayside temperature that fits the CHEOPS, TESS, and Spitzer data but not the HST/WFC3 data. The retrievals with {\pyratbay} and {\platon} also confirm the low albedo result (see Sect. \ref{sec:retrieval_eclipse}).

Two independent retrieval pipelines ({\pyratbay} and \textsc{platon}) yield consistent atmospheric results with each other, though they lead to different physical implications depending on the assumed modeling framework. Retrievals assuming thermochemical-equilibrium and free-chemistry abundances produce non-inverted and inverted thermal profiles, respectively. We note that the free-chemistry retrievals provide a better fit to the observations than the equilibrium-chemistry retrievals, even though the high brightness temperature of the occultations suggests that the atmosphere should be in thermochemical equilibrium. These inconclusive results may reflect limitations in both the physical models and data analysis methods. We also performed an independent reduction of the HST data using the method described in \citet{bruno2018}. This yields similar results to those published by \citet{Pluriel1}. Offsets between atmospheric spectra of the same planet obtained with different instruments are not uncommon \citep[see, e.g.,][]{murgas2020,wilson2020,yip2021, EdwardsEtal2024rastiAbsoluteDepthsKELT11b}. Furthermore, some observations may be affected by stellar activity in this case, which could significantly hinder the correct interpretation of the brightness temperature \citep{Tabernero1,saba2025}. Additionally, we note that stellar pulsations are reported for KELT-7b \citep{Zhou1, Stangret1, sicilia2025}, which might additionally contaminate the planetary signal.

With a very simple calculation using $R_\mathrm{s} = 1.712 \pm 0.037\,\mathrm{R_\odot}$, $v \sin I_\mathrm{s} = 71.4 \pm 0.2\,\mathrm{km.s^{-1}}$ \citep{Tabernero1}, and $I_\mathrm{s} = 86\pm25$ deg (see Sect. \ref{transitasymmetry}) we can find a stellar rotational period of $P_\mathrm{rot,s} = 1.22 \pm 0.36\,\mathrm{d}$, which is consistent with the maximum peak of the periodogram at 1.368 d (see middle panel of Fig. \ref{periodogram}). This means that -- given the planet's host star is an F2V-type star with a convective envelope -- in addition to the weak pulsations, mentioned earlier, the stellar variability (see Figs. \ref{tessallsec1} and \ref{tessallsec2}; left panels) is very probably dominated by star spots, co-rotating with the star surface. Unocculted star spots can reduce the apparent stellar brightness, potentially deepening the measured planetary eclipse signal and leading to an overestimated planet's dayside temperature \citep{Zellem2017}. The large inferred temperature variations of 400\,K for KELT-7b in a narrow wavelength range suggest, however, a more complex scenario. In such a fast-rotating host star ($P_{\mathrm{rot, s}}=1.368$~d), star spots may have rotated in and out of view during the HST/WFC3 observation. In addition, the oblateness of the rapidly rotating host star is expected to result in a stellar flux gradient from the equatorial (cooler) to the polar (hotter) regions according to the von Zeipel theorem \citep{Zeipel1, Zeipel2}. The appearance and disappearance of star spots at different stellar latitudes during the observation may thus lead to nontrivial changes in the HST/WFC3 eclipse depth. However, a detailed assessment of stellar activity’s impact on the HST/WFC3 measurement lies beyond the scope of this work.

In this study, we accounted for the relatively high stellar variability of the fast-rotating host star KELT-7 when analyzing the TESS and CHEOPS data (see Sect. \ref{stellaractivity}). Furthermore, our physically consistent atmospheric model fits the CHEOPS, TESS, and Spitzer data well, with the latter likely being least affected by stellar variability. We therefore consider our conclusions regarding the dayside temperature, low albedo, and inefficient heat transport -- based on the combined CHEOPS and TESS data -- to be robust. In any case, the example of the exoplanet KELT-7b underscores the need for a coherent modeling framework that incorporates physical noise sources, including stellar variability.

\section{Conclusions}
\label{concl}

In this work, we analyzed the exoplanet KELT-7b with the main scientific goal of characterizing the atmosphere of the planet. Furthermore, we aimed to provide a precise measurement of KELT-7b’s albedo and to search for a possible distortion in its transit light curve caused by the rapid rotation of the host star. To fulfill these aims, we performed several photometric observations of KELT-7b secondary eclipses and transits using the CHEOPS space observatory. Moreover, we also used TESS photometric observations from seven sectors and literature data, which include published occultation and transit depths from HST/WFC3 and Spitzer/IRAC.

We first processed and jointly fitted the CHEOPS and TESS photometric data to detect the secondary eclipse of the planet in these passbands and to probe a potential asymmetry in the CHEOPS transit light curve. We were able to measure the occultation depths of KELT-7b with a $\sim3\sigma$ and $\sim8\sigma$ significance in the CHEOPS and TESS passbands, respectively. Our analysis yields the occultation depths of $D_\mathrm{occ,CHEOPS} = 36 \pm 11\,\mathrm{ppm}$ and $D_\mathrm{occ,TESS} = 69 \pm 9\,\mathrm{ppm}$. We can conclude that the secondary eclipses of KELT-7b are relatively shallow in the CHEOPS and TESS passbands, which is characteristic of hot Jupiters \citep[see, e.g.,][]{Singh_K20, Pagano2024}. In the optical wavelength range, the measured occultation depths are typically below 100\,ppm. 

Our most interesting results are related to the atmosphere modeling of KELT-7b. Based on the HST and Spitzer dataset, we performed occultation retrievals, and based on the CHEOPS, TESS, HST, and Spitzer dataset, transmission retrievals. We used two retrieval tools by applying the same set of assumptions to both frameworks as closely as permitted by their respective constraints. In all cases, we tested two alternatives, one following the free-chemistry parameterization and another assuming thermochemical equilibrium. We also tested a range of configurations to explore the dependence of the retrieval results on our assumptions. We can conclude that both retrieval codes produce consistent results when subjected to the same set of assumptions. When adopting a thermochemical-equilibrium atmospheric composition, the occultation retrievals return a non-inverted T-P profile with a composition characterized by C/O > 1, a super-solar metallicity, and a relatively weak \ch{H2O} abundance. In contrast, when adopting a free-chemistry atmospheric parameterization, the occultation retrievals return an inverted T-P profile with -- likely unphysically -- high concentrations of TiO and VO. None of the additional tests resulted in qualitatively different retrieval results. The transmission retrievals in the thermochemical-equilibrium case support high metallicity, low C/O, and a cloud-free atmosphere; in the free-chemistry case, they detect \ch{H2O} and an additional optical absorber.

Adopting a free-chemistry parameterization, \citet{Pluriel1} and \citet{ChangeatEtal2022apjsHSTeclipseHotJupiters} find, via occultation retrievals on the HST/WFC3 and Spitzer data of KELT-7b, an inverted T-P profile, a nondetection of \ch{H2O}, and a detection of $\mathrm{H}^-$ absorption. Although these findings are in agreement with our results assuming free chemistry, and although the free-chemistry retrievals generally yield better fits to the observations, we can conclude that assuming free constant-with-altitude abundances risks adopting unphysical scenarios. The dayside atmosphere of ultrahot Jupiters similar to KELT-7b is hot enough, and the chemical reaction rates are fast enough to overcome the effect of disequilibrium chemistry. The equilibrium-chemistry approach was also previously applied by \citet{ChangeatEtal2022apjsHSTeclipseHotJupiters}, finding via occultation retrievals solar to super-solar metallicities, C/O > 1, and a different thermal structure from the ones found by the free-chemistry runs. The equilibrium-chemistry retrieval on the transit spectrum does not lead to strong constraints in the transmission retrieval performed by \citet{ChangeatEtal2022apjsHSTeclipseHotJupiters}, whereas only the free-chemistry retrieval returns \ch{H2O} and $\mathrm{H}^-$ absorption, in agreement with the results presented by \citet{Pluriel1} and our findings. We can therefore conclude that the choice of a free-chemistry approach or a thermochemical-equilibrium chemistry is the main factor determining the retrieval results. 

Given KELT-7b is an ultrahot Jupiter, although near the lower limit from the viewpoint of equilibrium temperature, the preferred non-inverted T-P profile does not support the predictions of \citet{Hubeny1} and \citet{Fortney2}, similarly to the case of WASP-12b \citep{Sing3, Akin1}. However, as we mentioned in Sect. \ref{intro}, and as we showed in Sect. \ref{sec:retrieval_eclipse}, the identification of T-P inversion is a model-dependent process. The 3D GCM results support a TiO-induced temperature inversion, in tension with the results obtained via the thermochemical-equilibrium-based atmospheric retrievals. We can conclude that this discrepancy is because the occultation-retrieval models focus mainly on the HST data, while the 3D GCM model fits the CHEOPS, TESS, and Spitzer data, but underestimates the HST observations. Based on the retrieval results, we can also conclude that there might be a problem with the HST/WFC3 data in the case of this particular planet, which could be due to contamination of the data by stellar activity. As a consequence, the HST observations of KELT-7b suggest a high brightness temperature gradient that is difficult to reconcile with self-consistent atmospheric models. To support this argument, we independently reduced the HST/WFC3 transmission spectrum. Our output is in full agreement with \citet{Pluriel1}'s, which means that with instrumental artifacts we cannot explain the poor thermochemical-equilibrium-retrievals fit of the HST/WFC3 transmission spectrum. We applied coherent stellar variability treatment on TESS and CHEOPS occultation measurements of KELT-7b, commensurate with the known stellar activity of the host star \citep{Tabernero1}. We conclude that HST/WFC3 observations of KELT-7b would also benefit from a coherent stellar variability treatment as proposed by \citet{saba2025}.     

Although our attempt to understand the atmosphere of the ultrahot Jupiter KELT-7b with CHEOPS, TESS, and additional data ended with a discrepancy in the T-P profiles, depending on which approach and data are used, we report for the exoplanet KELT-7b a very low geometric albedo of $A_\mathrm{g} = 0.05 \pm 0.06$ in the CHEOPS and TESS passbands, which corresponds to a brightness temperature of $T_\mathrm{day} =2387^{+123}_{-159}\,\mathrm{K}$. This supports previous observations that hot Jupiters have, in general, low geometric albedos. The very low geometric albedo is consistent with heat distribution $\epsilon$ being close to zero, and also consistent with the occultation retrieval results and with a 3D GCM simulation that includes magnetic drag ($\tau_\mathrm{drag}=10^4\,\mathrm{s}$). Utilizing the eclipse and transit spectra, we also derived upper limits for the Bond albedo, assuming $\epsilon = 0$, finding values consistent with zero within the given uncertainties. We can conclude that the planet effectively absorbs nearly all incoming stellar irradiation to heat its dayside atmosphere.  

Given the rapid rotation of the host star, we also probed the precise CHEOPS transit light curves from the viewpoint of transit asymmetry. Unfortunately, several astrophysical effects in such systems remain poorly understood. The von Zeipel theorem \citep{Zeipel1, Zeipel2} is not strictly valid, and hence it needs further investigation. For example, \citet{Claret2} found significant deviations from the von Zeipel theorem at the upper layers of a distorted star in radiative equilibrium. Based on the CHEOPS photometry, we are unable to place any meaningful constraint on the sky-projected orbital obliquity. The obtained value is $\lambda = 8 \pm 105\,\mathrm{deg}$. On the other hand, we find that the stellar inclination is $I_\mathrm{s} = 86 \pm 25\,\deg$. We can conclude that additional CHEOPS observations would be necessary to put significant constraints on the sky-projected orbital obliquity via photometric methods.

\section*{Data availability}

Photometry data of KELT-7 used in this work are available in electronic form at the CDS via anonymous ftp to \url{cdsarc.u-strasbg.fr} (130.79.128.5) or via \url{http://cdsweb.u-strasbg.fr/cgi-bin/qcat?J/A+A/}.

\begin{acknowledgements}
We thank Quentin Changeat for the constructive comments and suggestions. CHEOPS is an ESA mission in partnership with Switzerland with important contributions to the payload and the ground segment from Austria, Belgium, France, Germany, Hungary, Italy, Portugal, Spain, Sweden, and the United Kingdom. The CHEOPS Consortium would like to gratefully acknowledge the support received by all the agencies, offices, universities, and industries involved. Their flexibility and willingness to explore new approaches were essential to the success of this mission. CHEOPS data analysed in this article will be made available in the CHEOPS mission archive (\url{https://cheops.unige.ch/archive_browser/}). This paper includes data collected with the TESS mission, obtained from the MAST data archive at the Space Telescope Science Institute (STScI). Funding for the TESS mission is provided by the NASA Explorer Program. STScI is operated by the Association of Universities for Research in Astronomy, Inc., under NASA contract NAS 5-26555. This research made use of the open source Python package exoctk, the Exoplanet Characterization Toolkit \citep{bourque2021}. The authors thank M. Zhang for his help with the configuration of \textsc{platon}. ZG was supported by the VEGA grant of the Slovak Academy of Sciences No. 2/0031/22 and by the Slovak Research and Development Agency - the contract No. APVV-20-0148. P.E.C. is funded by the Austrian Science Fund (FWF) Erwin Schroedinger Fellowship, program J4595-N. LBo, GBr, VNa, IPa, GPi, RRa, GSc, VSi, and TZi acknowledge support from CHEOPS ASI-INAF agreement n. 2019-29-HH.0. TWi acknowledges support from the UKSA and the University of Warwick. ABr was supported by the SNSA. MNG is the ESA CHEOPS Project Scientist and Mission Representative, and as such also responsible for the Guest Observers (GO) Programme. MNG does not relay proprietary information between the GO and Guaranteed Time Observation (GTO) Programmes, and does not decide on the definition and target selection of the GTO Programme. LCa and CHe acknowledge financial support from the Österreichische Akademie der Wissenschaften and from the European Union H2020-MSCA-ITN-2019 under Grant Agreement no. 860470 (CHAMELEON). Calculations were performed using supercomputer resources provided by the Vienna Scientific Cluster (VSC). ML acknowledges the support of the Swiss National Science Foundation under grant number PCEFP2\_194576. This work was supported by FCT - Funda\c{c}\~{a}o para a Ci\^{e}ncia e a Tecnologia through national funds and by FEDER through COMPETE2020 through the research grants UIDB/04434/2020, UIDP/04434/2020, 2022.06962.PTDC. O.D.S.D. is supported in the form of a work contract (DL 57/2016/CP1364/CT0004) funded by national funds through FCT. YAl acknowledges support from the Swiss National Science Foundation (SNSF) under grant 200020\_192038. We acknowledge financial support from the Agencia Estatal de Investigación of the Ministerio de Ciencia e Innovación MCIN/AEI/10.13039/501100011033 and the ERDF “A way of making Europe” through projects PID2019-107061GB-C61, PID2019-107061GB-C66, PID2021-125627OB-C31, PID2021-125627OB-C32, and PID2023-150468NB-I00, from the Centre of Excellence “Severo Ochoa” award to the Instituto de Astrofísica de Canarias (CEX2019-000920-S), from the Centre of Excellence “María de Maeztu” award to the Institut de Ciències de l’Espai (CEX2020-001058-M), and from the Generalitat de Catalunya/CERCA programme. C.B. acknowledges support from the Swiss Space Office through the ESA PRODEX program. This work has been carried out within the framework of the NCCR PlanetS supported by the Swiss National Science Foundation under grants 51NF40\_182901 and 51NF40\_205606.  ACC acknowledges support from STFC consolidated grant number ST/V000861/1, and UKSA grant number ST/X002217/1. ACMC acknowledges support from the FCT, Portugal, through the CFisUC projects UIDB/04564/2020 and UIDP/04564/2020, with DOI identifiers 10.54499/UIDB/04564/2020 and 10.54499/UIDP/04564/2020, respectively. This project was supported by the CNES. B.-O. D. acknowledges support from the Swiss State Secretariat for Education, Research and Innovation (SERI) under contract number MB22.00046. A.C., A.D., B.E., K.G., and J.K. acknowledge their role as ESA-appointed CHEOPS Science Team Members. This project has received funding from the Swiss National Science Foundation for project 200021\_200726. It has also been carried out within the framework of the National Centre of Competence in Research PlanetS supported by the Swiss National Science Foundation under grant 51NF40\_205606. The authors acknowledge the financial support of the SNSF. MF and CMP gratefully acknowledge the support of the Swedish National Space Agency (DNR 65/19, 174/18). DG gratefully acknowledges financial support from the CRT foundation under Grant No. 2018.2323 “Gaseous or rocky? Unveiling the nature of small worlds”. M.G. is an F.R.S.-FNRS Senior Research Associate. CHe acknowledges the European Union H2020-MSCA-ITN-2019 under Grant Agreement no. 860470 (CHAMELEON), and the HPC facilities at the Vienna Science Cluster (VSC). KGI is the ESA CHEOPS Project Scientist and is responsible for the ESA CHEOPS Guest Observers Programme. She does not participate in, or contribute to, the definition of the Guaranteed Time Programme of the CHEOPS mission through which observations described in this paper have been taken, nor to any aspect of target selection for the programme. K.W.F.L. was supported by Deutsche Forschungsgemeinschaft grants RA714/14-1 within the DFG Schwerpunkt SPP 1992, Exploring the Diversity of Extrasolar Planets. This work was granted access to the HPC resources of MesoPSL financed by the Region Ile de France and the project Equip@Meso (reference ANR-10-EQPX-29-01) of the programme Investissements d'Avenir supervised by the Agence Nationale pour la Recherche. PM acknowledges support from STFC research grant number ST/R000638/1. This work was also partially supported by a grant from the Simons Foundation (PI Queloz, grant number 327127). NCSa acknowledges funding by the European Union (ERC, FIERCE, 101052347). Views and opinions expressed are however those of the author(s) only and do not necessarily reflect those of the European Union or the European Research Council. Neither the European Union nor the granting authority can be held responsible for them. A. S. acknowledges support from the Swiss Space Office through the ESA PRODEX program. S.G.S. acknowledges support from FCT through FCT contract nr. CEECIND/00826/2018 and POPH/FSE (EC). The Portuguese team thanks the Portuguese Space Agency for the provision of financial support in the framework of the PRODEX Programme of the European Space Agency (ESA) under contract number 4000142255. GyMSz acknowledges the support of the Hungarian National Research, Development and Innovation Office (NKFIH) grant K-125015, a PRODEX Experiment Agreement No. 4000137122, the Lend\"ulet LP2018-7/2021 grant of the Hungarian Academy of Science and the support of the city of Szombathely. V.V.G. is an F.R.S-FNRS Research Associate. JV acknowledges support from the Swiss National Science Foundation (SNSF) under grant PZ00P2\_208945. EV acknowledges support from the ‘DISCOBOLO’ project funded by the Spanish Ministerio de Ciencia, Innovación y Universidades under grant PID2021-127289NB-I00. NAW acknowledges UKSA grant ST/R004838/1. TZi acknowledges NVIDIA Academic Hardware Grant Program for the use of the Titan V GPU card and the Italian MUR Departments of Excellence grant 2023-2027 “Quantum Frontiers”. This project has received funding from the HUN-REN Hungarian Research Network. 

\end{acknowledgements}

\bibliographystyle{aa} 
\bibliography{Yourfile}

\begin{appendix}

\section{Supplementary light curves}
\label{addfigs}

Here we present supplementary light curves related to the joint fit of the CHEOPS and TESS data for eclipse detection (see Sect. \ref{sececl}).

\begin{@twocolumnfalse}
\begin{figure}[h]
\parbox{\textwidth}{
\centering
\centerline{
\includegraphics[width=0.95\columnwidth]{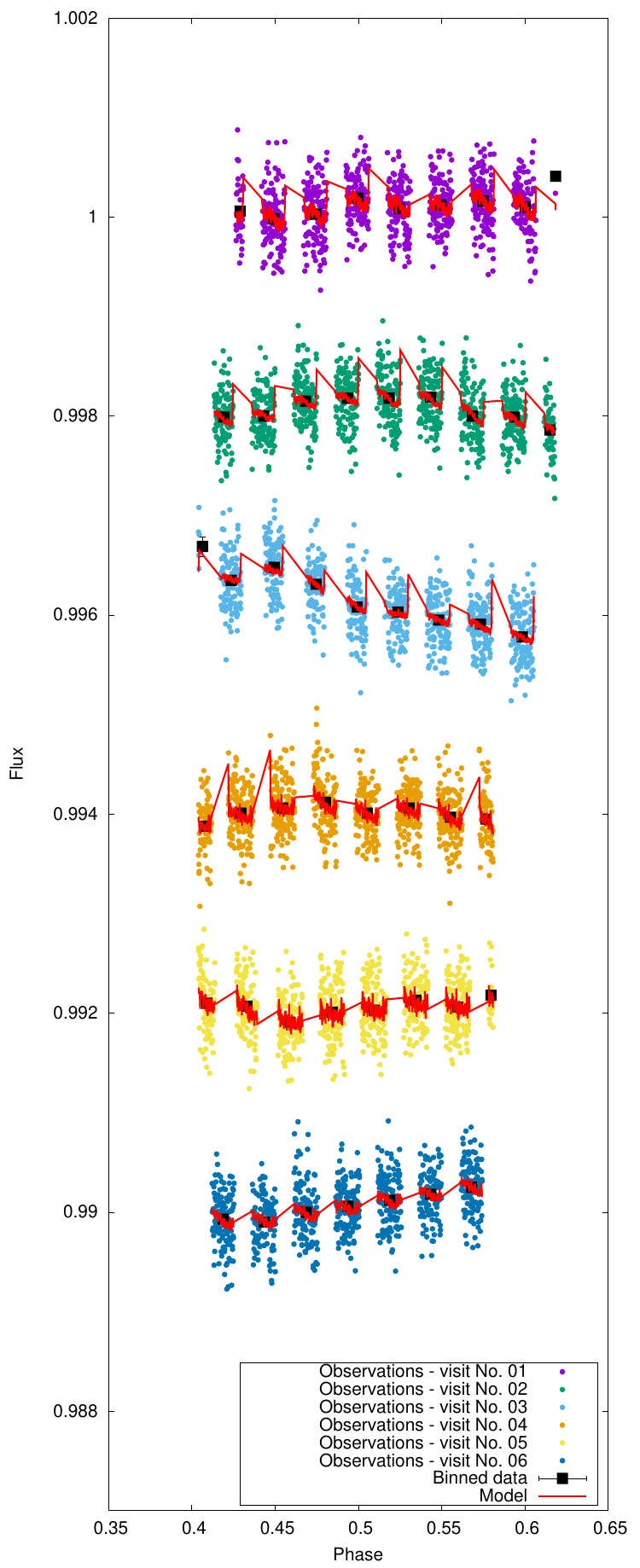}
\includegraphics[width=0.95\columnwidth]{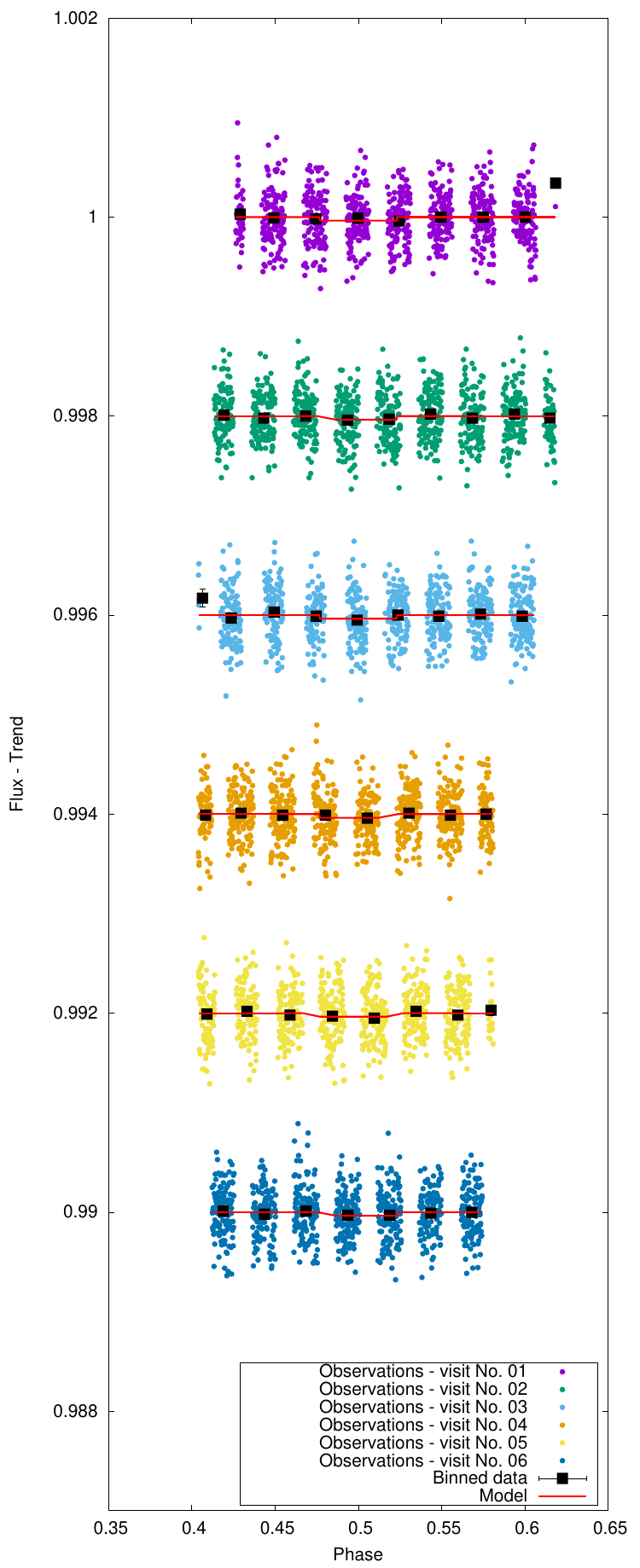}}
\caption{Phase-folded, and binned individual CHEOPS secondary eclipse observations of KELT-7b from visits $1 - 6$, overplotted with the best-fitting \texttt{CONAN3} model and arbitrarily shifted in flux for clarity. The left panel shows the nondetrended data overplotted with the full model, while the right panel shows the detrended data overplotted with the occultation model.}
\label{cheopsindocc1}} 
\end{figure}
\end{@twocolumnfalse}
\clearpage

\begin{@twocolumnfalse}
\begin{figure}[h]
\parbox{\textwidth}{
\centering
\centerline{
\includegraphics[width=0.95\columnwidth]{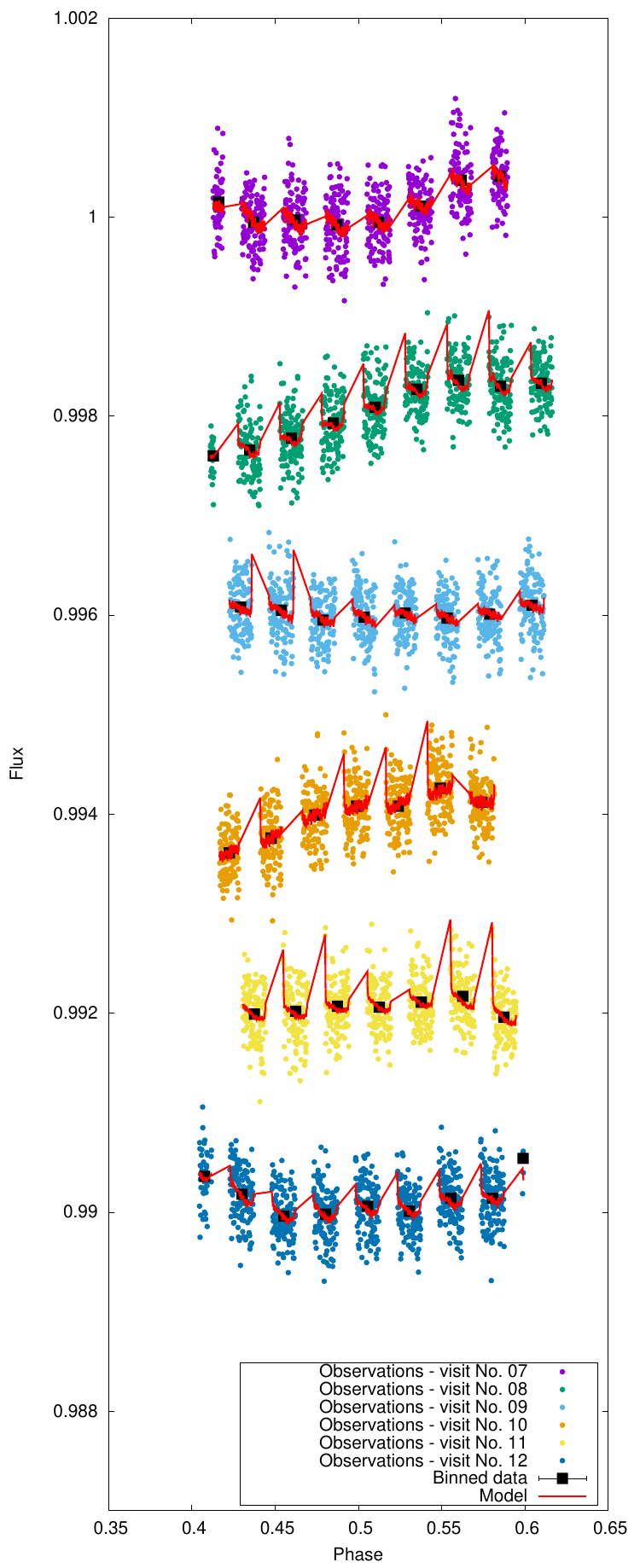}
\includegraphics[width=0.95\columnwidth]{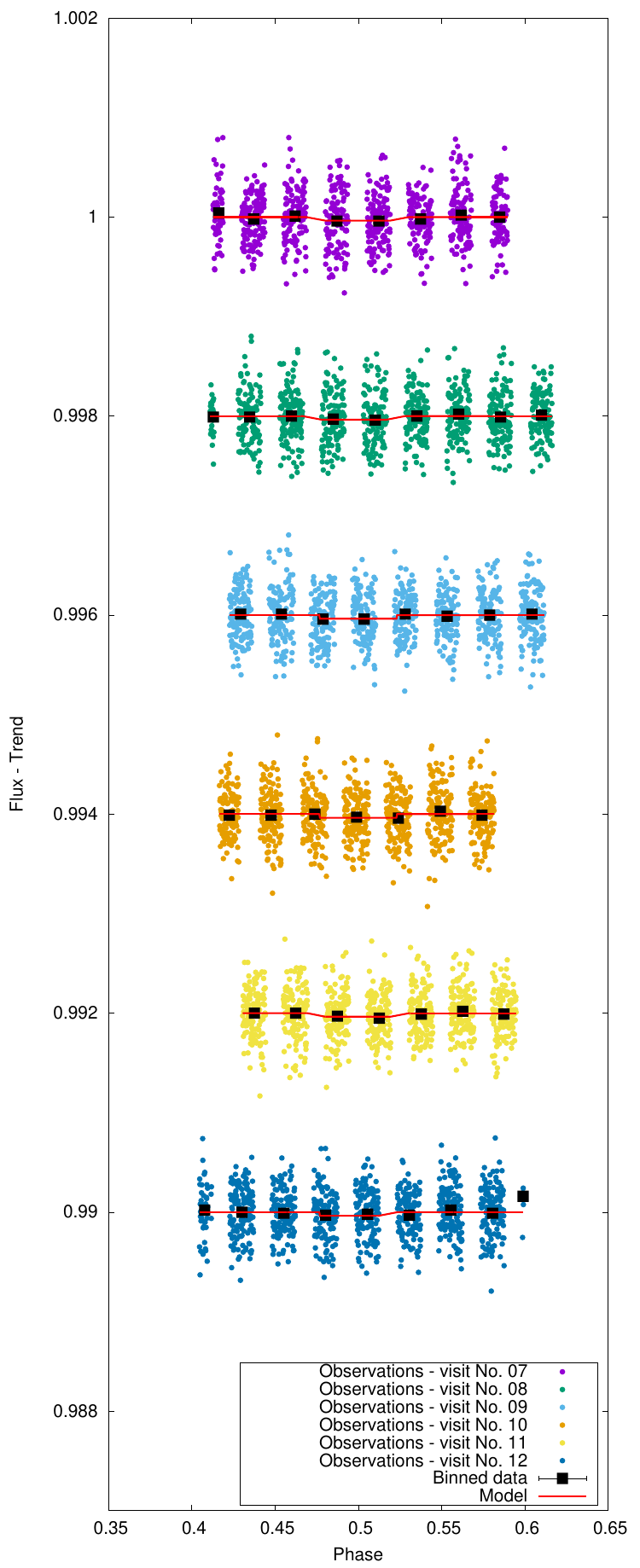}}
\caption{As in Fig. \ref{cheopsindocc1}, but for CHEOPS visits $7 - 12$.}
\label{cheopsindocc2}}
\end{figure}
\end{@twocolumnfalse}

\begin{figure*}
\centering
\centerline{
\includegraphics[width=0.85\columnwidth]{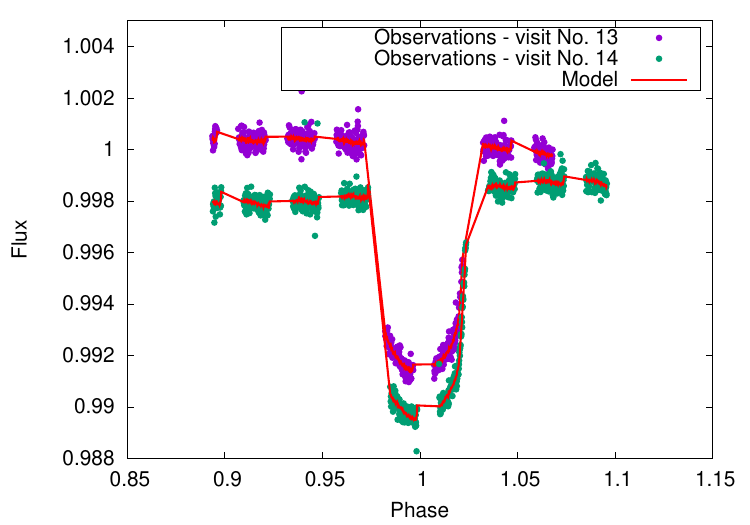}
\includegraphics[width=0.85\columnwidth]{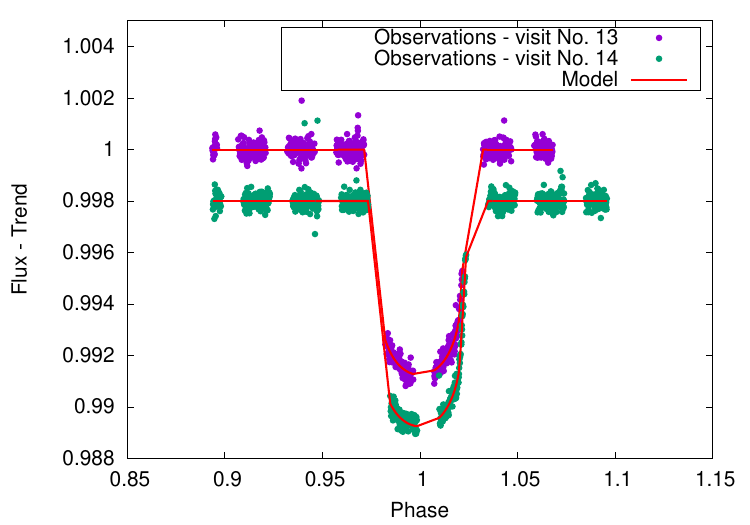}}
\caption{Phase-folded, individual CHEOPS transit observations of KELT-7b from visits $13 - 14$, overplotted with the best-fitting \texttt{CONAN3} model and arbitrarily shifted in flux for clarity. The left panel shows the nondetrended data overplotted with the full model, while the right panel shows the detrended data overplotted with the transit model.}
\label{cheopsobstraall} 
\end{figure*}

\begin{figure*}
\centering
\centerline{
\includegraphics[width=0.85\columnwidth]{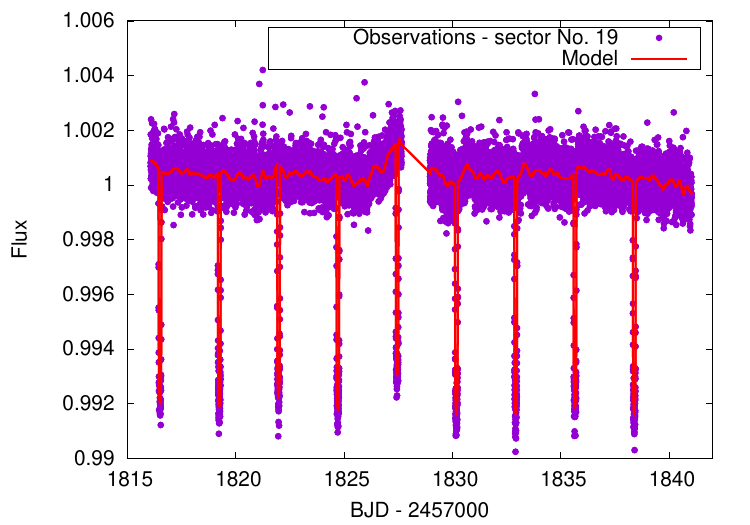}
\includegraphics[width=0.85\columnwidth]{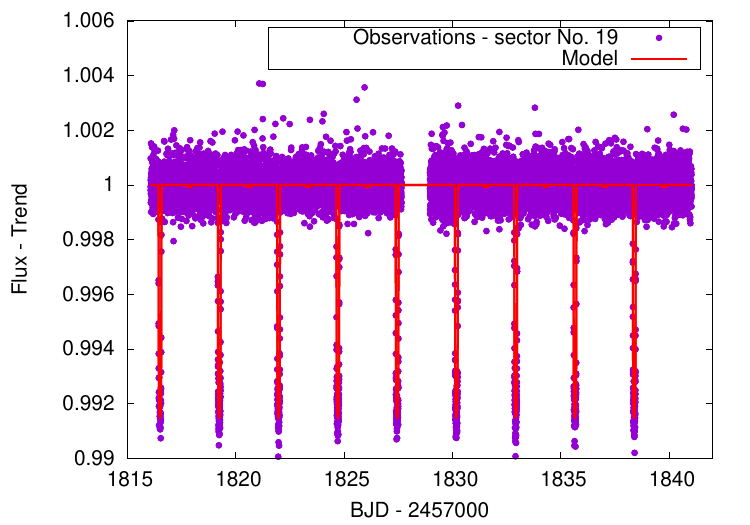}}
\centerline{
\includegraphics[width=0.85\columnwidth]{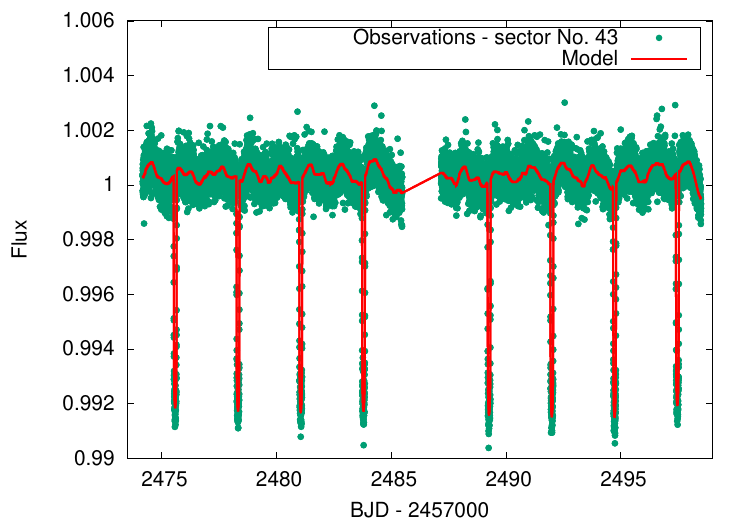}
\includegraphics[width=0.85\columnwidth]{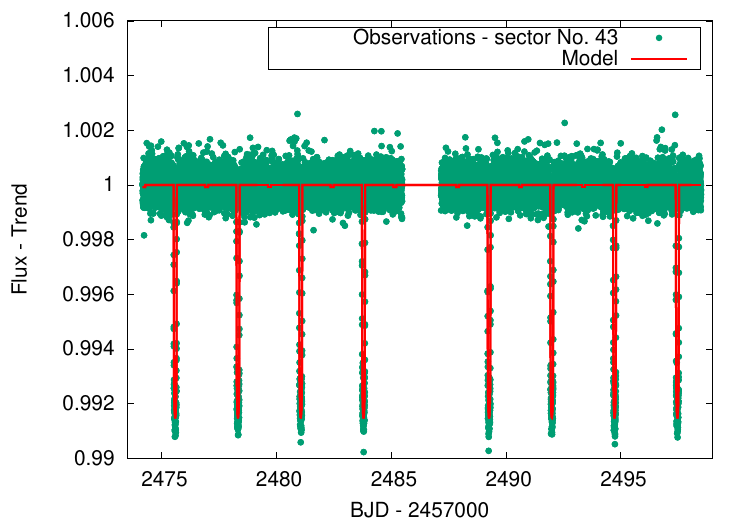}}
\centerline{
\includegraphics[width=0.85\columnwidth]{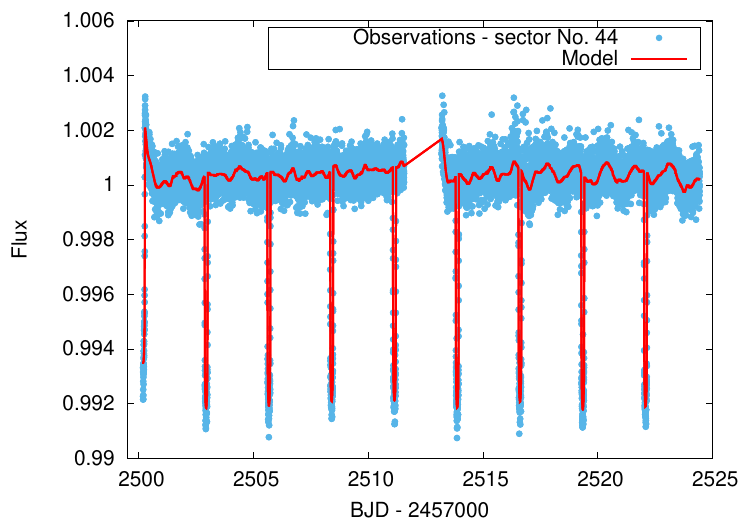}
\includegraphics[width=0.85\columnwidth]{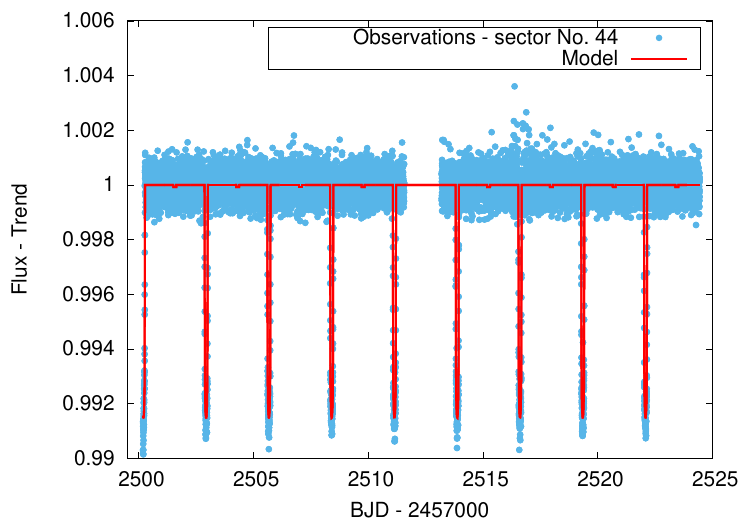}}
\caption{TESS observations of KELT-7b from Sectors 19, 43, and 44, overplotted with the best-fitting \texttt{CONAN3} model. The left panels show the nondetrended data overplotted with the full model, while the right panels show the detrended data overplotted with the transit and occultation model.}
\label{tessallsec1} 
\end{figure*}

\begin{figure*}
\centering
\centerline{
\includegraphics[width=0.85\columnwidth]{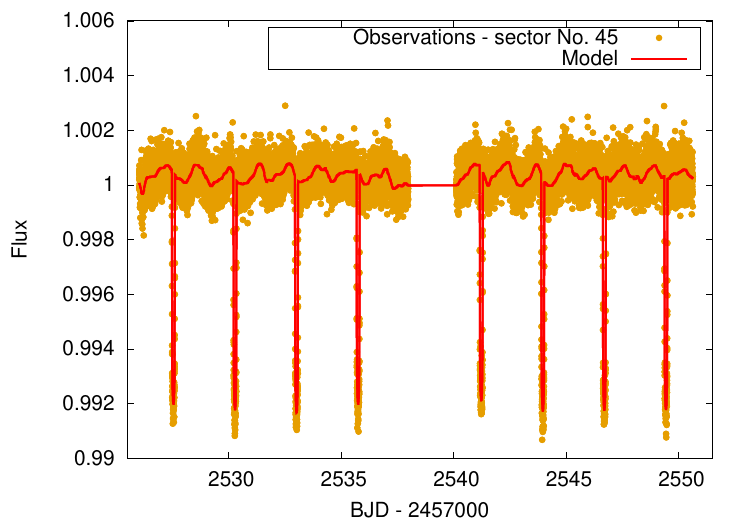}
\includegraphics[width=0.85\columnwidth]{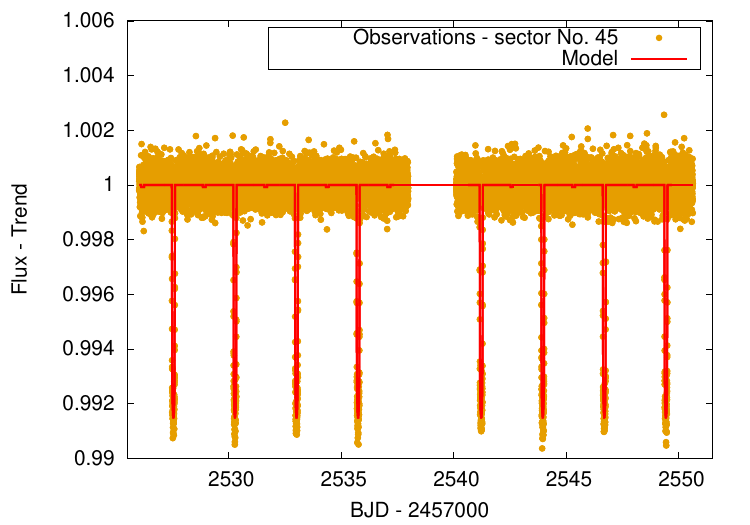}}
\centerline{
\includegraphics[width=0.85\columnwidth]{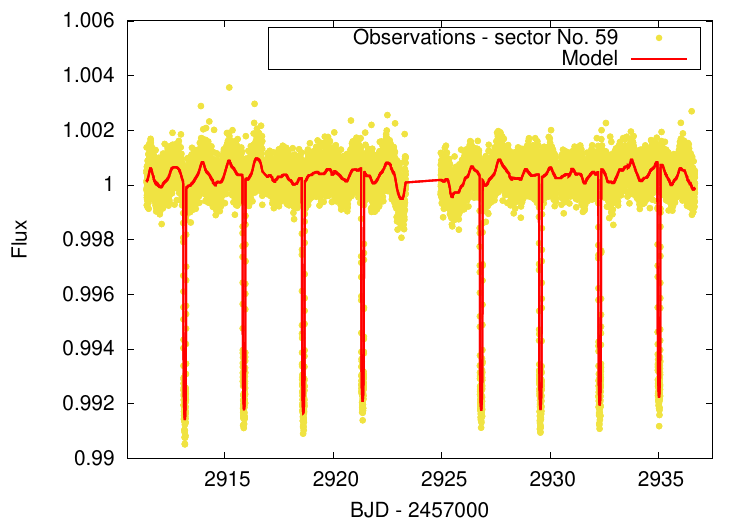}
\includegraphics[width=0.85\columnwidth]{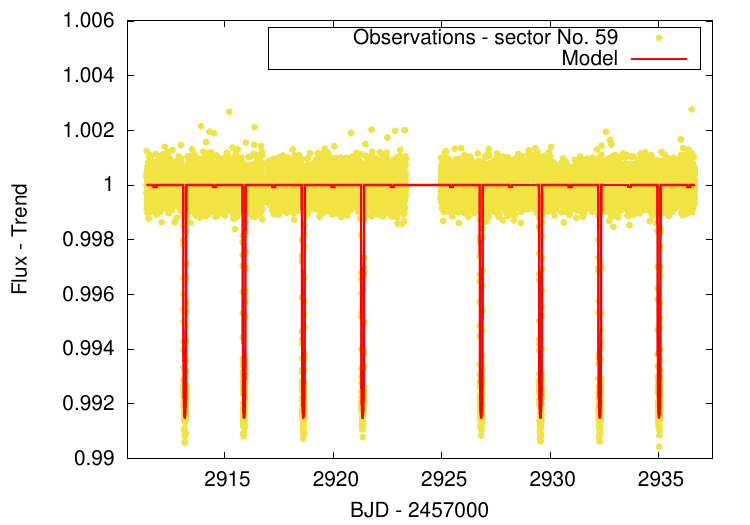}}
\centerline{
\includegraphics[width=0.85\columnwidth]{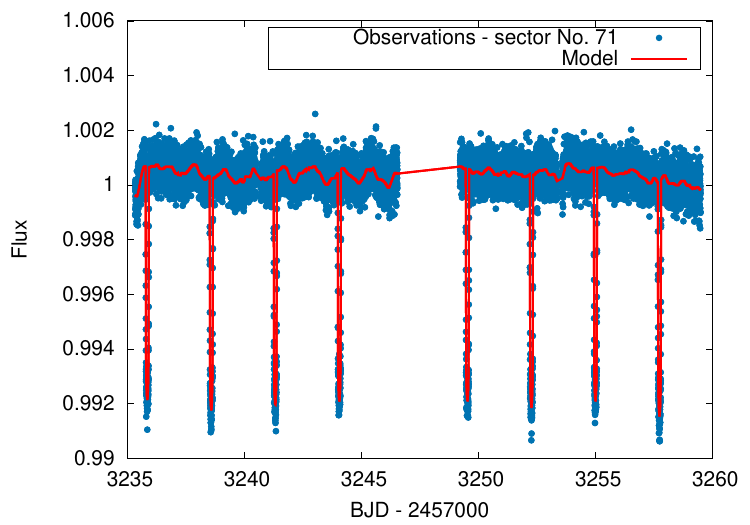}
\includegraphics[width=0.85\columnwidth]{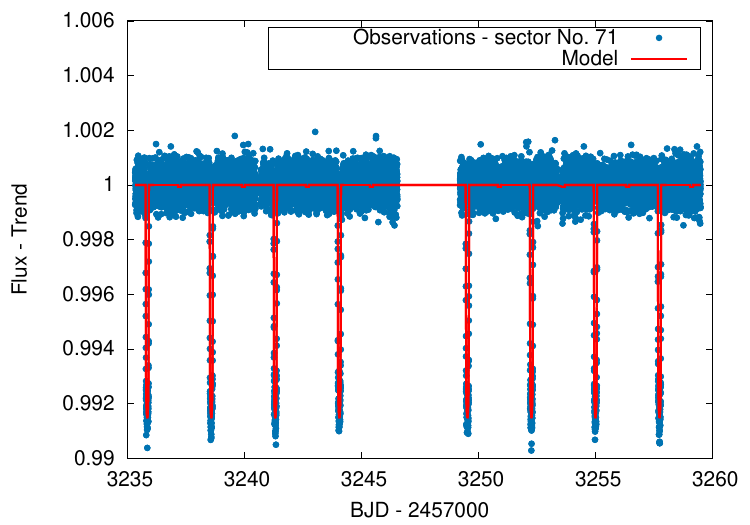}}
\centerline{
\includegraphics[width=0.85\columnwidth]{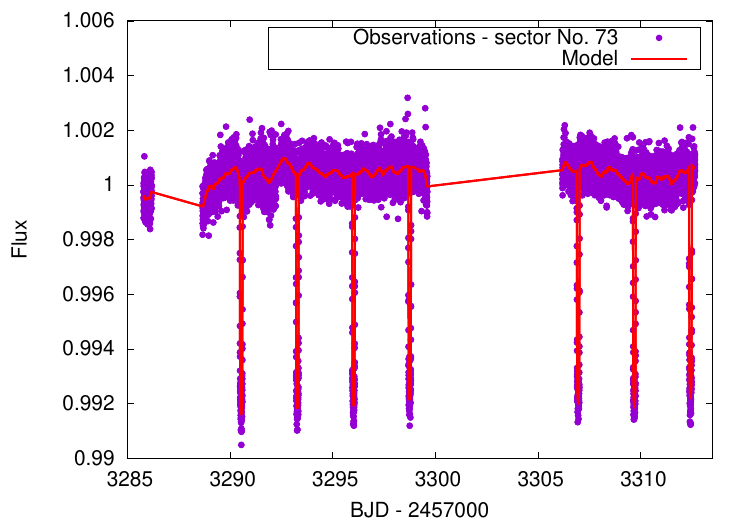}
\includegraphics[width=0.85\columnwidth]{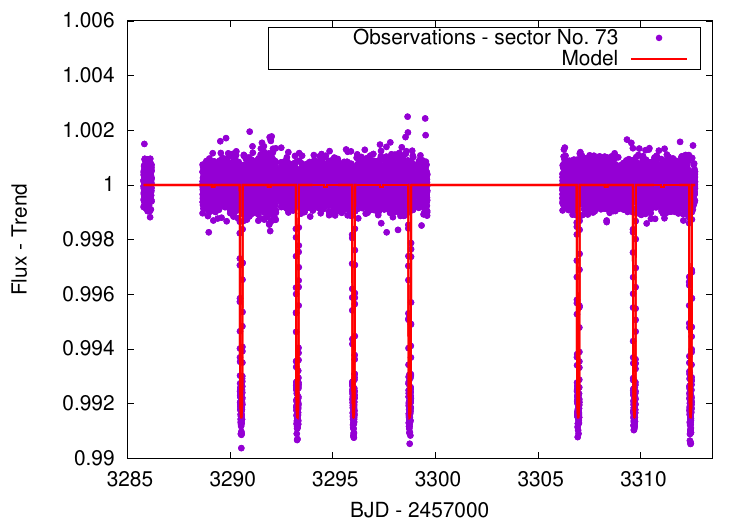}}
\caption{As in Fig. \ref{tessallsec1}, but for the TESS Sectors 45, 59, 71, and 73.}
\label{tessallsec2}
\end{figure*}
\clearpage

\FloatBarrier

\section{Independent HST/WFC3 transit data reduction}

Here we present diagnostic plots of the HST/WFC3 data reduction discussed in Sect. \ref{sec:wfc3_reduction}. Each plot, corresponding to one transit, shows all spectra of that visit divided by the median of the visit's first ten spectra.

\begin{figure}[hb]
\centering
\includegraphics[width=\columnwidth]{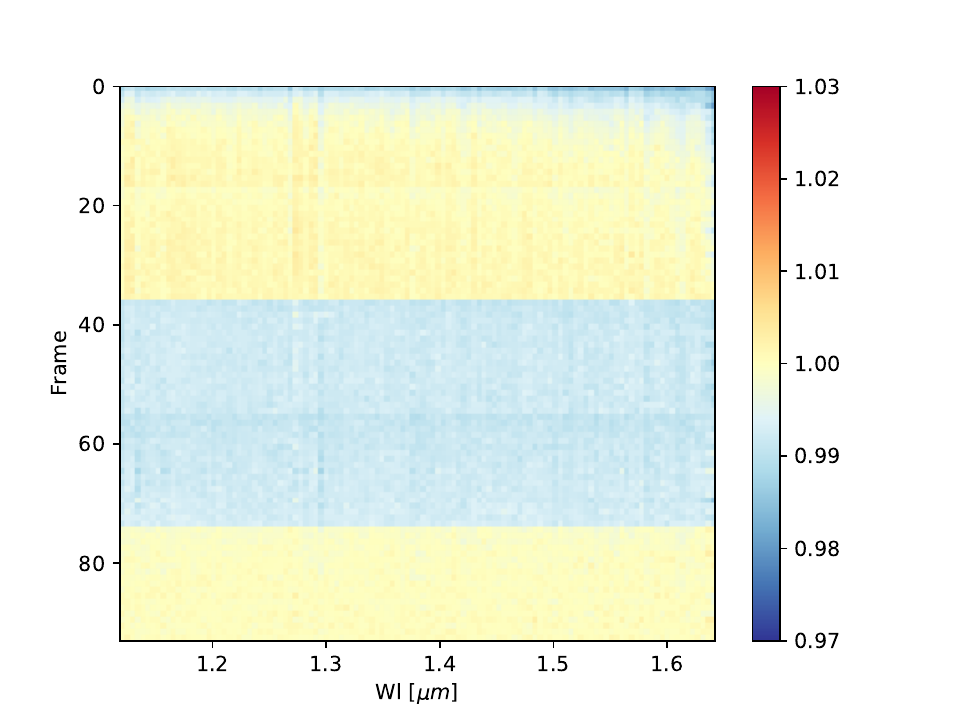}\\
\includegraphics[width=\columnwidth]{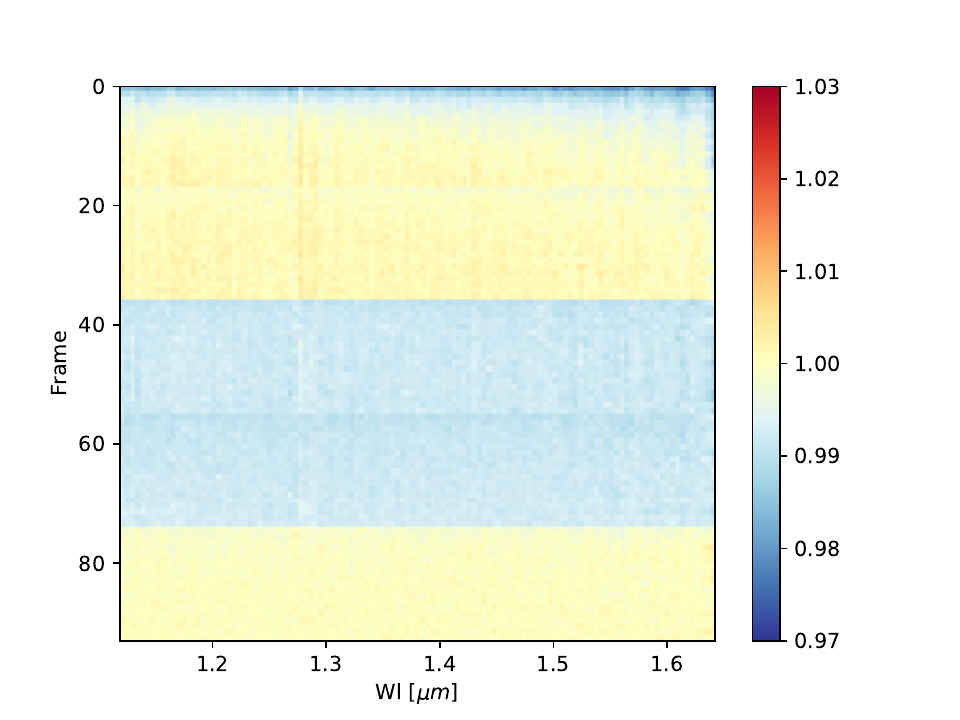}
\caption{Normalized time sequence of the HST/WFC3 spectra corresponding to the two transits reduced as in Sect. \ref{sec:wfc3_reduction}. The $y$-axis represents increasing time with increasing frame number.}
\label{fig:spectra_top}
\end{figure}

\FloatBarrier

\vfill\break

\section{Detailed GCM set-up, chemistry and climate diagnostics}
\label{Detail_GCM}

The GCM \texttt{expeRT/MITgcm} \citep{Schneider2022} used in this work as part of the \texttt{ExoRad} climate framework \citep{Carone2020} uses the \texttt{MITgcm} dynamical core that solves the hydrostatic primitive equations \citep[see, e.g.,][]{Showman2009} in an Arakawa C-type cubed-sphere \citep{Adcroft2004}. The horizontal grid comprises $128 \times 64$ cells in longitude and latitude, respectively. The vertical grid consists of 47 cells in total: 41 logarithmically spaced cells between $10^{-5}\,\mathrm{bar}$ and 100\,bar, and six linearly spaced cells between 100\,bar and 700\,bar. 
        
A fourth-order Shapiro filter is applied to the horizontal wind field, which suppresses small grid-scale noise. The damping timescale of the filter is set to $\tau_\mathrm{shap}=25\,\mathrm{s}$, equal to the dynamical time step $\Delta t=25\,\mathrm{s}$.\footnote{The damping timescale $\tau_\mathrm{shap}$ corresponds to the dissipation time $\tau_{\nu}$ used in \citet{Heng2} to compare horizontal dissipation in different dynamical cores.} The GCM is further stabilized against gravity wave reflection on top of the modeling domain with a sponge layer between $10^{-4}$ and $10^{-5}$. In this layer, the zonal horizontal velocity $u$ is damped towards its longitudinally averaged mean $\bar{u}$ via the following relation:

\begin{equation}
\frac{\mathrm{d}u}{\mathrm{d}t} = - k \left(u-\bar{u}\right). 
\end{equation}
        
\noindent Here, $t$ is time and $k$ is the strength of the Rayleigh prescription applied in the sponge layer, depending on pressure $p$ as

\begin{equation}
k = k_{\rm top} \cdot \mathrm{max} \left[0,1 - \left(\frac{p}{p_{\rm sponge}}\right)^2 \right]^{2}.
\end{equation}
        
\noindent The control parameters $p_{\rm sponge}$ and $k_{\rm top}$ determine the position and strength of friction in the sponge layer. In this paper, the default values of $k_{\rm top} = 20$\,days$^{-1}$ and $p_{\rm sponge} = 10^{-4}\,\mathrm{bar}$ are used \citep{Carone2020}.
        
To stabilize the model against shear flow instabilities at the bottom boundary, basal drag is applied to the zonal wind velocity $u$ and meridional wind velocity $v$ in pressure layers deeper than 400\,bar via:

\begin{align}
&\frac{\mathrm{d}u}{\mathrm{d}t} = -k_{\rm deep}\cdot u,\\
&\frac{\mathrm{d}v}{\mathrm{d}t} = -k_{\rm deep}\cdot v,
\end{align}

\noindent where the control parameter $k_{\rm deep}$ is defined as 
        
\begin{equation}
k_{\rm deep} = k_{\rm bottom} \cdot \mathrm{max} \left[0, \frac{p-490\,\rm bar}{700\,\rm bar - 490\,\rm bar} \right]
\end{equation}

\noindent with $k_{\rm bottom}=1$~day$^{-1}$.
        
The model is started without wind flow and an initial analytical 1D temperature profile, following the formalism of \citet{Parmentier2015} with the host star's parameters as listed in Sect.~\ref{host}, a semi-major axis of 0.0442\,AU, and an interior temperature of $T_{\rm int}=663\,\mathrm{K}$, following the fit of \citet{Thorngren2019}. The model is run with the dynamical timestep of $\Delta t = 25\,\mathrm{s}$, where fluxes are recalculated every fourth dynamical timestep. The model is run for a 1000-day simulation time to ensure that the temperature structure does not evolve further in the modeling domain in the last 100 days. The final model output is derived by time averaging over the last 100 days of the whole simulation run. We note that there is some discrepancy in the host star parameters because some data were reduced based on the parameters from \citet{Pluriel1}. However, we note that the GCM here is used to provide a physical background to the possible climate state of this planet. A small change in the host star temperature will not strongly shift the climate states, for example, the need to include TiO/VO and magnetic drag to explain the overall flux. Furthermore, we note that we match CHEOPS, TESS, and Spitzer data sufficiently well to give us confidence that we capture the basic thermodynamics of the planetary atmosphere. Performance and stability tests for the sponge layer and basal drag are presented in \citet{Carone2020}. A more detailed description of the radiative transfer implementation and performance tests can be found in \citet{Schneider2022}.

\begin{figure*}
\centering
\includegraphics[width=\textwidth]{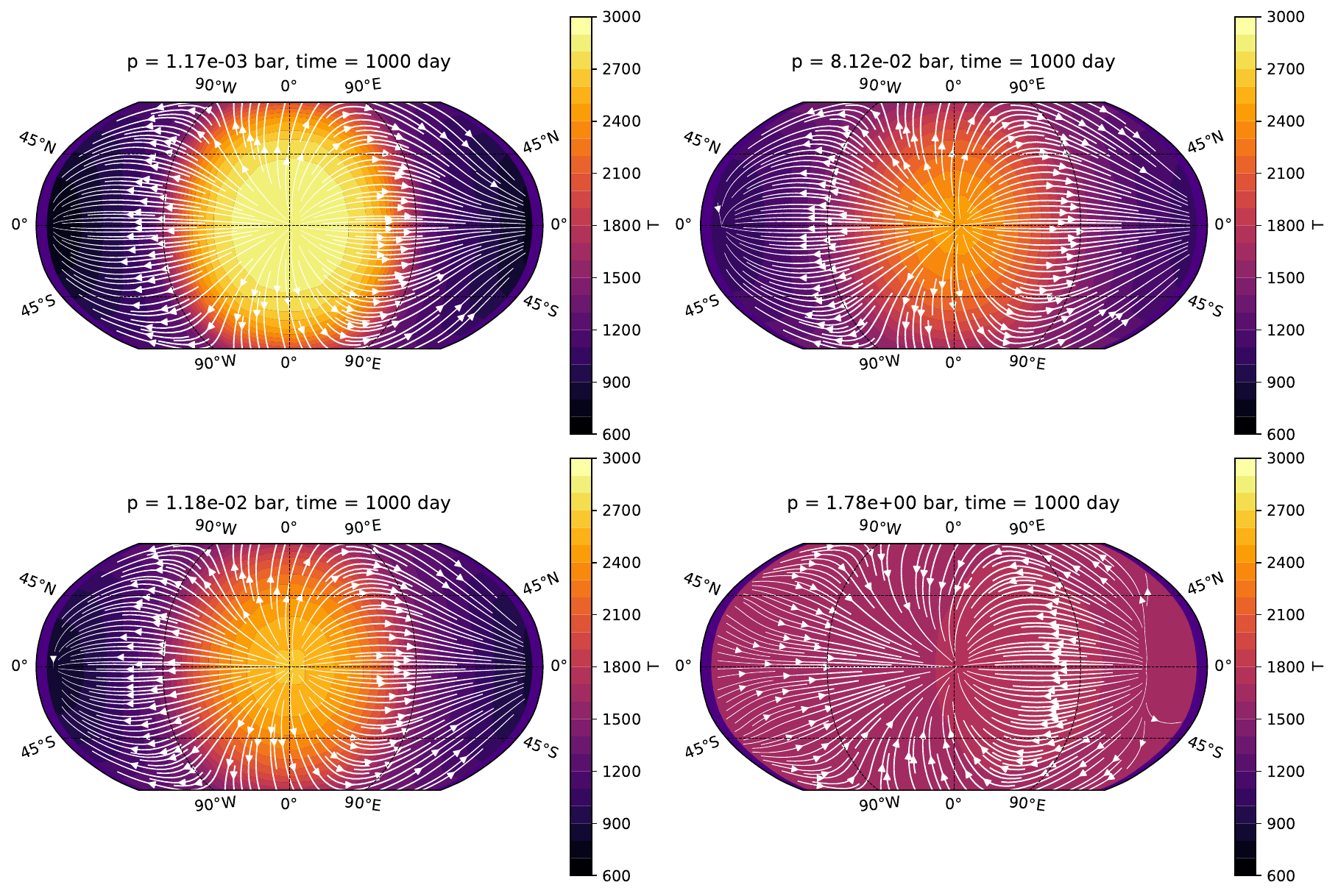}
\caption{Horizontal slices in the GCM with TiO and strong magnetic drag of $\tau_{\rm drag}=10^4\,\mathrm{s}$ across four representative pressure levels. Colors denote the local gas temperatures, white lines and arrows depict the horizontal wind flow, including regions of strong divergence/convergence at the substellar point (lat.: $0^\circ$, lon.: $0^\circ)$.}
\label{3DTempwDragwTio}
\end{figure*}

\begin{figure*}
\centering
\includegraphics[width=\textwidth]{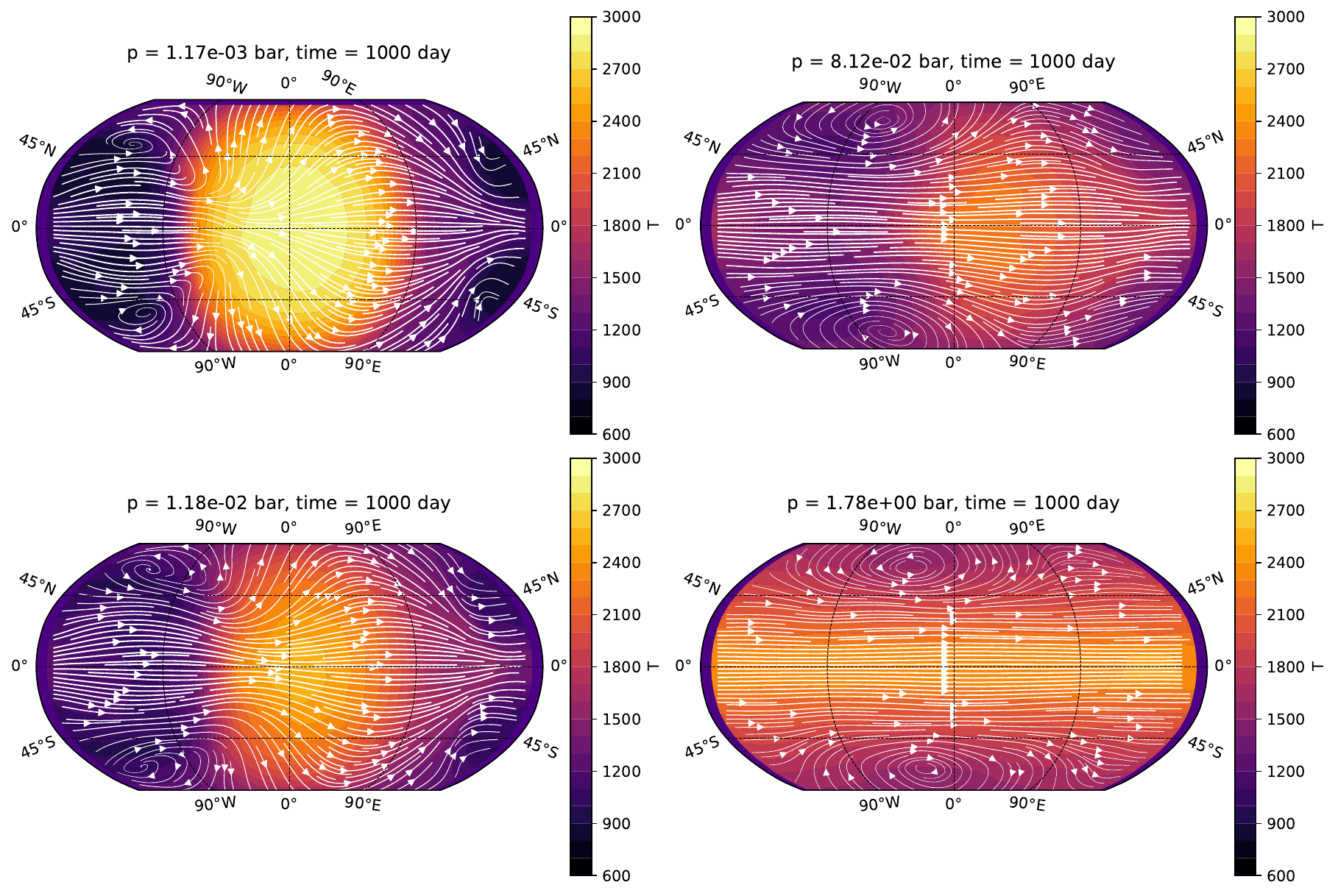}
\caption{Horizontal slices in the GCM with TiO and with no magnetic drag across four representative pressure levels. Colors denote the local gas temperatures, white lines and arrows depict the horizontal wind flow.}
\label{3DTempwDragwoTio}
\end{figure*}

In addition, a more detailed diagnostic of the climate state is useful to understand the impact of model choice on the dayside emission spectrum based on the GCM (see Fig. \ref{3DEmisison}). Figure \ref{3DTempwDragwTio} shows that for a climate model with TiO and magnetic drag, the dayside is heated up strongly between  $p=10^{-2} - 10^{-3}$\,bar (see also Fig. \ref{3DEmisison}, right panel). Due to strong magnetic drag, superrotation (fast eastward wind jet along the equator) is effectively suppressed. Instead, direct radial flow from the day to the nightside is present for $p < 10^{-2}\,\mathrm{bar}$. This climate state leads to a strong divergence in the horizontal wind flow field, which is closed by wind flow convergence in deeper layers ($p > 1$\,bar). The choice of $\tau_{\rm drag}=10^4\,\mathrm{s}$ has been shown in several GCM simulations for similar ultrahot Jupiters such as WASP-76b \citep{Demangeon2024} and WASP-18b (Deline et al., in review) to consistently lead to a climate state with no superrotation, and consequently, no eastward hot spot shift, as also shown here. In particular, for WASP-18b it becomes apparent that $\tau_{\rm drag}=10^4\,\mathrm{s}$, which effectively prohibits the formation of superrotation, appears to yield dayside emission that agrees remarkably well even with the JWST data. In contrast, a climate model containing TiO and no magnetic drag shows that the dayside hot spot is partly shifted eastwards due to the presence of a superrotating jet (see Fig. \ref{3DTempwDragwoTio}). Here, the regions of wind flow convergence are the Rossby gyres located at the morning terminator. Thus, with these two extreme choices in drag, we can explore two possible climate states or regimes that shape the dayside 3D temperature and chemistry that are measured with the eclipse spectra. Thus, a simulation with $\tau_{\rm drag} = 10^4\,\mathrm{s}$ represents a markedly different climate regime, dominated by radial wind flow rather than eastward jets. Such extreme scenarios can be used to meaningfully discuss efficient versus inefficient day-to-nightside heat transport. Higher $\tau_{\rm drag}$ (not explored here) typically leads to a climate with some remnant of superrotation, which would be an intermediate step, and here of limited usability. 

\begin{figure*}
\centering
\includegraphics[width=0.63\textwidth]{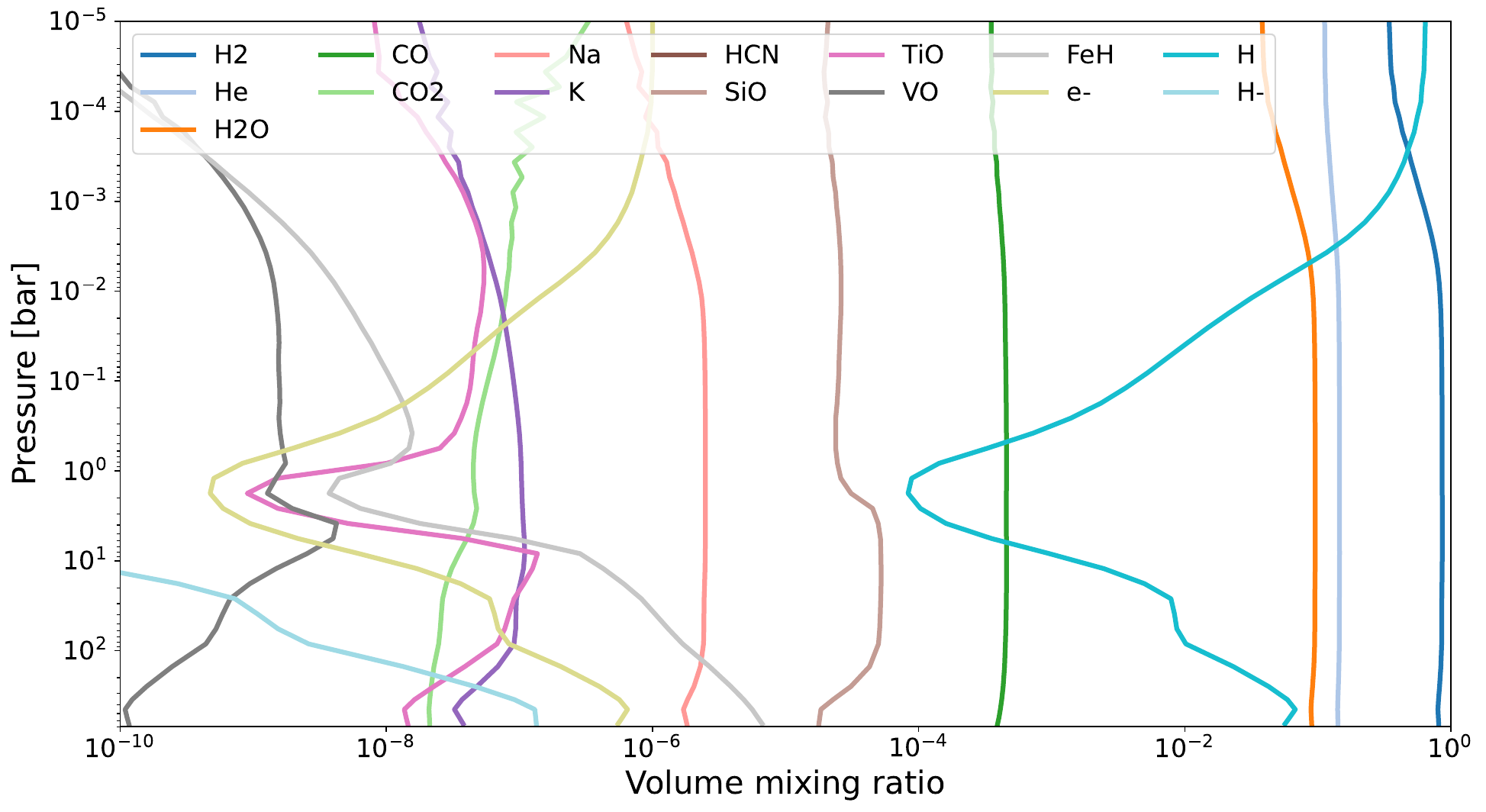}
\includegraphics[width=0.3\textwidth]{KELT-7/Kelt7b_TiO_with_drag_Tday.pdf}
\caption{Left panel: Dayside averaged equilibrium gas-phase chemistry from the GCM with TiO and VO and with strong magnetic drag. Right panel: Dayside average T-P profile.}
\label{3DTempwDragChem}
\end{figure*}

\begin{figure*}
\centering
\includegraphics[width=0.64\textwidth]{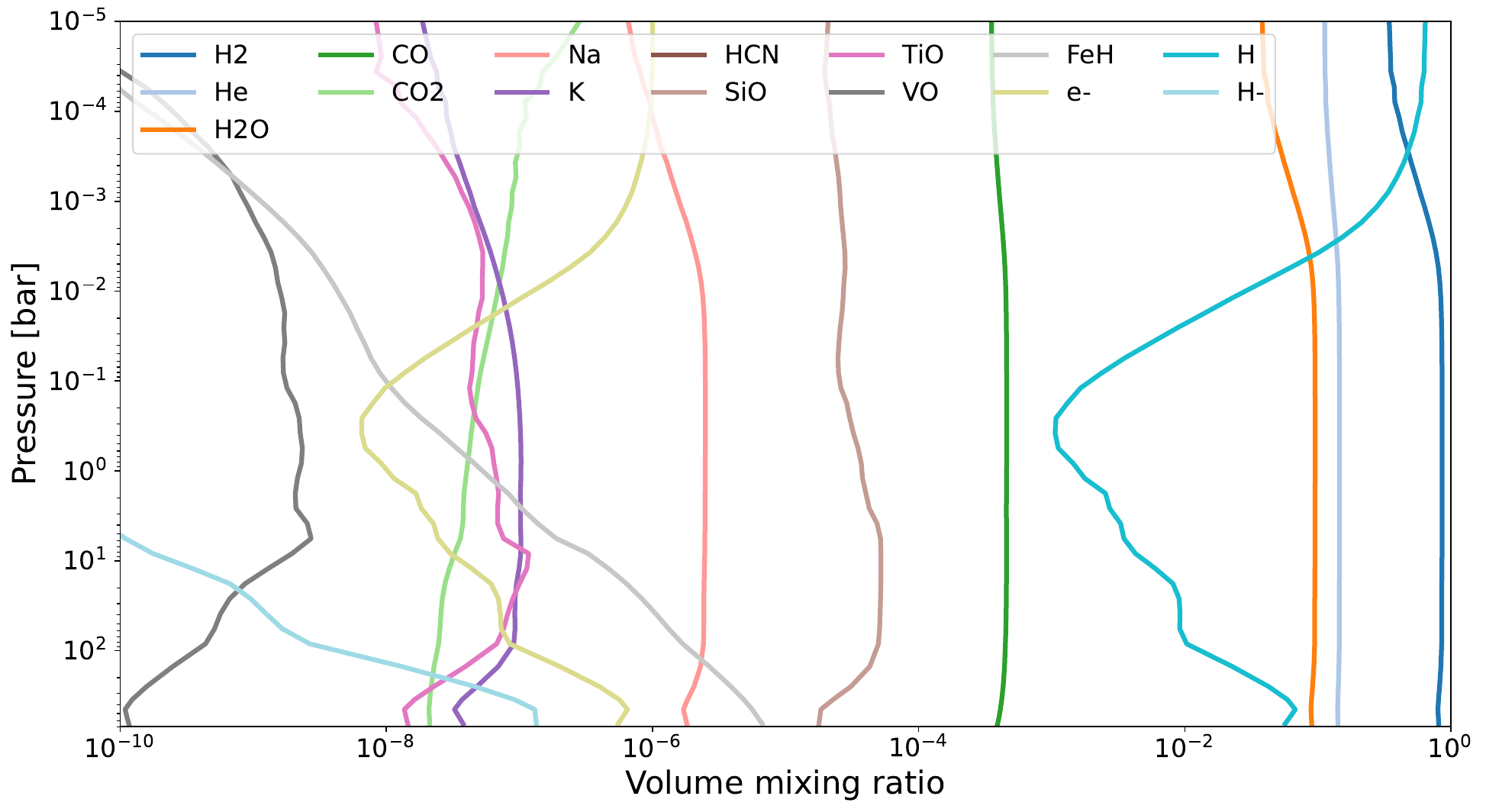}
\includegraphics[width=0.3
\textwidth]{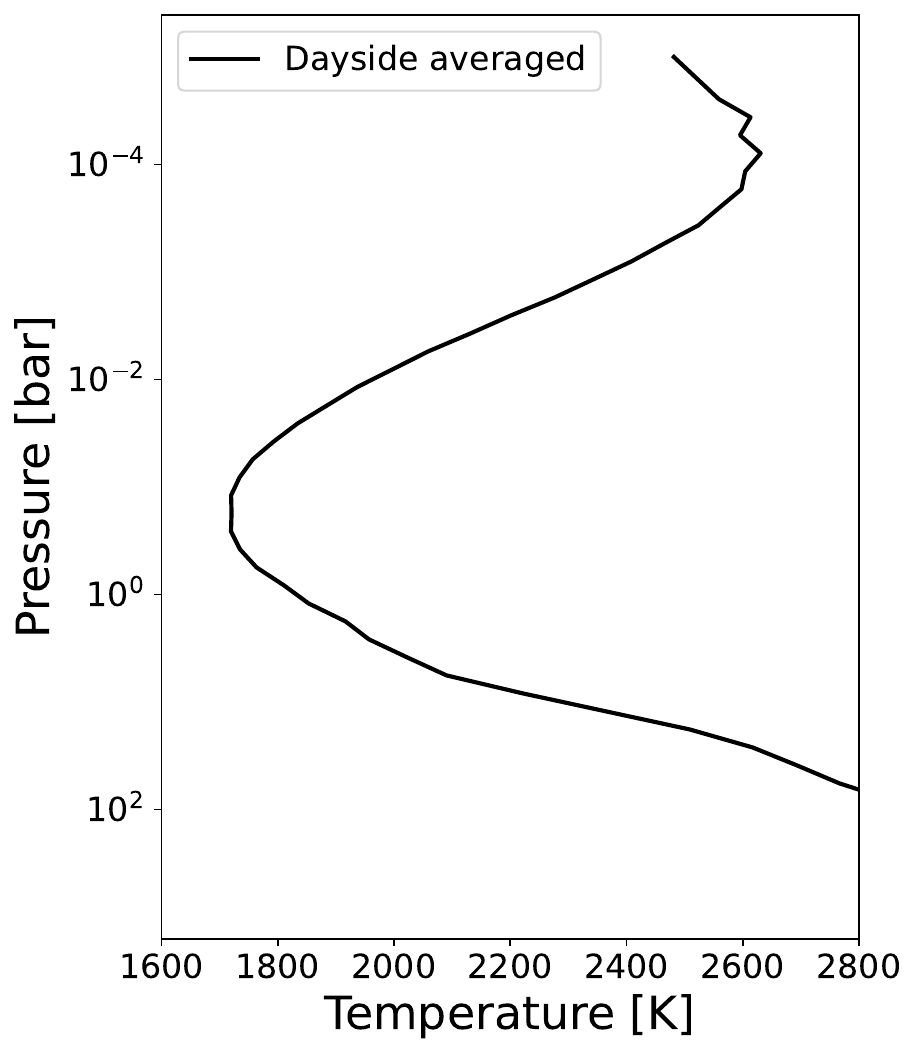}
\caption{As in Fig \ref{3DTempwDragChem}, but with no magnetic drag.}
\label{3DTempwoDragChem}
\end{figure*}

Figures \ref{3DTempwDragChem} and \ref{3DTempwoDragChem} show the dayside-averaged abundances and T-P profile for the GCM with and without drag, respectively. The model \texttt{expeRT/MITgcm} uses the equilibrium-chemistry package of \texttt{petitRADTRANS} \citep{Molliere2019}. Thus, we also use this package to derive the chemical abundances of species on the dayside of KELT-7b. It is evident that in the model with strong magnetic drag, the temperature inversion occurs at deeper pressures compared to the model with no magnetic drag. Furthermore, in both models, partial ionization of the atmosphere is evident in the upper atmosphere ($p < 10^{-2}\,\mathrm{bar}$) as the abundances of H$_2$ drop. Interestingly, the magnetic-drag model exhibits a temperature regime between 1 and 10\,bar, where TiO and VO are less thermally stable, resulting in a local minimum for these two species. Because the region between 1 and 10\,bar in the GCM with drag also represents the coldest part of the dayside atmosphere, the degree of ionization is particularly low there, resulting in a minimum of electron and atomic hydrogen abundances. It should be noted that in the model without drag, the coldest temperature regions lie comparatively higher at $p\approx 10^{-1}\,\mathrm{bar}$. Consequently, a local minimum in electron and atomic H abundances at the same pressure levels can be found. TiO and VO are, however, more thermally stable at these lower pressures. In fact, the TiO and VO abundances do not vary strongly on the dayside in the vertical direction in this drag-free simulation. 

The comparison of temperatures and equilibrium chemistry in the GCMs with and without magnetic drag reveals that there is no direct causal connection between electron abundance minima and TiO and VO abundance minima. The question of whether and where TiO and VO may condense out of the atmosphere in hot Jupiters is, however, an ongoing research question \citep{Roth2024, Beatty2017, Evans2016, Parmentier2013}. In this work, the presence of TiO and VO may be favored due to the relatively high eclipse depths measured in the optical with CHEOPS, TESS, HST, and Spitzer. 

\end{appendix}

\end{document}